\documentclass[preprint,nofootinbib,aps,superscriptaddress,eqsecnum]{revtex4-1}
\usepackage{float}
\usepackage{graphicx}
\usepackage{amsmath}
\usepackage{amsfonts}
\usepackage{amssymb}
\usepackage[bookmarks=true,bookmarksopen=true,
                    bookmarksnumbered=true,
                    colorlinks=true,linkcolor=black,
                    citecolor=red]{hyperref}
\usepackage{subfigure}
\usepackage{slashed}
 \usepackage{setspace} 
\usepackage{ulem}
\usepackage{caption}
\usepackage{color}
\allowdisplaybreaks
\def\bea{\begin{eqnarray}}
\def\eea{\end{eqnarray}}
\def\be{\begin{equation}}
\def\ee{\end{equation}}
\def\nn{\nonumber}

\begin{document}

\title{{Flavon Signatures at the HL-LHC}}

\author{\small M. A. Arroyo-Ure\~na }
\email{marco.arroyo@fcfm.buap.mx}
\affiliation{\small Facultad de Ciencias F\'isico-Matem\'aticas, Benem\'erita Universidad Aut\'onoma de Puebla, C.P. 72570, Puebla, M\'exico,}
\affiliation{\small Centro Interdisciplinario de Investigaci\'on y Ense\~nanza de la Ciencia (CIIEC), Benem\'erita Universidad Aut\'onoma de Puebla, C.P. 72570, Puebla, M\'exico.}

\author{\small Amit Chakraborty }
\email{amit.c@srmap.edu.in }
\affiliation{\small Department of Physics, School of Engineering and Sciences, SRM University AP, Amaravati, Mangalagiri 522240, India.}
 
\author{\small J. Lorenzo D\'iaz-Cruz }
\email{jldiaz@fcfm.buap.mx }
\affiliation{\small Facultad de Ciencias F\'isico-Matem\'aticas, Benem\'erita Universidad Aut\'onoma de Puebla, C.P. 72570, Puebla, M\'exico,}
\affiliation{\small Centro Interdisciplinario de Investigaci\'on y Ense\~nanza de la Ciencia (CIIEC), Benem\'erita Universidad Aut\'onoma de Puebla, C.P. 72570, Puebla, M\'exico.}

\author{\small Dilip Kumar Ghosh }
\email{tpdkg@iacs.res.in} 
\affiliation{\small School of Physical Sciences, Indian Association for 
the Cultivation of Science, 2A $\&$ 2B Raja S.C. Mullick Road, 
Kolkata 700032, India.}

\author{\small Najimuddin Khan }
\email{najimuddinkhan@hri.res.in }
\affiliation{\small Harish-Chandra Research Institute, A CI of Homi 
Bhabha National Institute, Chhatnag Road, Jhunsi, Prayagraj 211019, India.}

\author{\small Stefano Moretti }
\email{s.moretti@soton.ac.uk; stefano.moretti@physics.uu.se }
\affiliation{\small   School of Physics $\&$ Astronomy, University of Southampton, Highfield, Southampton SO17 1BJ, UK }
\affiliation{\small   Department of Physics $\&$ Astronomy, Uppsala University, Box 516, SE-751 20 Uppsala, Sweden.}
\hspace*{-2cm}
\begin{abstract}
{\small 
The detection of a single Higgs boson at the Large Hadron Collider (LHC) has allowed one to probe some properties of it, including the Yukawa and gauge couplings. However, in order to probe the Higgs potential, one has to rely on new production mechanisms, such as Higgs pair production. In this paper, we show that such a channel is also sensitive to the production and decay of a so-called `Flavon' field ($H_F$), a new scalar state that arises in models that attempt to explain the hierarchy of the Standard Model (SM) fermion masses. 
Our analysis also focuses on the other decay channels involving the Flavon particle, specifically the decay of the Flavon to a pair of $Z$ bosons ($H_F \to Z Z$) and the concurrent production of a top quark and charm quark ($H_F\to tc$), having one or more leptons in the final states.
In particular, we show that, with 3000 fb$^{-1}$ of accumulated data at 14 TeV (the Run 3 stage) of the LHC an heavy Flavon $H_F$ with mass $M_{H_F} \simeq 2m_t $ can be explored with $3\sigma -5\sigma $ significance through these channels.}
\end{abstract}

\maketitle

\section{Introduction}
The discovery of a Higgs boson~\cite{ATLAS:2012yve, CMS:2012qbp, Giardino:2013bma} with mass $M_h=125.5$ GeV has provided a firm evidence for the mechanism of Electro-Weak Symmetry Breaking (EWSB) based on a Higgs potential \cite{Sirunyan:2017khh, Sirunyan:2018koj} pointing towards the minimal  realization of it that defines the Standard 
Model (SM). So far, the corresponding studies have relied on the four standard single Higgs production mechanism, i.e.,
gluon-gluon fusion, vector boson fusion, Higgs-strahlung and associated production with top-quark pairs (see Ref. \cite{Kunszt:1996yp}), which have permitted to extract the Higgs boson couplings with quarks ($b$ and $t$), leptons ($\tau$ and $\mu$) and gauge bosons ($W$ and $Z$) as well as the effective interaction with photon and gluon pairs. However, there still remains the task of probing the Higgs self-coupling. Moreover, we still do not understand the origin of the Yuwaka couplings, the flavor coupling. 
{The Higgs boson pair $(hh)$ production serves as a direct means of investigating the 
self-interactions of the Higgs boson, which play a crucial role in determining the Higgs potential 
of the SM. Additionally, the magnitude of the $hh$ production rate is directly proportional to
the square of this self-coupling. Within the SM, the non-resonant production of 
Higgs boson pairs represents the only direct method for measuring the Higgs boson self-coupling. 
Nonetheless, due to the limited size of the cross section, accurately determining this coupling 
presents a significant challenge.}
Next-to-Leading Order effects help somewhat to improve the situation~\cite{Dawson:1998py,Baglio:2018lrj, Baglio:2020ini}. 
The production of SM-like Higgs boson pairs at the LHC provides a valuable avenue to probe
various scenarios Beyond the SM (BSM) that contain particles having couplings with the Higgs boson 
\cite{ATLAS:2017tlw, Baglio:2013toa,Adhikary:2017jtu}. 
These new particles could be (pseudo)scalars, fermions and gauge bosons. Thus di-Higgs production offers
insights into the properties of the Higgs boson itself and can potentially shed  light on the Higgs 
self-interactions as well as its interactions with other particles in the model. For example,
in the dominant production mode for di-Higgs bosons at the LHC,  through fusion of gluons,
mediated by top quark loops that couple to both gluons and Higgs boson. Any additional 
heavy coloured fermions that couple with the Higgs boson can contribute to the di-Higgs 
production mode too. Similarly, in some BSM scenarios, Higgs production can be associated 
with other coloured particles in the loops, such as squarks in Supersymmetry.

%
Studies have shown that, in all generality, in scenarios with an  extended Higgs sector, new heavy resonances, supersymmetric theories, effective field theories with modified top Yukawa coupling, etc., di-Higgs $(hh)$ and di-gauge boson($W^+W^-/ZZ$) production receives additional BSM contributions along with the SM ones~\cite{ATLAS:2020tlo,Baglio:2012np,Barger:2013jfa, Kumar:2014bca,Adhikary:2018ise,Adhikary:2017jtu,Adhikary:2020fqf,Baglio:2014nea,Hespel:2014sla,Lu:2015qqa,Kribs:2012kz,Bian:2016awe,Dawson:2012mk,Pierce:2006dh,Kanemura:2008ub,Ellwanger:2013ova,Chen:2014xra,Liu:2014rba,Goertz:2014qta,Azatov:2015oxa,Dolan:2012ac,Barger:2014taa,Crivellin:2016ihg,Sun:2012zzm,Costa:2015llh,Cheung:2020xij,Alves:2019igs,Englert:2019eyl,Basler:2018dac,Heng:2018kyd,Das:2020ujo,Abouabid:2021yvw,Dasgupta:2021fzw,Huang:2022rne,Li:2019uyy,Cao:2015oaa,Cao:2016zob,Cao:2014kya,Cao:2013si,Lu:2015jza}.  These effects make the study of these two production processes particularly interesting then and, at the same time, also very challenging. However, the possibility to produce  Higgs  {and gauge bosons} pairs in the decay of a new heavy particle that belongs to the spectrum of 
 those models offers some hope to achieve detectable signals at current and future colliders.
{It is also to be noted that  flavor-violating Higgs decays, i.e., those violating the conservation of flavor quantum numbers~\cite{Kopp:2016rzs} can be 
possible. This phenomenon is of great interest as it can provide evidence for BSM physics and shed light on the origin of flavor mixing and hierarchy in the fermion sector. The study of flavor-violating Higgs decays thus offers a unique opportunity to explore new physics and deepen our understanding of fundamental interactions in the universe~\cite{Arganda:2019gnv,Altunkaynak:2015twa,Dorsner:2016wpm}. }

Specifically, we will study  the interactions of the discovered Higgs boson with the so-called `Flavon' field $H_F$ which appears in 
models that attempt to explain the hierarchy of quark and lepton masses using the Froggatt-Nielsen (FN) mechanism~\cite{Davidson:1983fy,Davidson:1981zd,Froggatt:1978nt}.  This mechanism assumes that, above some scale $M_F$ roughly corresponding to the Flavon mass,   there is a symmetry,  
perhaps of Abelian type $U(1)_F$,  with the SM fermions being charged under it, which then forbids the  appearance of Yukawa couplings at the renormalizable level. However,   Yukawa matrices can arise  through  
non-renormalizable operators.  The Higgs spectrum of these models includes a light  $H_F$ state,  which could mix effectively with
the SM Higgs boson when the flavor scale is of the order 1 TeV or lower.
Recently,   the phenomenology  of Higgs vs Flavon interactions at particle colliders has  been the focus of some attention~\cite{Bolanos:2016aik,Bauer:2016rxs,Huitu:2016pwk,  Berger:2014gga,Diaz-Cruz:2014pla,Arroyo-Urena:2018mvl,Arroyo-Urena:2019fyd,Tsumura:2009yf}. In particular,   within this framework, it is possible to have a coupling of this new scalar with Higgs {and gauge bosons}  pairs,
which can then provide interesting signals to be searched for at the LHC. {Another characteristic to highlight is the emergence of Flavor Changing Neutral Currents (FCNCs) mediated by the Flavon, which allows the $H_F\to tc$ decay at tree level. Our study could thus not only serve as a strategy for the Flavon search, but it can also be helpful to assess the order of magnitude of flavor violation mediated by such a particle, which is an indisputable signature of BSM physics.}

{In this paper,  we are interested in studying the detection of the Flavon signal emerging from the production and decay processes $pp \to H_F\to h h~(h\to \gamma \gamma, h\to b \bar{b})$,\, $pp \to H_F\to ZZ~(Z\to  \ell \bar{\ell})$ and the FCNC process $pp\to H_F\to tc\,(t\to \ell\nu_{\ell}b)$ at future stages of the LHC,  namely, Run 3 and  the High-Luminosity LHC (HL-LHC) \cite{Gianotti:2002xx,Apollinari:2015wtw}.} 
%
%
{In this analysis, we do not take into consideration the $pp\to H_F\to hh~(h\to \tau^+\tau^- , 
h\to b\bar{b})$ channel, which has the potential to be competitive with our selected signal. 
We have opted to exclude this channel from the current study and instead reserve it for a 
future publication. Additionally, we will not be presenting other channels such as 
$pp\to H_F\to hh~(h\to b\bar{b})$ and $pp\to H_F\to WW~(W\to \ell\nu_\ell)$ because these channels 
are highly suppressed by large SM backgrounds.}

The ATLAS and CMS collaborations at the LHC have already performed several studies of non-resonant di-Higgs  and di-boson($W/Z$)  production with various possible final states using both Run 1 and the Run 2 dataset. None of these searches have observed a statistically significant excess over the SM background, therefore, upper limits on the di-Higgs production cross section are placed~\cite{ATLAS:2019qdc,ATLAS:2018ili,ATLAS:2018fpd,ATLAS:2018uni,CMS:2017hea,ATLAS:2018hqk,ATLAS:2018dpp,ATLAS:2018rnh,CMS:2018ipl,ATLAS:2019pbo, ATLAS:2023dbw,ATLAS:2023dnm,Zubov:2023bck}.
 {We focus here on the `2 $\gamma$ plus  2 $b$-jets',  `2 pairs of same flavor opposite sign (SFOS) leptons' and `2 jets plus a charged lepton with its neutrino' (with one of the jets labeled as a $b$-jet)  signatures. These particular (and comparatively clean) final 
states are obtained through $pp \to H_F\to h h$, $pp \to H_F\to ZZ$ and $pp\to H_F \to tc$ production followed by $h\to\gamma\gamma$, $h\to b\bar b$, $Z\to \ell\bar{\ell}$ and $t\to \ell\nu_{\ell}b$ decays. We will show that these channels have large 
significances in specific parameter space regions in the context of the LHC operated at $\sqrt{s}=14$ TeV of energy with integrated luminosity  3000 ${\rm fb}^{-1}$. 
Besides these future energies and luminosities, we also present our results based on the data set accumulated to date, i.e., with a 
luminosity of $139~{\rm fb}^{-1}$ at the 13 TeV LHC (Run 2).}

The advocated signature of SM di-Higgs ($hh$) and di-boson ($ZZ$) processes have been explored 
earlier in the literature, albeit in 
different scenarios~\cite{ATLAS:2020tlo,Baglio:2012np,Barger:2013jfa, Kumar:2014bca,Adhikary:2018ise,Adhikary:2017jtu,Adhikary:2020fqf,Baglio:2014nea,Hespel:2014sla,Lu:2015qqa,Kribs:2012kz,Bian:2016awe,Dawson:2012mk,Pierce:2006dh,Kanemura:2008ub,Ellwanger:2013ova,Chen:2014xra,Liu:2014rba,Goertz:2014qta,Azatov:2015oxa,Dolan:2012ac,Barger:2014taa,Crivellin:2016ihg,Sun:2012zzm,Costa:2015llh,Cheung:2020xij,Alves:2019igs,Englert:2019eyl,Basler:2018dac,Heng:2018kyd,Das:2020ujo,Abouabid:2021yvw,Dasgupta:2021fzw,Huang:2022rne},
 {while the processes tackled here, $pp \to H_F\to h h~(h\to \gamma \gamma, h\to b \bar{b})$ 
and $pp \to H_F\to ZZ~(Z\to  \ell \bar{\ell})$ and $pp\to H_F\to tc$ in the context of the present model have not been
discussed in any depth~\cite{Bolanos:2016aik,Bauer:2016rxs,Huitu:2016pwk,  Berger:2014gga,Diaz-Cruz:2014pla,Arroyo-Urena:2018mvl}.}
Our analysis of these final states give promising results as a 
discovery channel for a heavy CP-even $H_F$ boson in the aforementioned FN framework. 
In order to prove this, we first choose three sets of reference points for three heavy Higgs masses 
$800,900$ and $1000$ GeV. A signal region 
(a set of different kinematic cuts) is then defined to maximize signal 
significances in the presence the SM backgrounds having the same final state.
In our cut-based analysis, we further use the same signal region for 
different combinations of the 
singlet scalar Vacuum Expectation Value (VEV)  $v_s$ and  heavy Higgs mass $M_{H_F}$ to compute the 
signal significances. The latter are  only  
mildly affected (at the $5-10\%$ level) by incorporating a realistic $5\%$ 
systematic uncertainty in the SM background estimation.
We find a large number of signal events that have 
significances exceeding $2\sigma$ and  they can be explored with $3000~{\rm fb}^{-1}$ of data at LHC runs using $\sqrt{s} = 14 $ TeV.

The rest of the paper is organized as follows. In  sec.~\ref{se:model}, we present the details of the model  and derive expressions for the masses and relevant interaction couplings for all the particles.
Afterwards, we introduce the constraints acting on it from both the theoretical and experimental side in sec.~\ref{se:conts}. Sec. \ref{se:col_an} is focused on the analysis of the signals arising from the decay of the Flavon.Finally, we conclude in sec.~\ref{se:concl}.

\section{The model} 
\label{se:model}
We now focus on some relevant theoretical aspects of what we will refer to as the  FN singlet Model (FNSM). In Ref.~\cite{Bonilla:2014xba},  a comprehensive theoretical analysis of the Higgs potential therein  is presented  along with the constraints on the parameter space from the Higgs boson signal strengths and the oblique parameters,  including presenting a few benchmark scenarios amenable to phenomenological investigation. 
(See Ref.~\cite{ Barradas-Guevara:2017ewn} for the effects of Lepton Flavor Violation (LFV).)

\subsection{The scalar sector} 
The scalar sector of this model consists of the SM Higgs doublet $\Phi$ ane and 
one SM singlet complex FN scalar $S_F$. 
In the unitary gauge, we parameterize these fields as: 
\begin{eqnarray} 
	& \Phi = \left( \begin{array}{  c} 0 \\ \frac{  v + \phi^0}{\sqrt 2}\\
	\end{array}  \right), \label{dec_doublets}&\\ 
	& S_F = \frac {(v_s + S_R + i S_I )}{ \sqrt 2 }   , \label{dec_Singlet} &
\end{eqnarray}
where $v$ and $v_s$ represent the VEVs of the SM Higgs doublet and 
 FN  singlet, respectively. The scalar potential should be 
invariant under the FN $U(1)_F$ flavor symmetry. Under this symmetry, the SM Higgs doublet $H$ and  FN 
singlet  $S_F$ transform as $\Phi \to \Phi $ and $S_F \to e^{i\theta }
S_F$, respectively.

In general, such a scalar potential admits a complex VEV, 
$\langle S_F\rangle_0=\frac{v_s}{\sqrt{2}}e^ {i\xi} $, but in this work we 
consider the special case in which the Higgs potential is CP-conserving, 
{by setting the phase $\xi = 0$.}
Such a CP-conserving Higgs potential is then given by:
\begin{eqnarray} \label{potential} 
	V_0=-\frac {1}{2} m_1^2\Phi^ \dagger \Phi-\frac{1}{2} m_{2}^2 
S_F^*S_F +\frac {1}{2} \lambda_1 \left(\Phi^ \dagger \Phi\right)^2+\lambda_2
\left(S_F^*S_F\right)^2
	+\lambda_ {3} \left(\Phi^ \dagger \Phi\right)\left(S_F^* S_F\right). 
\end{eqnarray}
The $U(1)_F$ flavor symmetry of this scalar potential is 
spontaneously broken by the VEVs of the spin-0 fields $(\Phi, S_F)$ and this
leads to a massless Goldstone boson in the physical spectrum. In order to
give a mass to it, we add the following soft $U(1)_F$ breaking term to the
potential: 
\begin{eqnarray}
V_{\rm soft} = -\frac{m_3^2}{2} \left (S_F^{2} + S_F^{*2} \right).  
\end{eqnarray}
The full scalar potential is thus:
\begin{eqnarray}
V = V_0 + V_{\rm soft}. 
\end{eqnarray}
The presence of the $\lambda_3$ term allows mixing between the
Flavon and the Higgs fields after both the $U(1)_F$ flavor and EW symmetry
breaking and contributes to the mass parameters for both the Flavon and Higgs field, as
can be seen below. The soft $U(1)_F$ flavor symmetry breaking term 
$V_{\rm soft}$ is responsible for the pseudoscalar Flavon $(S_I)$ mass.
Once the minimization conditions for the potential $V$ are applied, we obtain 
the following relations between the parameters of $V$:
\begin{eqnarray} 
	m_{1} ^2  &=&  v^2 \lambda_1 + v_s^2 \lambda_{3},   \\
	m_{2} ^2 &=& -2 m^2_{3} + 2 v_s^2 \lambda_2 + v^2 \lambda_{3}.
\end{eqnarray}
All the parameters of the scalar potential are real and therefore the real 
and imaginary parts of $V$ do not mix. The CP-even mass matrix can be 
written in the $(\phi_0, S_R)$ basis as:
\begin{equation} 
	M^2_S =
	\left( \begin{array}{cc} 
		\lambda_1 v^2      &  \lambda_{3} v v_s \\
		\lambda_{3}v v_s   &  2 \lambda_2 v_s^2
	\end{array}  \right).
\end{equation} 
The corresponding mass eigenstates are obtained via the standard $2\times 2$ 
rotation:
\begin{eqnarray} 
	\phi^0   &=& \  \  \  \  \cos \  \alpha \  h + \sin \  \alpha  \  H_F,   \\
	S_R    &=& -\sin \  \alpha \  h + \cos \  \alpha \  H_F, 
\end{eqnarray}
with $\alpha$ a mixing angle. Here $h$ is identified with the SM-like Higgs 
boson  with mass $M_h$=125.5 GeV whereas the mass eigenstate $H_F$ is the 
CP-even Flavon. The corresponding CP-odd Flavon
$A_F \equiv S_I$ will have a mass such that $M^2_{A_F} = 2m_3^2 $. 
Both $H_F$ and $A_F$ are considered to be  heavier than $h$.
In this model, we will work with the mixing angle $\alpha$ and physical masses $M_h, M_{H_F}$ and $M_{A_F}$, which are related to 
the quartic couplings of the scalar potential in Eq.~(\ref{potential}) as follows:
\begin{eqnarray}\label{eq:relate}
\lambda_1&=& \frac{ \cos\alpha^2 M_h^2+\sin\alpha^2 M_{{H_F}}^2}{v^2},\nn\\
\lambda_2&=& \frac{M_{{A_F}}^2+{\cos\alpha }^2 M_{{H_F}}^2+{\sin\alpha }^2 M_h^2}{2 v_s^2},\\
\lambda_3&=& \frac{ \cos\alpha \, \sin\alpha }{ v v_s} \, ( M_{{H_F}}^2 -  M_h^2).\nn
\end{eqnarray} 
We consider  the mixing angle $\alpha$, the FN singlet VEV $v_s$ and its (pseudo)scalar field masses $M_{H_F,A_F}$ as free parameters in this work.

\subsection{The Yukawa sector} 
The effective $U(1)_{F}$ invariant Yukawa Lagrangian, 
\'{a} la FN, is given by
\cite{ Froggatt:1978nt}:
\begin{align} 
\mathcal{L}_ Y &= \rho^d_{ ij } \left( \frac{S_F}{\Lambda_F} 
\right)^{q_{ ij }^d}  \bar{Q}_i d_j  \tilde \Phi 
+ \rho^u_{ ij } \left(\frac{S_F}{\Lambda_F }\right)^{q_{ij }^u}\bar{Q}_i u_j 
\Phi \nonumber\\&+ \rho^\ell_{ij}\left(\frac{S_F}{\Lambda_F}\right)^{q_{ij}^l}
\bar{L}_i \ell_j \Phi  + \rm h.c.,  
\end{align} 
where $\rho^{u/d/\ell}$ are dimensionless couplings seemingly of order one. 
This will lead to  Yukawa couplings once the $U(1)_F$ flavor 
symmetry is spontaneously broken. The integers 
$q_{ij}^f$ $(f=u,\, d ,\, \ell)$ are the combination of $U(1)_{F}$ 
charges of the respective fermions.
In order to generate the Yukawa couplings, one spontaneously 
breaks both the $U(1)_{F}$ and EW symmetries. In the unitary 
gauge one can make the following first order expansion of the neutral 
component of the heavy Flavon field $S_F$ around its VEV $v_s$:
\begin{align} 
	\Bigg(\frac{S_F}{\Lambda_F}\Bigg)^{q_{ ij }} &=\left(\frac{v_s+S_R+iS_I}{  \sqrt 2\Lambda_F}  \right)^{q_{ ij }} \nonumber\\&
	\simeq \left(\frac{v_s}{ \sqrt 2\Lambda_F}  \right)^{q_{ ij }} \left[1+q_{ ij }\left(\frac{S_R+iS_I}{v_s}\right)\right],
\end{align}
which leads to the  following fermion couplings after replacing the  mass 
eigenstates in $\mathcal{L}_Y$:
\begin{eqnarray} \label{Yukalagrangian} 
	\mathcal {L}  _Y &=& \frac 1  v [\bar{U}  M^u U+\bar{D}  M^d D+\bar{L} M^ \ell L](c_ \alpha h+s_ \alpha H_F) \nonumber\\
	&+&\frac{v }{ \sqrt 2 v_s } [\bar{U}_i\tilde Z_{ij}^u U_j+\bar{D}_i\tilde Z_{ij}^d D_j+\bar{L}_i\tilde Z_{ij}^ \ell  L_j]\nonumber\\&\times&
	(-s_ \alpha h+c_ \alpha H_F+iA_F)+ \rm h.c.,  
\end{eqnarray} 
where we define $\sin\alpha \equiv  s_\alpha$ {and} $\cos\alpha\equiv  c_\alpha$. Here, $M^f$ stands for the diagonal fermion mass matrix while the intensities of the Higgs-Flavon couplings are encapsulated in the $\tilde{Z}_{ij}^f=U_L^f Z_{ij}^f U_L^{f\dagger}$ matrices. In the flavor basis,   the $Z_{ij}^f$ matrix elements are given by:
\begin{equation}
Z_{ij}^f= \rho_{ij}^f \left(\frac{v_s}{\sqrt 2\Lambda_F}
\right)^{q_{ij}^f}q_{ij}^f,
\end{equation}
which remains non-diagonal even after diagonalizing the mass matrices, thereby giving rise to FV scalar couplings.
In addition to the Yukawa couplings  we  also need  the $\phi VV$ ($V=W, Z$)  couplings for our calculation  which can be extracted from the kinetic terms of the Higgs doublet and  complex singlet.
In Tab.~\ref{couplings}  we show the coupling constants for the interactions of  the SM-like Higgs boson and the Flavon to  fermions and gauge bosons.
\begin{table}[t!]   
	\begin{centering} 
		\begin{tabular}{ |c|c|} 
			\hline
			Vertex ($\phi XX$) &Coupling constant ($g_ \phi XX $) \tabularnewline
			\hline
			$ hf_ i \bar{f} _ j  $ & $\frac{ c_ \alpha}{   v} \tilde M _{ ij }^ f -s_ \alpha r_ s \tilde Z ^ f _{ ij }$\tabularnewline
			$ H_ F f_ i \bar{f }_ j  $ & $\frac{ s_ \alpha }{  v} \tilde M _{ ij }^ f +c_ \alpha r_ s \tilde Z ^ {f _{ ij }}$\tabularnewline
			$ A_ F f_ i \bar{f} _ j  $ & $i \, r_ s \tilde Z ^{ f _{ ij }}$\tabularnewline
			$ hZZ $ & $i\,\frac{ g M_Z}{ c_W}      c_\alpha $\tabularnewline
			$ hWW $ & $i\,g M_W  c_\alpha $\tabularnewline
			$ H_ F ZZ $ & $i\,\frac{ g M_Z}{ c_W}    s_\alpha $\tabularnewline
			$ H_ F WW $ & $i\,g M_W  s_\alpha $\tabularnewline
			$ H_ F h h $ & $-i \{c_\alpha^3 \lambda_{3}v_s +c_\alpha^2 s_\alpha  v  (3 \lambda_{1}-2 \lambda_{3})-2 c_\alpha s_\alpha^2 v_s  (\lambda_{3}-3 \lambda_{2})+\lambda_{3} s_\alpha^3  v \} $\tabularnewline
			  & $\equiv -i\, \{c_\alpha s_\alpha (3 M_{A_F}^2 s_\alpha v + (M_{H_F}^2 + 2 M_h^2) (s_\alpha v + 
      c_\alpha v_s))\}/(v v_s)$\tabularnewline
			$ A_ F h h $ & $0 $\tabularnewline
			$ A_ F ZZ $ & $0 $\tabularnewline
			\hline
		\end{tabular}
		\caption{Tree-level couplings of the SM-like Higgs boson $h$ and the Flavons $H_F$ and $A_F$ to fermion and gauge boson pairs in the FNSM. 
Here, $r_s=v/\sqrt 2 v_s$.}
		\label{couplings}
		\par\end{centering} 
\end{table} 

\begin{figure}[h!]
	\begin{center}
		\subfigure[]{\includegraphics[scale=0.5]{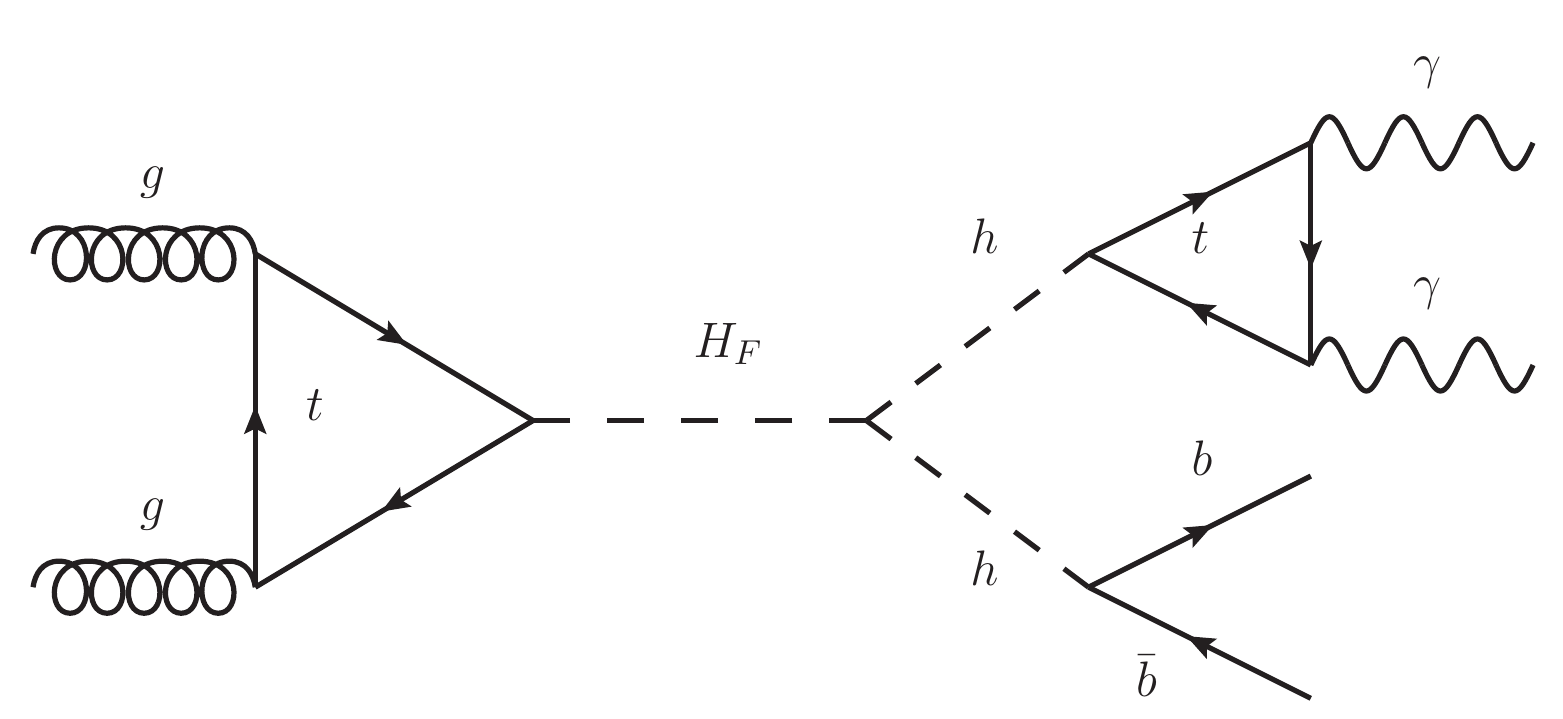}\label{hh}} 
		\subfigure[]{\includegraphics[scale=0.5]{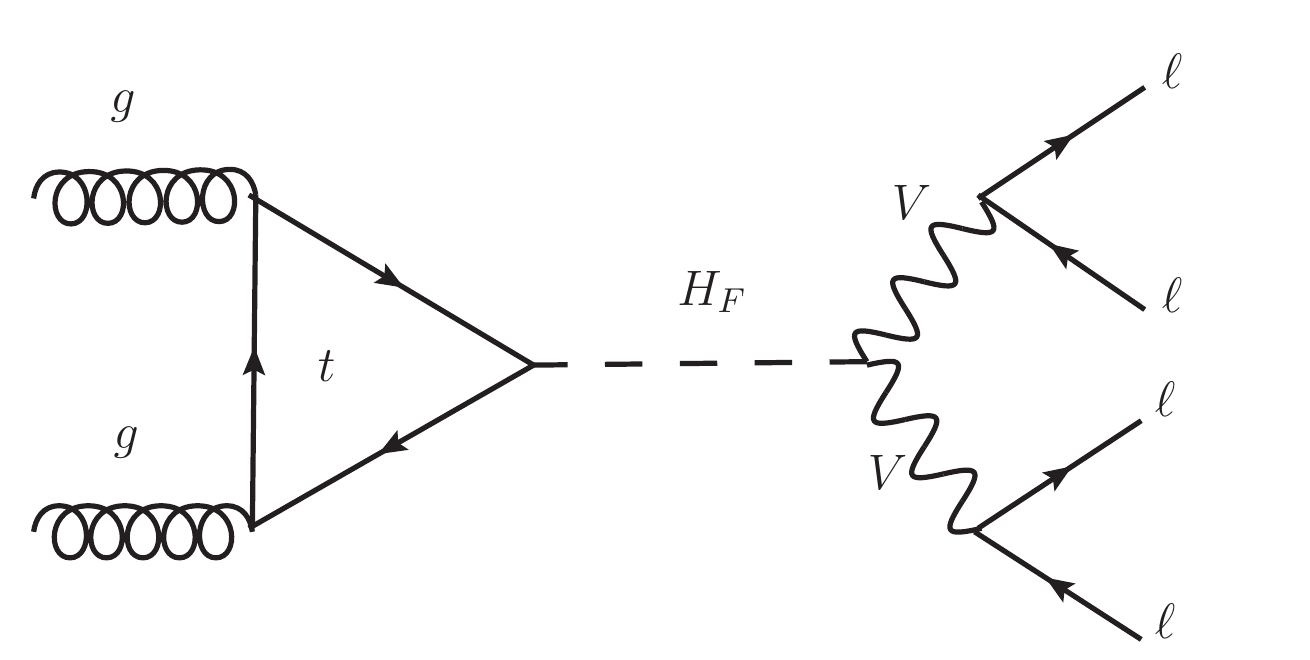}\label{ZZ}}
		\subfigure[]{\includegraphics[scale=0.5]{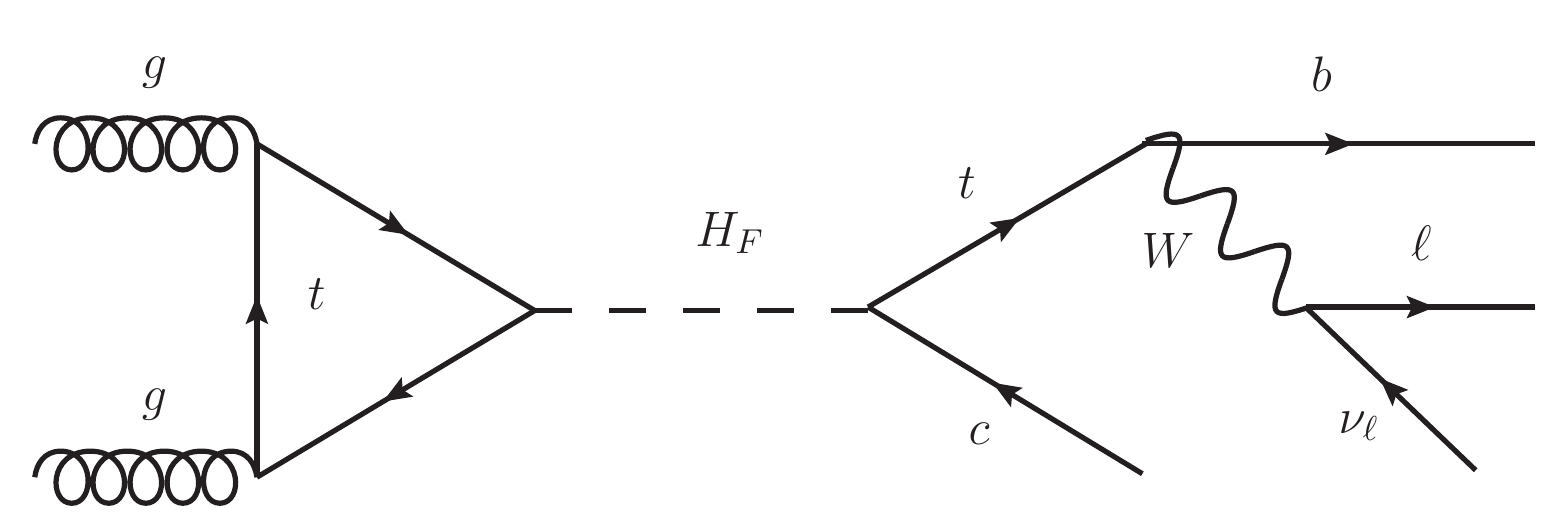}\label{tc}}
	\end{center}
	\caption{Representative Feynman diagram of the signal: (a) $g g \to H_F\to hh~(h\to b\bar{b}, h\to\gamma\gamma)$, (b)  $g g \to H_F\to ZZ~(Z\to \ell\bar{\ell})$ and (c) $pp\to H_F\to tc\,(t\to \ell\nu_{\ell}b)$.}
	\label{FeynDiag}
\end{figure}

\section{Constraints on the FNSM parameter space}
\label{se:conts}
In order to perform a realistic numerical analysis of the signals analyzed in this work, i.e., $pp\to H_F\to hh~(h\to b\bar{b}, h\to\gamma\gamma)$, $pp\to H_F\to ZZ\,(Z\to\ell \bar{\ell})$ and $pp\to H_F\to tc\,(t\to \ell\nu_{\ell}b)$  (see Fig. \ref{FeynDiag}),
we need to constrain the free FNSM parameters, i.e.: $(i)$ the mixing angle $\alpha$ of the real components of the doublet $\Phi$ and the FN singlet $S$, $(ii)$ FN singlet VEV $v_s$,  $(iii)$ the heavy scalar(pseudo) field masses $M_{H_F,A_F}$, $(iv)$ the diagonal $\tilde{Z}_{33}^u\equiv\tilde{Z}_{tt}$, $\tilde{Z}_{22}^u\equiv\tilde{Z}_{bb}$ and the non-diagonal $\tilde{Z}_{32}^u\equiv\tilde{Z}_{tc}$ matrix elements which will be used to evaluate both the production cross section of the Flavon $H_F$ and the decay of the Higgs boson to a pair of $b$ quarks; all of which have an impact on the upcoming calculations.
These parameters are constrained by various kinds of theoretical bounds like absolute vacuum stability, triviality, perturbativity and unitarity of scattering matrices and different experimental data, chiefly, LHC Higgs boson coupling modifiers, null results for additional Higgs states plus the muon and electron anomalous magnetic (dipole) moments $\Delta a_{\mu}$ and $\Delta a_{e}$, respectively. The various LFV processes $\tau\to 3\mu$, $\mu\to 3e$, $\tau\to \mu\gamma$, $\mu\to\ e\gamma$, $B_s^0\to\mu^+\mu^-$ and the total decay width of the Higgs boson 
($\Gamma_T^h$) are also modified in the presence of these
new Yukawa couplings, so they have also been tested against available data.
In the following, we discuss the various constraints on the model parameters in turn.

\subsection{Stability of the scalar potential}
The absolute stability of the scalar potential in Eq.~(\ref{potential}) requires that the potential should not become unbounded from below, i.e., it should not approach negative infinity along any direction  of the field space ($h,H_F,A_F$) at large field values.
Since in this limit the quadratic terms in the scalar potential are negligibly small as compared to the quartic terms, the absolute stability conditions are \cite{Khan:2014kba}:
\begin{equation}
\lambda_1(\Lambda) > 0, \quad \lambda_2(\Lambda) > 0 \quad {\rm and} \quad \lambda_3(\Lambda) + \sqrt{2 \lambda_1(\Lambda) \lambda_2(\Lambda)} > 0,
\end{equation}
wherein these quartic couplings are evaluated at a scale $\Lambda$ using Renormalization Group Evolution (RGE) equations. If the the scalar potential in Eq.~(\ref{potential}) has a metastable EW vacuum, then these conditions are modified \cite{Khan:2014kba}.
One can then use Eq. \eqref{eq:relate} to translate these limits into those on the free parameters such as scalar fields' mass and mixing angles.

\subsection{Perturbativity and unitarity constraints}

\begin{figure}[!t]
	\begin{center}
		\includegraphics[scale=0.155]{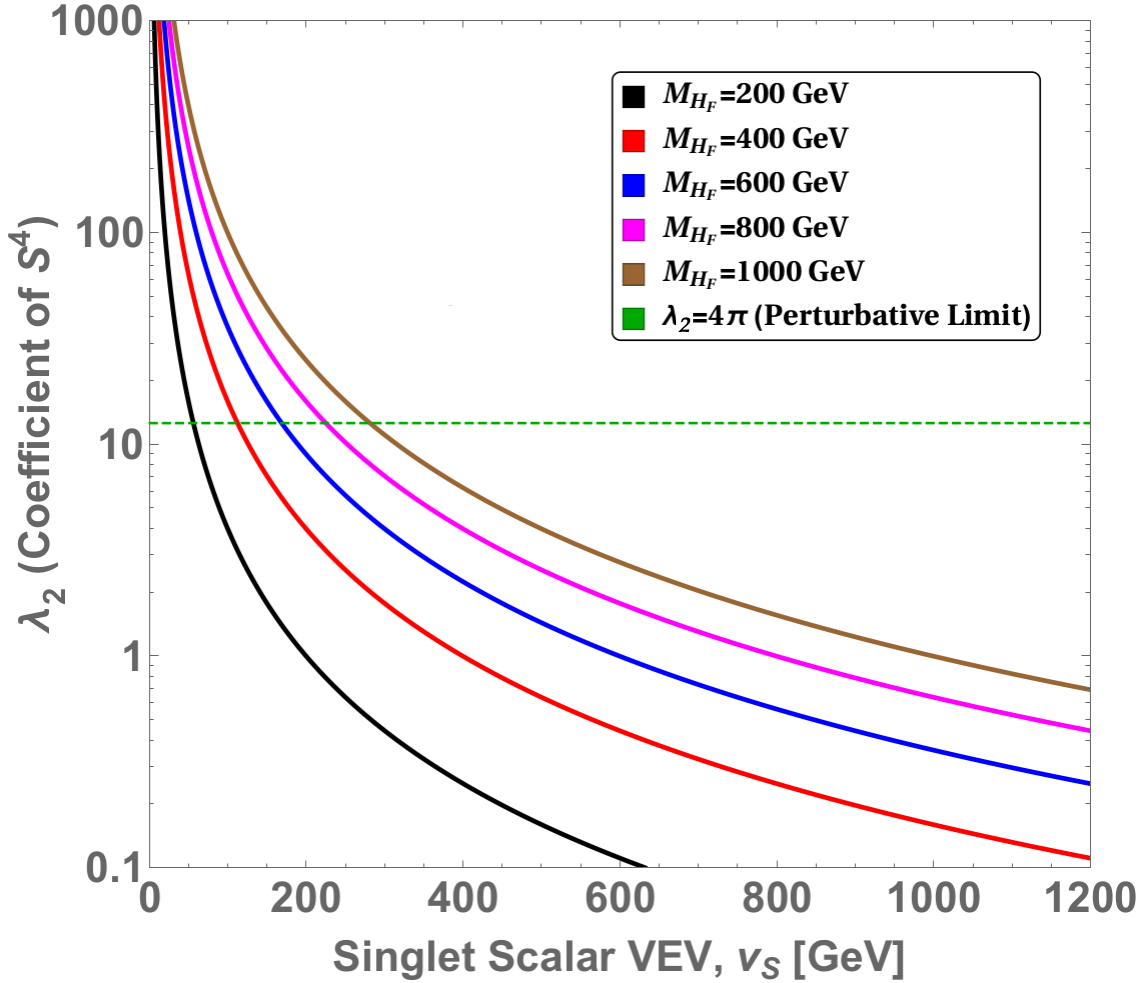}
		\includegraphics[scale=0.155]{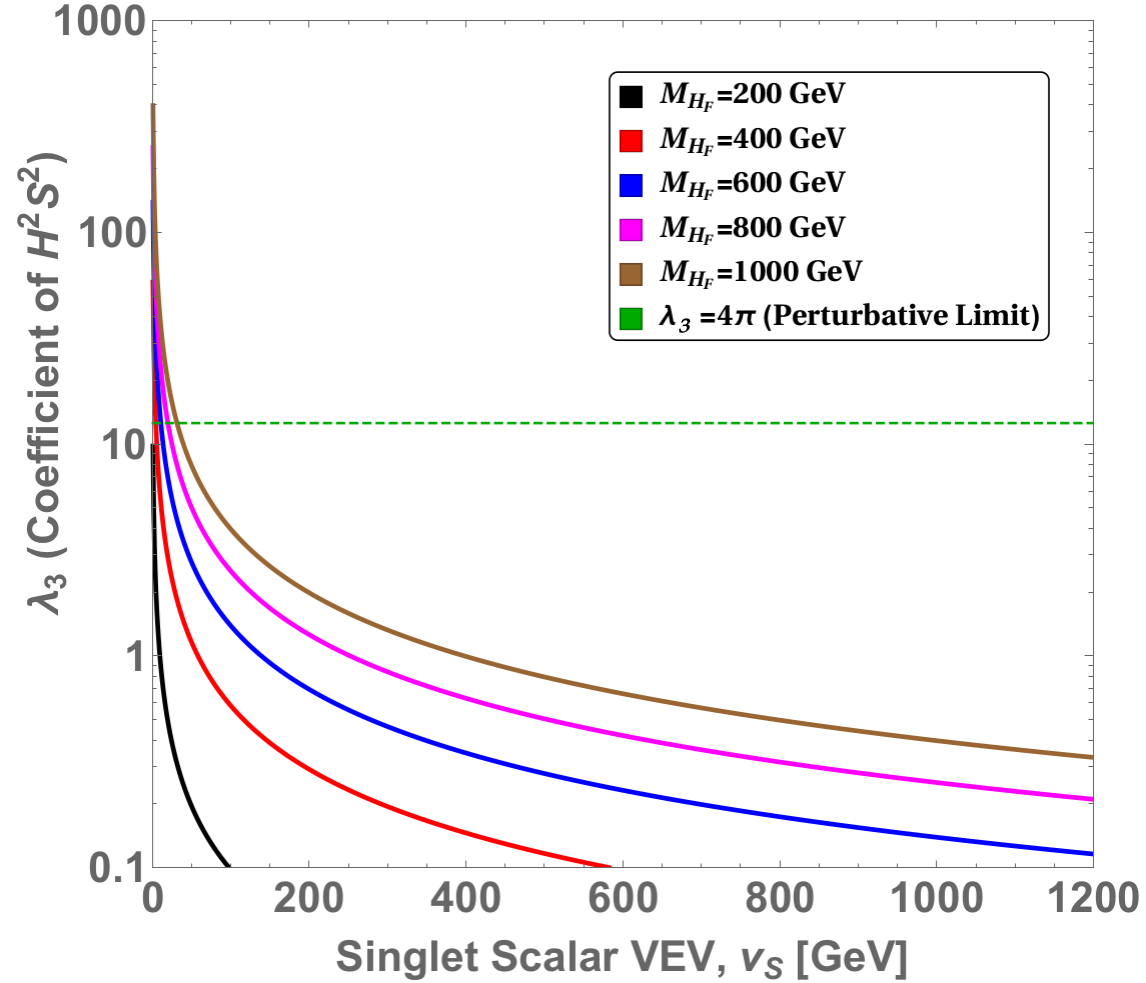}
		\includegraphics[scale=0.155]{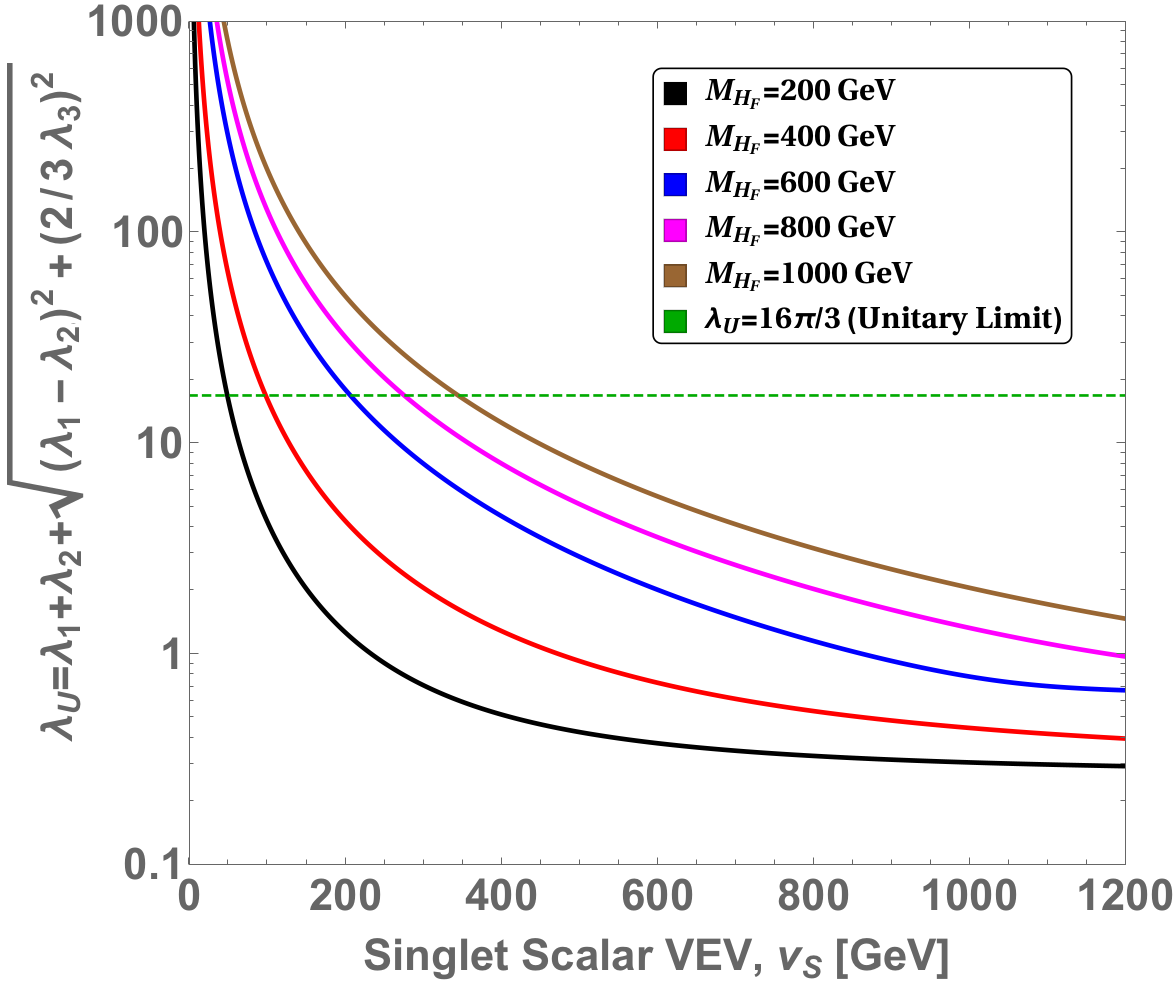}
	\end{center}
	\caption{ In the first two plots we show the perturbative bounds on the quartic couplings $\lambda_{2,3}$ while the third plot shows the stringent unitary bounds on $\lambda_U$.}
	\label{PLimit}
\end{figure}

To ensure that the radiatively improved scalar potential of the FNSM remains perturbative at any given energy scale, one must impose the following upper bounds on the quartic couplings:
\begin{equation}
\mid \lambda_1(\Lambda), \lambda_2(\Lambda), \lambda_3(\Lambda)\mid \leq 4 \pi.
\end{equation}

The quartic couplings in the scalar potential of our scenario are also severely constrained by the unitarity of the Scattering matrix ($S$-matrix). At very large field values, one can get the $S$-matrix by using various (pseudo)scalar-(pseudo)scalar, gauge boson-gauge boson and (pseudo)scalar-gauge boson interactions in $2\to2$ body processes. The unitarity of the $S$-matrix demands that the eigenvalues of it  should be less than $8\pi$ \cite{Cynolter:2004cq,Khan:2014kba}. In the FNSM, the unitary bounds are obtained from the $S$-matrix (using the
equivalence theorem) as:
\begin{eqnarray}
\lambda_1 (\Lambda) \leq 16 \pi  \quad {\rm and} \quad \Big| {\lambda_1(\Lambda)}+{\lambda_2(\Lambda)} \pm \sqrt{ ({\lambda}_1(\Lambda)-{\lambda_2}(\Lambda))^2+(2/3 \lambda_3(\Lambda))^2}\Big| \leq 16/3 \pi.
\end{eqnarray}
We now use the relation in Eq.~(\ref{eq:relate}) to display theoretical  bounds on the scalar singlet VEV $v_s$ for various values of the heavy Higgs masses, $M_{H_F}$ and $M_{A_F}$.
In Fig.~\ref{PLimit} we display the constraints on 
scalar quartic couplings coming from the perturbativity (Fig.~\ref{PLimit}(left) \& (middle)) 
and unitarity (Fig.~\ref{PLimit}(right)) of the $S$-matrix.
Here, we assume $M_{H_F}=M_{A_F}$ and  $\cos\alpha=0.995$, which agrees with the constraints from the Higgs boson coupling modifiers from the LHC measurements, which 
we will discuss in some detail  later.
Fig.~\ref{PLimit}(left) shows the $v_s-\lambda_{2}$ plane for $M_{H_F}=200,\,400,\,600,\,800$ and $1000$ GeV whereas in 
Fig.~\ref{PLimit}(middle) the $v_s-\lambda_{3}$ plane is presented.
The plane $v_s-\lambda_U(\equiv  {\lambda_1}+{\lambda_2} + \sqrt{ ({\lambda}_1-{\lambda_2})^2+(2/3 \lambda_3)^2})$ in 
Fig.~\ref{PLimit}(right) shows the unitary bounds.
We find that $|\lambda_U| \leq 16\pi/3 $ is the most stringent upper bound for the scalar quartic couplings. From these plots, we can see that
 the lower limit on the scalar singlet VEV $v_s$ is, for $M_{H_F}= (200,\,400,\,600,\,800,\,1000)$ GeV,  $v_s\geq (69,\,
138,\,207,\,276,\,345)$ GeV. Note that we are working at the EW scale only, as detailed RGE analysis is beyond the scope of this work.
We also choose the parameters in such a way that the scalar potential remains absolutely stable in all the directions of the scalar fields $h,\,H_H,\,A_F$. (Further details can be found in Ref.~\cite{Khan:2014kba}.)

\subsection{Experimental constraints}
\begin{figure}[htb!]
	\begin{center}
		\subfigure[]{\includegraphics[scale=0.26]{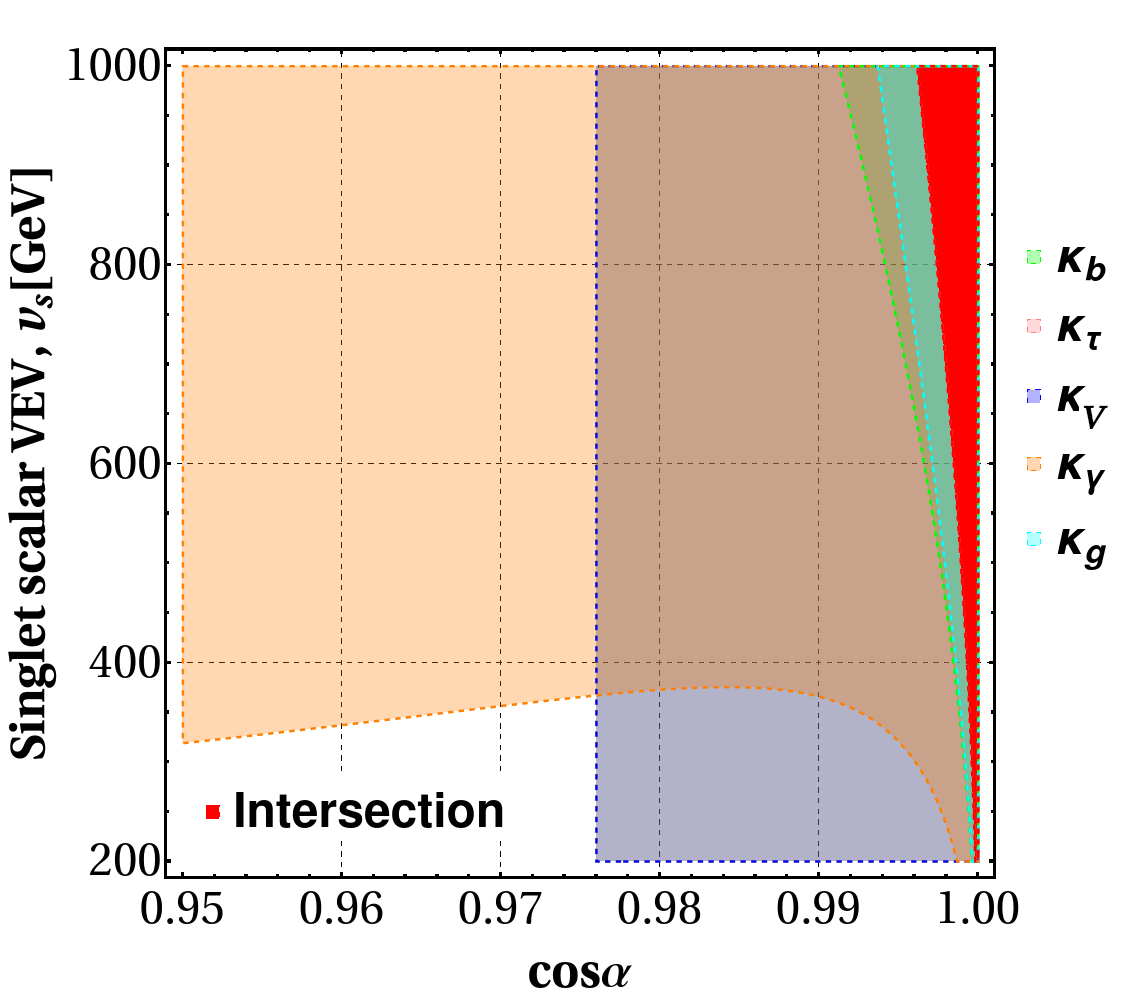}\label{fig:kappas}
}
		\subfigure[]{\includegraphics[scale=0.245]{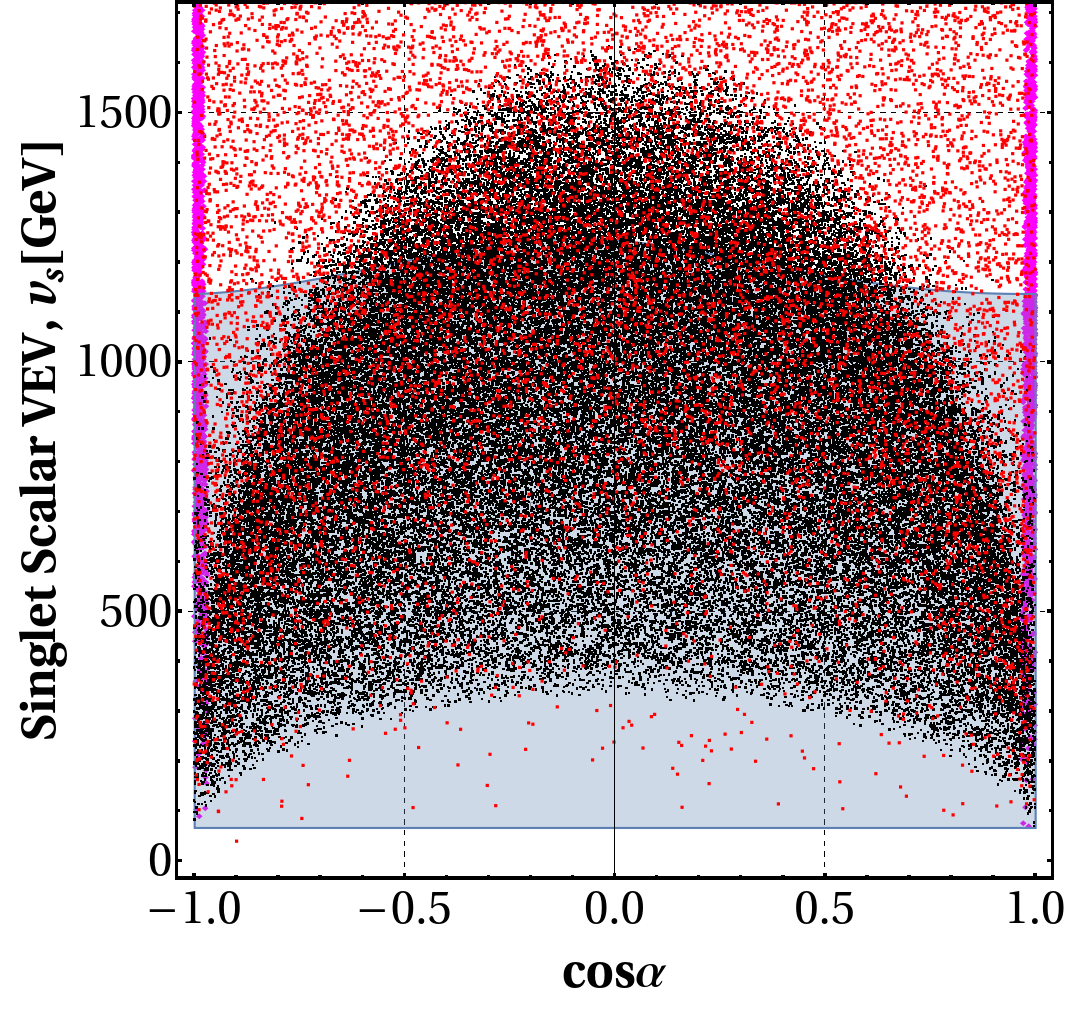}\label{fig:LFVp}
}
	\end{center}
\caption{VEV of the FN singlet $v_s$ as a function of  the 
cosine of the mixing angle $\alpha$: constraints are from (a) the SM-like Higgs boson coupling modifiers and (b) flavor observables (as described in the text).}
\end{figure}
To constrain the mixing angle $\alpha$ and the VEV of the FN singlet $v_s$, we use HL-LHC projections for the Higgs boson coupling modifiers $\kappa_i$ at a CL of $2\sigma$ \cite{Cepeda:2019klc}, as this machine configuration is the one with highest sensitivity among those we will consider
in the analysis section. For a production cross section $\sigma(pp\to \phi)$ or a decay width $\phi\to X$ ($\phi=h,\,h^{\text{SM}}$), we introduce:
\begin{equation}
	\kappa_{pp}^2=\frac{\sigma(pp\to h)}{\sigma(pp\to h^{\text{SM}})},\,\,\,\,\,\,\kappa_X^2=\frac{\Gamma(h\to X)}{\Gamma(h^\text{SM}\to X)},
\end{equation} 
where $X=b\bar{b},\,\tau^-\tau^+,\,W^-W^+,\,ZZ,\,\gamma\gamma$. 
Fig. \ref{fig:kappas} shows all the regions complying with the aforementioned 
projections  for each  channel in the 
$\cos\alpha-v_s$ plane: here,  the green, pink, blue, orange and cyan area 
corresponds to $\kappa_b$, $\kappa_{\tau},\, \kappa_V,\,\kappa_{\gamma}$ and
$\kappa_g$, respectively,  while the red area represents  the intersection of all the 
 areas allowed by all the individual channels. We consider 
$\tilde{Z}_{bb}=0.01$ and $\tilde{Z}_{tt}=0.4$ in the evaluations 
for the $\kappa_X$.  Such 
values are well motivated because they simultaneously accommodate all the $\kappa_X$'s.  In fact, values in the $0.01\leq \tilde{Z}_{bb}\leq 0.1$ and $0.1\leq \tilde{Z}_{tt}\leq 1$ intervals have no important impact on the coupling  modifiers, however, in the case when $\tilde{Z}_{bb} \geq 0.1$ and $\tilde{Z}_{tt}\geq 2$, a large reduction of allowed values in the $\cos\alpha-v_s$ plane is found\cite{Arroyo-Urena:2018mvl,Arroyo-Urena:2019fyd}.

\begin{figure}[htb!]
	\begin{center}
		\includegraphics[scale=0.22]{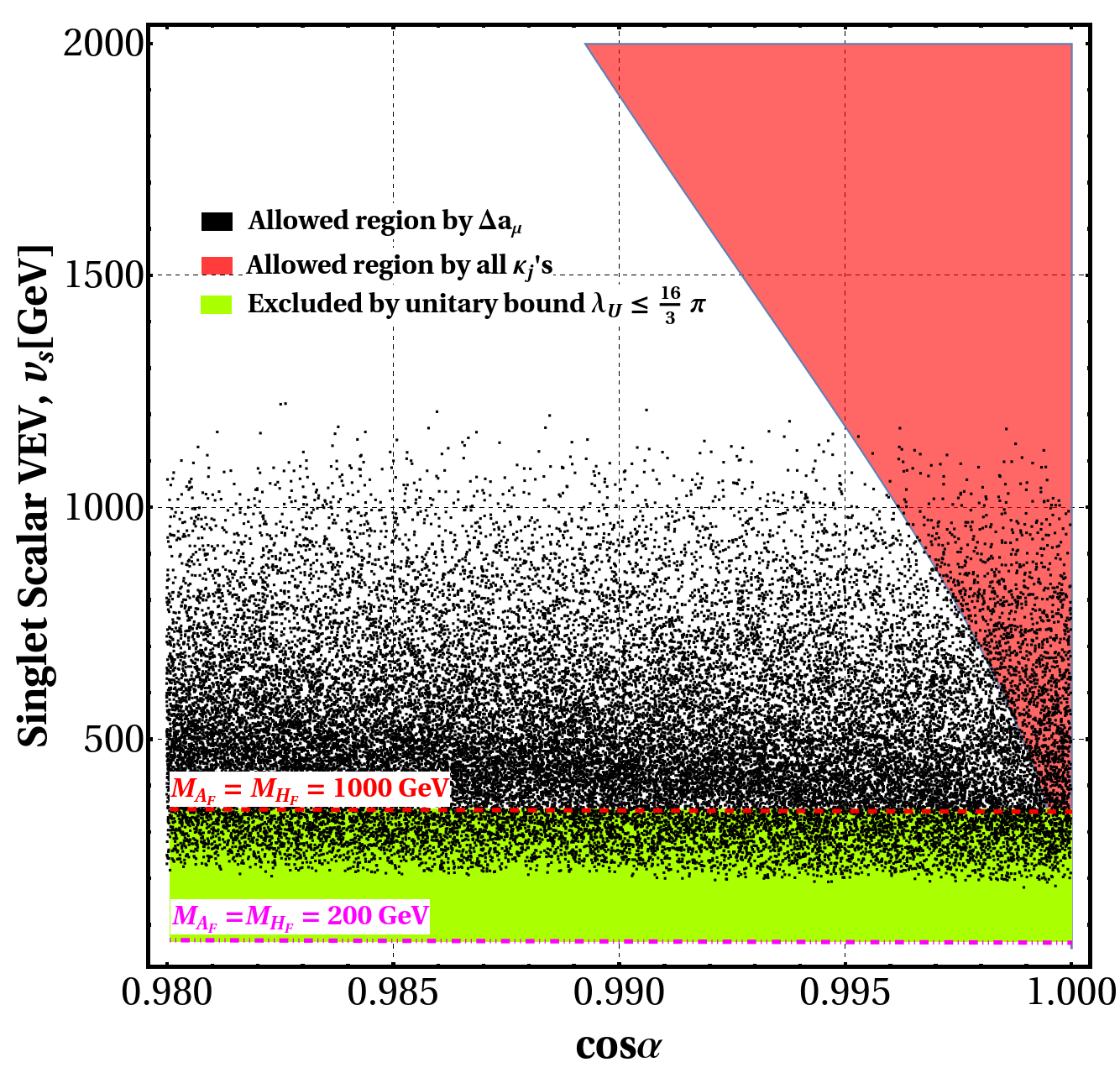}
	\end{center}
	\caption{VEV of the FN singlet $v_s$ as a function of cosine of the mixing angle $\alpha$ in the presence of the most stringent ones among all 
theoretical and experimental constraints considered.}
\label{fig:intersection}
\end{figure}

\begin{figure}[t!]
	\begin{center}
		\includegraphics[scale=0.25]{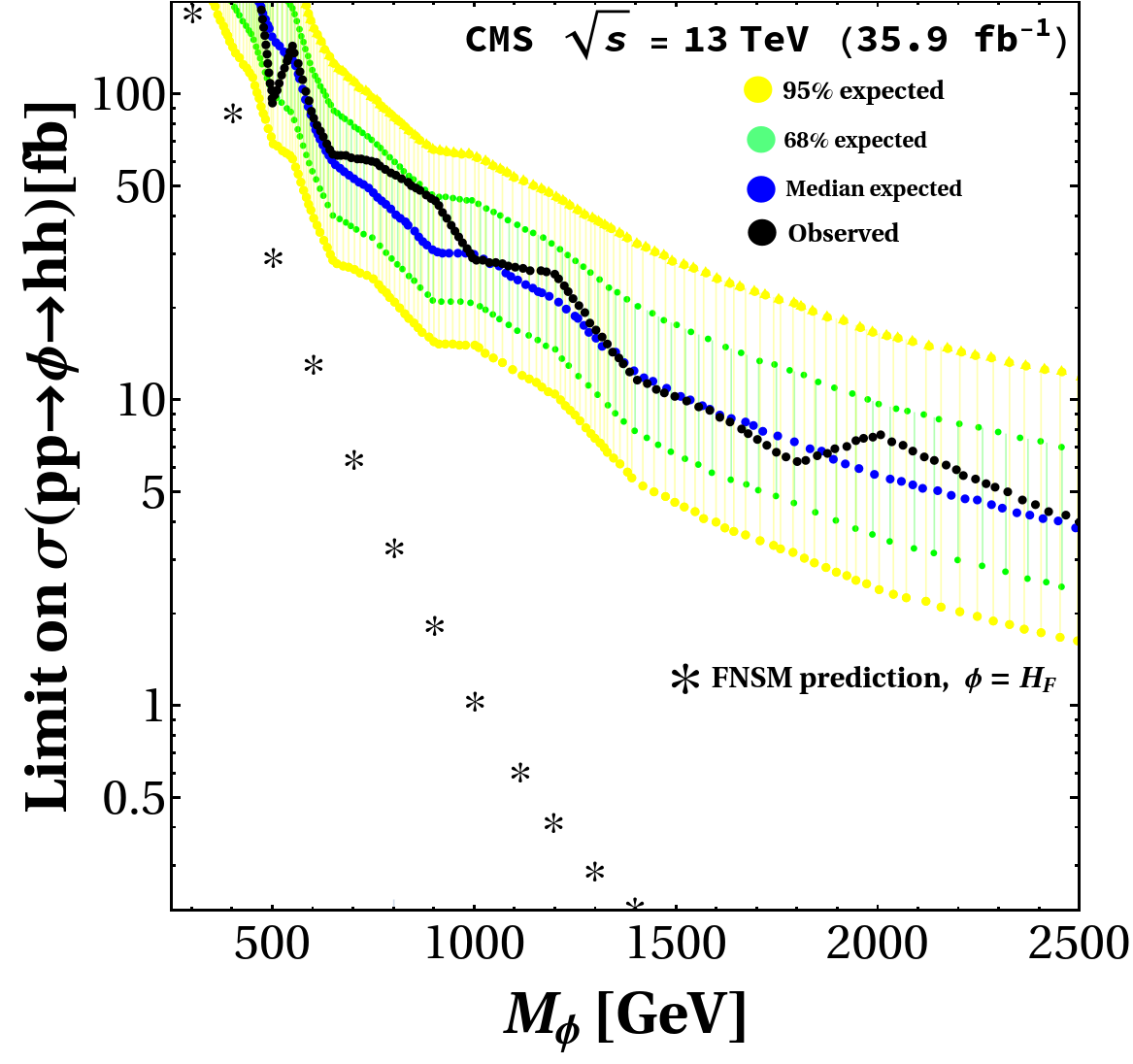}
	\end{center}
	\caption{Expected (blue points) and observed (black points) $95\%$ CL exclusion limits on the production of a narrow, spin-0  resonance ($\phi$) decaying into a pair of SM-like Higgs bosons at the LHC. The inner (green fill)  and  outer (yellow fill) bands indicate the regions containing $68$ and $95\%$ CL, respectively, results on the  limit applicable to the $pp\to \phi\to hh$ cross section expected under the background-only hypothesis. The starred points are predictions in the FNSM for a selection of heavy Higgs masses ($\phi\equiv H_F$) containing  BPs used in our analysis.}
	\label{XSvsMphi}
\end{figure}  
Furthermore, we present in Fig. \ref{fig:LFVp} the  $\cos\alpha-v_s$ plane regions allowed by $\Delta a_{\mu}$ (black points), $\Delta a_{e}$ (magenta points),  $\mu\to 3e$ (red points) and  $B_s^0\to\mu^+\mu^-$ (blue area). We have also analyzed the decays $\tau\to 3\mu$, $\tau\to \mu\gamma$, $\mu\to\ e\gamma$, however, these processes are not very restrictive in the FNSM. 
{{This is mainly due to the choice we made for the matrix elements $\tilde{Z}_{\mu\mu}$ and $\tilde{Z}_{\tau\tau}$, as they  play a subtle role in the couplings (see Tab. \ref{couplings}) $\phi\mu^-\mu^+$ and $\phi\tau^-\tau^+$ ($\phi=h,\,H_F,\,A_F$), which have a significant impact on the observables $\tau\to 3\mu$, $\tau\to \mu\gamma$, $\mu\to\ e\gamma$. In fact, we use $\tilde{Z}_{\tau\tau}=0.2$ and $\tilde{Z}_{\mu\mu}=10^{-4}$ (hence, a strong hierarchy), otherwise the SM $h\mu^-\mu^+$ coupling would be swamped by  new corrections due to the FNSM\footnote{Such a choice was adopted in the evaluation of $\kappa_{\tau\tau}$ and $\kappa_{\mu\mu}$, respectively, and then we scanned on the $\cos\alpha-v_s$ plane, as shown in Fig. \ref{fig:kappas}.}.}} So the bounds coming from the processes $\tau\to 3\mu$, $\tau\to \mu\gamma$, $\mu\to\ e\gamma$ are not included in Fig. \ref{fig:LFVp}.

 Then, in Fig. \ref{fig:intersection}, we display the result of applying all discussed theoretical and experimental constraints, limitedly to the 
reduced interval  $0.98\leq\cos\alpha\leq 1$, since it is the region in which all the analyzed observables converge. Here, we only show the most restrictive bounds so as to not overload the plot.  Among the latter,  the unitarity bound plays a special role, as it helped us to find a lower limit for the singlet scalar VEV, $v_s$, depending on the Flavon mass, e.g., for $M_{H_F}=1000$ GeV one has $v_s\geq 345$ GeV. By comparison, the intersection of all $\kappa_i$'s and $\Delta a_{\mu}$ imposes a less stringent upper limit of $v_s\leq 1200$ GeV\footnote{Notice that, to generate Figs.~\ref{fig:kappas}, \ref{fig:LFVp} and \ref{fig:intersection}, we have used our own \texttt{Mathematica} package, so-called \texttt{SpaceMath} \cite{Arroyo-Urena:2020qup}, which is available upon request.}.

\begin{figure}[h!]
	\begin{center}
		\subfigure[]{\includegraphics[scale=0.15]{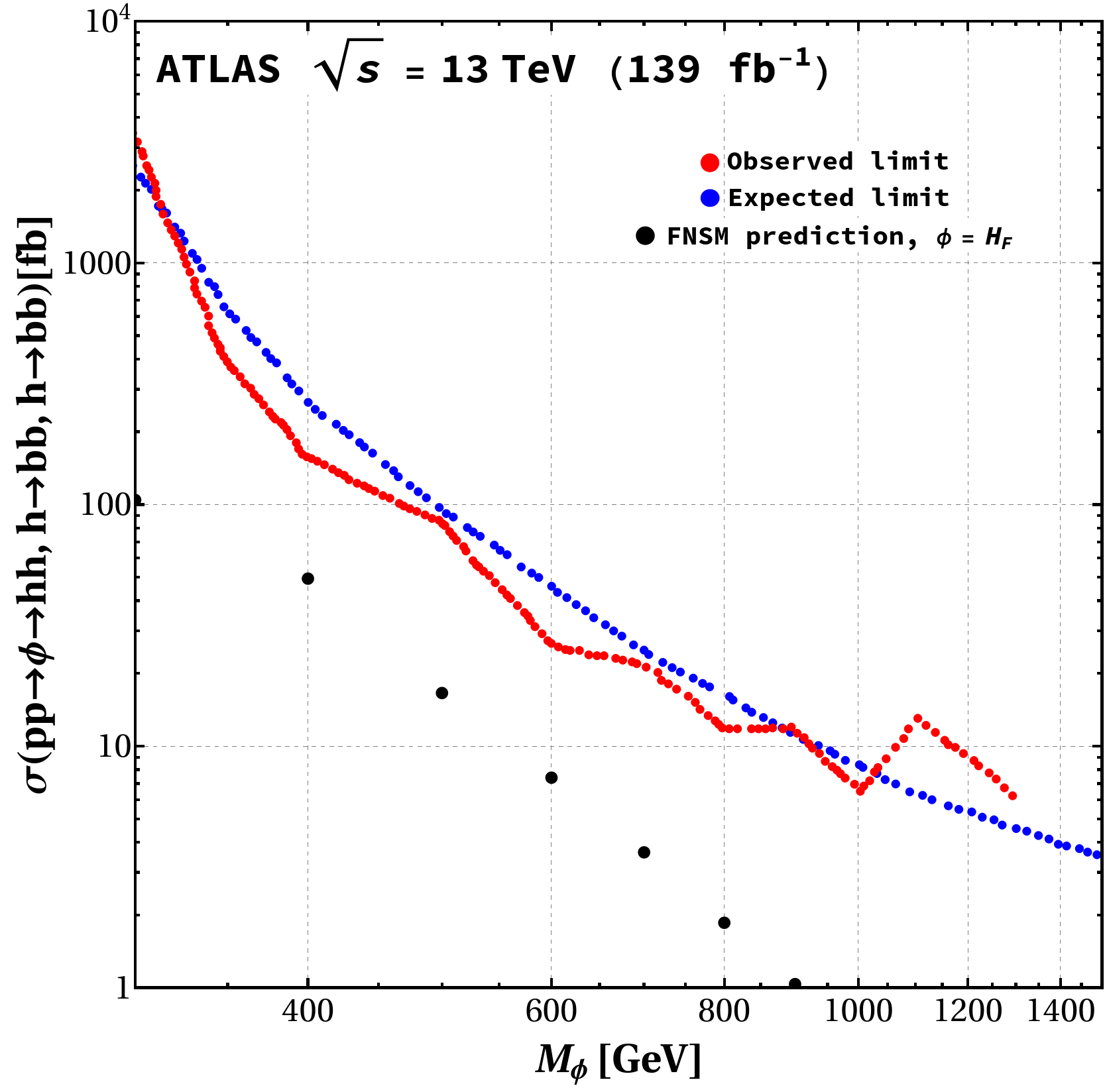}\label{fig:bbbb}
}
		\subfigure[]{\includegraphics[scale=0.15]{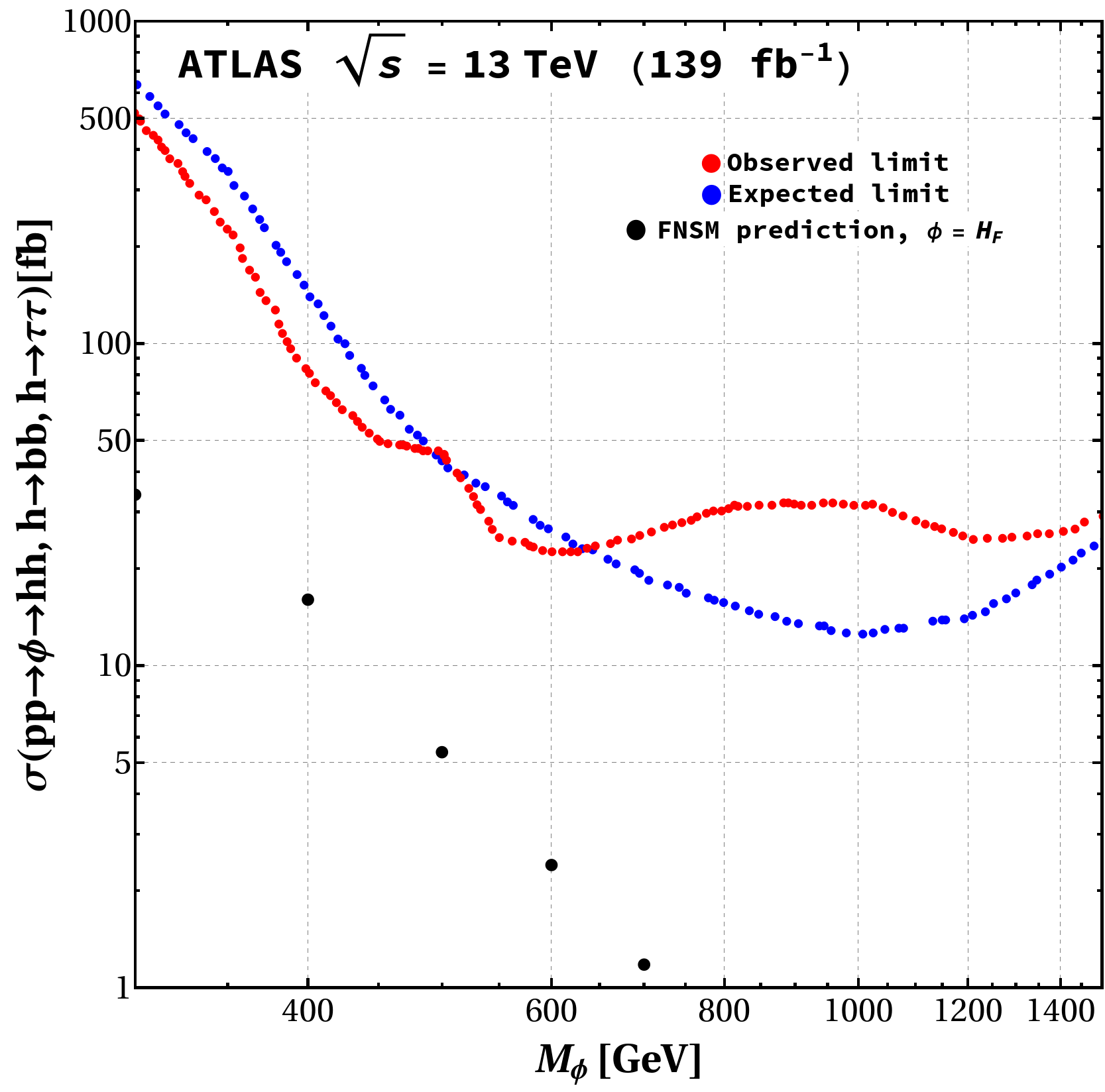}\label{fig:bbtata}
}
	\end{center}
\caption{Upper limits (observed and expected) on the cross section for di-Higgs production through an intermediate heavy particle $\phi$ as a function of the particle mass $M_{\phi}$ as obtained through the processes $pp\to H_F\to hh~(h,\to b\bar b, h\to b\bar b)$ (left) and $pp\to H_F\to hh$ $(h,\to b\bar b$, $h\to \tau^+\tau^-)$ (right).}
\label{IndChannels}
\end{figure}

\begin{figure}[h!]
	\begin{center}
	{\includegraphics[scale=0.45]{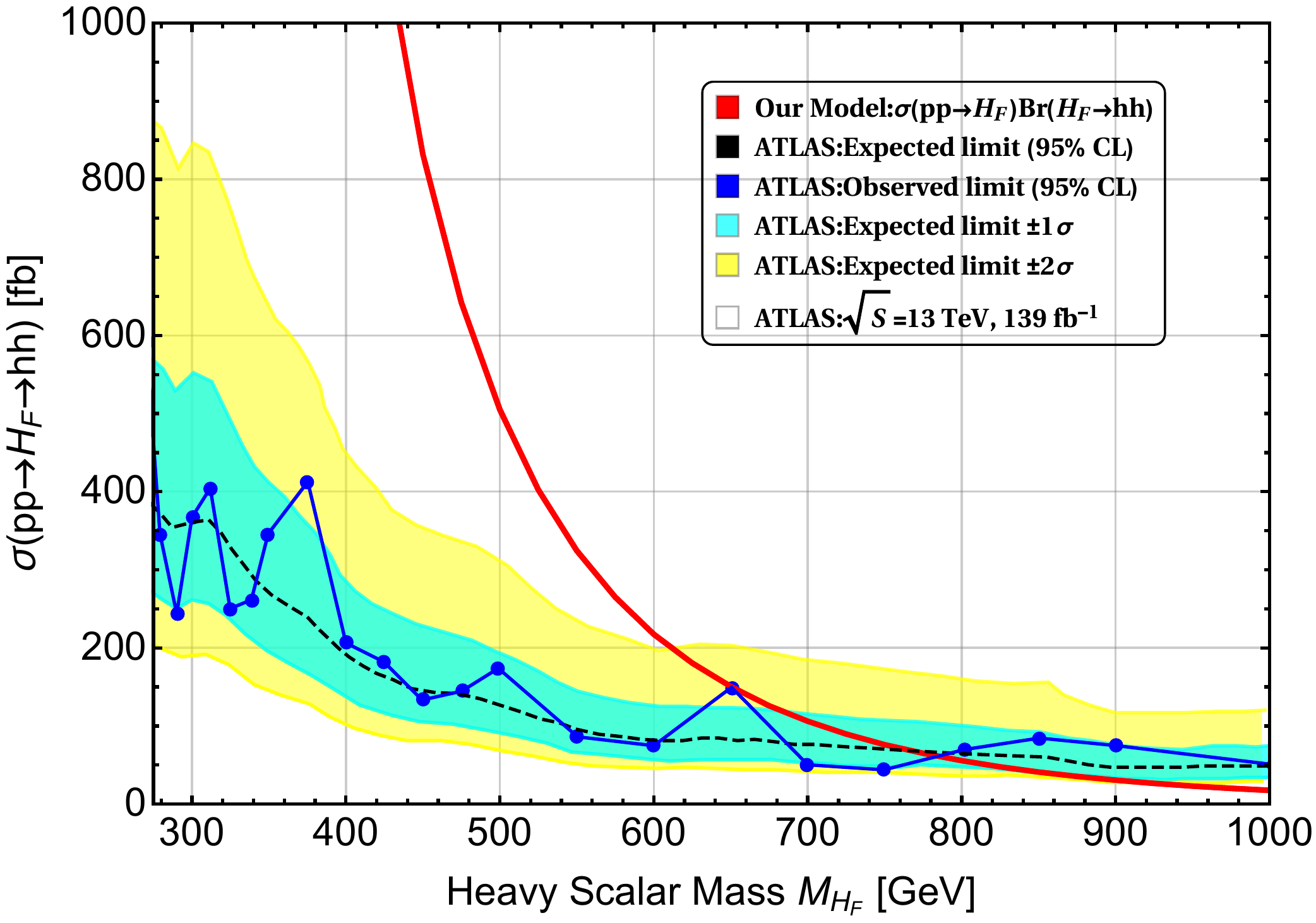}}
	\end{center}
\caption{Upper limits (observed and expected) on the cross section for di-Higgs production~\cite{ATLAS:2021ifb} through an intermediate heavy particle $\phi$ as a function of the particle mass $M_{\phi}$ as obtained through the process $pp\to H_F\to hh~(h,\to b\bar b, h\to \gamma \gamma)$.}
\label{fig:bbaaC}
\end{figure}

As far as the CP-even Flavon mass $M_{H_F}$ is concerned, to constrain it, we use the limit on the cross section of the process $pp\to\phi\to hh$ from \cite{CMS:2018ipl}, in which a combination of searches for SM-like Higgs boson pair production in proton-proton  collisions at $\sqrt{s}=$13 TeV and 35.9 fb$^{-1}$  is reported. We present in Fig. \ref{XSvsMphi} the cross section of the process $\sigma(pp\to H_F\to hh)$ in the FNSM as a function of $M_{H_F}$ and its comparison with the limit on $\sigma(pp\to\Phi\to hh)$, where $\phi$ stands for a generic spin-0 resonance. 
Furthermore, we show in Figs. \ref{fig:bbbb} and \ref{fig:bbtata} a 
comparison between the FNSM predictions and the ATLAS Collaboration 
limits \cite{atlas}, now for individual channels with final states 
$b\bar{b}b\bar{b}$ and $b\bar{b}\tau^-\tau^+$, respectively.  The most stringent constraints~\cite{ATLAS:2021ifb} come from $b\bar{b}\gamma \gamma$ production channel as shown in Fig.~\ref{fig:bbaaC}. In obtaining such limits, 
we have evaluated the inclusive cross section of our signal process, wherein we have used $v_s=1000$ GeV ans $\cos\alpha=0.995$.
 It is observed that the $M_{H_F}=300-1000$ GeV interval satisfies the bounds imposed, so we will define Benchmark Points (BPs) with $H_F$ masses herein.    
The model parameter space in this analysis is also consistent from the other search channels $pp\to H_F\to ZZ$ at ATLAS~\cite{ATLAS:2020tlo} and $pp\to H_F\to WW$ at CMS~\cite{CMS:2019bnu}.

\subsection{Constraints on $\tilde{Z}_{tc}$ ~from flavor-violating Higgs decays}	
\begin{figure}[h!]
\centering
 \subfigure{\includegraphics[scale = 0.40]{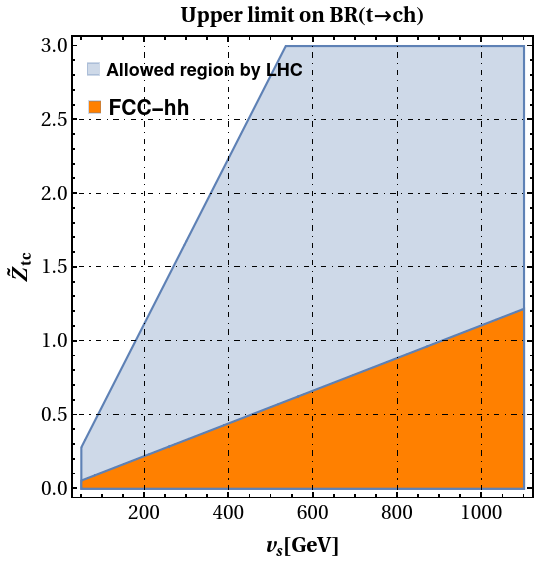}}
 \caption{Allowed region in the $v_s$-$\tilde Z_{tc}$ plane from the current bound on ${\rm BR}(t\to ch)<1.1\times 10^{-3}$ (blue color) and the projection at the FCC-hh (orange color).}
 \label{c}
	\end{figure}

Finally, because the $g_{H_F tc}$ coupling is proportional to the $\tilde{Z}_{tc}$ matrix element, 
we need a bound on it in order to evaluate the $H_F\to tc$ decay. Currently, there no specific 
processes that provide a stringent limit $\tilde{Z}_{tc}$, but we can estimate its order of magnitude 
by considering the upper limit on the Branching Ratio (BR) of $t\to ch$ at $<1.1\times 10^{-3}$~\cite{Workman:2022ynf}. 
{We also consider the prospects for  ${\rm BR}(t\to ch)<4.3\times 10^{-5}$ searches 
at the FCC-hh~\cite{Mandrik:2018yhe}. The resulting allowed region in the $v_s-\tilde{Z}_{tc}$ 
plane is illustrated in Fig.~\ref{c}. It is worth noting here that the behavior of the $\tilde{Z}_{tc}$ 
matrix element shows an increasing (decreasing) trend as $v_s$ increases (decreases). This 
observation is expected since the $g_{htc}$ coupling is governed by $\tilde{Z}_{tc}/v_s$. In order 
to have a realistic evaluation of the observables studied here, we adopt conservative values for 
$\tilde{Z}_{tc}$ and $v_s$.}


\section{Collider analysis }
\label{se:col_an}
Following our discussions on various model parameters 
and their constraints, we now study the collider signature 
emerging in the FNSM in the form of a  singlet-like 
CP-even heavy Higgs scalar $H_F$ decaying into SM-like Higgs $h$, neutral gauge bosons $Z$ and top-charm quark pairs at Run 3 of the LHC as well as the HL-LHC, assuming 
 $\sqrt{s}=14 $ TeV for both and a luminosity of 3000 fb$^{-1}$.
In our analysis, we adopt  
$ c_\alpha =0.995$ (i.e., a small mixing angle $\alpha$ between the CP-even part of the doublet and singlet scalar fields) and assume for the cut-off scale $\Lambda_F=10$ TeV, in order to easily avoid theoretical as well as experimental bounds (as discussed in the previous section).

Specifically, at the LHC, we consider the resonant production of the 
$H_F$ state via gluon-gluon fusion, followed by its decay into two on-shell SM-like Higgs bosons $(h)$, neutral gauge bosons $Z$ and a top-charm quark pair. For $hh$ production, one of the Higgs $h$ decays into a  pair of $b$-tagged jets while the other  decays into two photons, i.e., 
$pp \to H_F\to h h ~(h \to b \bar{b},~h \to \gamma \gamma)$:  
recall Fig.\ref{FeynDiag}. For the $ZZ$ channel, a $Z$ decays into a SFOS pair; while for $tc$ channel, the top quark decays into $\ell\nu_{\ell}b$, with $\ell=e^-,\,e^+,\mu^-,\,\mu^+$. Hence, we have three separate final states. The first one has two photons $(\gamma)$ and two $b$-jets, the second one has four leptons, and the third one contains a charged lepton plus its corresponding neutrino and two jets (one of them is a $b$-jet and the other is a $c$-jet). They all have some amount of hadronic activity generated from the initial state. Here, we only analyze the channels  $H_F \to h h, ZZ, tc$, since it is to be noted that the $A_F h h $ and $ A_F ZZ$ couplings are zero because of CP conservation, hence the twin production processes $pp\to A_F\to hh, ZZ$ via gluon-gluon fusion is not possible. The $A_F \to tc$ decay is dedicated for future analysis.

We use {\tt FeynRules} \cite{Alloul:2013bka}  to built the FNSM model and produce  the UFO  files for  {\tt MadGraph-2.6.5} \cite{Alwall:2014hca}. Using the ensuing particle spectrum into  {\tt MadGraph-2.6.5}, we calculate the production cross section of the aforementioned production and decay process.
The \texttt{$\rm MadGraph\_aMC@NLO$} \cite{Alwall:2014hca}  framework has been used to generate the background events in the SM.
Subsequent showering and hadronization have been performed with \texttt{Pythia-8} \cite{Sjostrand:2014zea}.
The detector response has been emulated using \texttt{Delphes-3.4.2}  \cite{deFavereau:2013fsa}. The default ATLAS configuration card which comes along with the \texttt{Delphes-3.4.2}  package has been used in the entirety of this analysis. For both the signal and background processes, we consider the Leading Order (LO) cross sections computed by \texttt{$\rm MadGraph\_aMC@NLO$},  unless stated otherwise. 

\begin{figure}[h!]
\begin{center}
\includegraphics[scale=0.30]{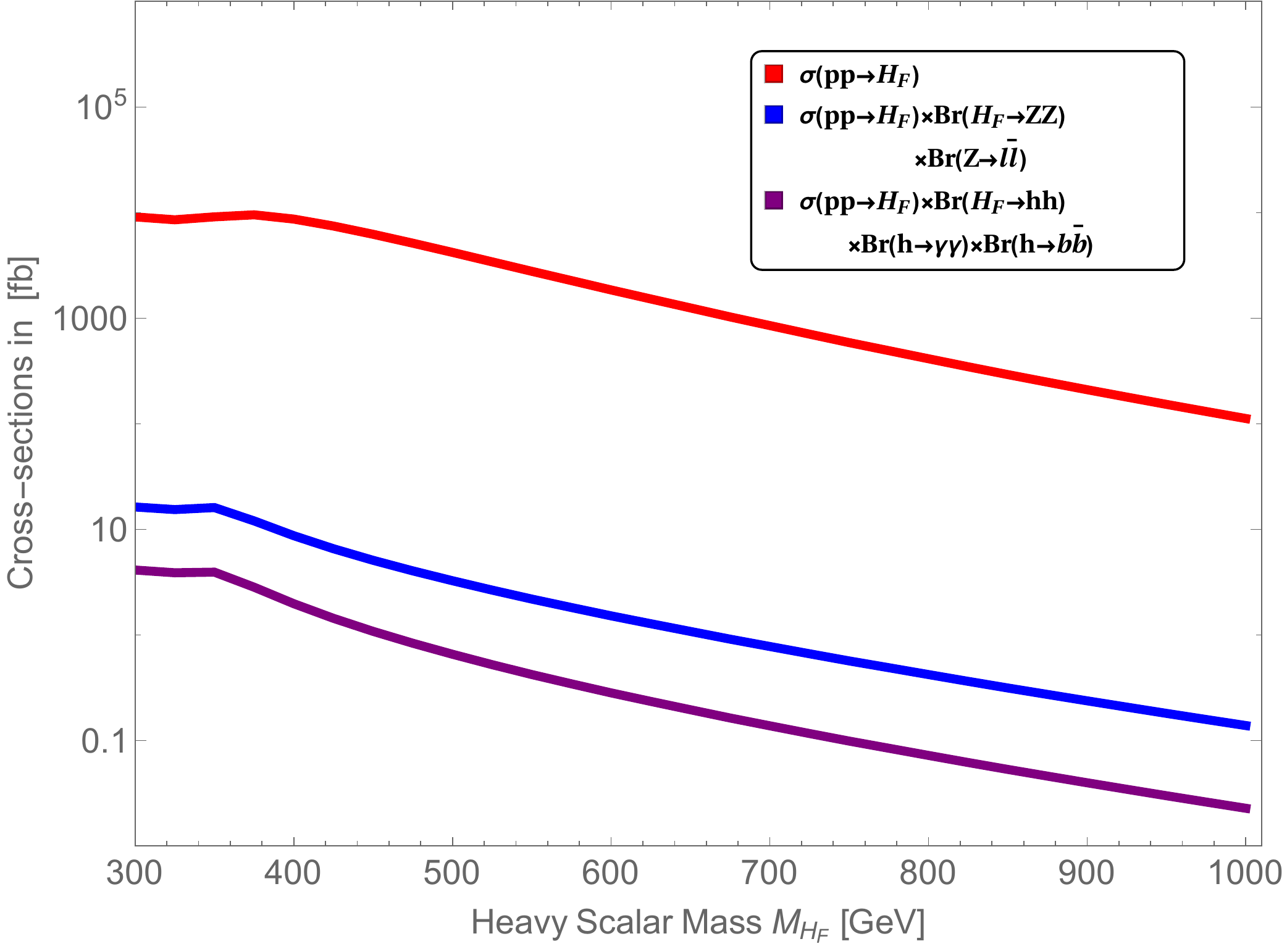}
\includegraphics[scale=0.305]{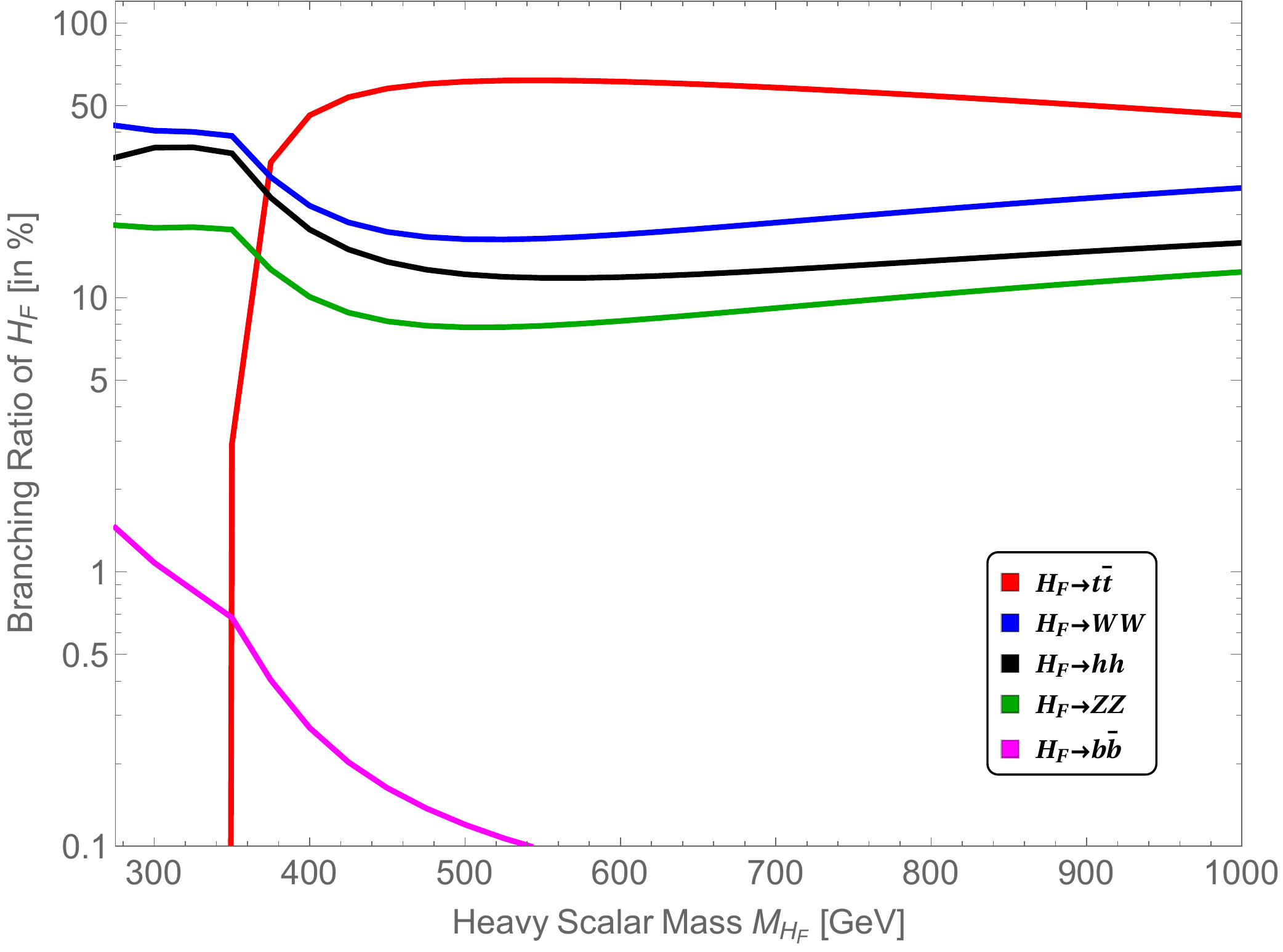}
\end{center}
\caption{The red (blue and purple) line on the left plot stands for the cross section of the processes $pp \rightarrow  H_F$ ($pp \to H_F\to h h ~(h \to \gamma \gamma, h \to b \bar{b})$ and $pp \to H_F\to ZZ ~(Z \to \ell \bar{\ell})$) at $14$ TeV. The variation in the {\rm BR}s of the heavy CP-even Flavon mass $M_{H_F}$ is displayed in the right plot. The heavy Higgs Flavor-violating decay is absent here, i.e., $\tilde{Z}_{ij}=0\,(i\neq j)$.  }
\label{fig:crossMe1}
\end{figure}

In the previous processes, we focus on the complete $FN$ diagonal basis, meaning no heavy Higgs Flavor-violating decay is present. This choice allow us to explore the large BRs to other channels, which could potentially provide a large signal significance in our study. We discuss the details now. Afterwards, we consider the  $FN$ off-diagonal basis to have new signals. This modification enables us to investigate the effects of heavy Higgs Flavor-violating interactions, which can have significant implications for our understanding of the $FN$ Higgs sector.

We first generate the signal events for various heavy CP-even Flavon masses, $M_{H_F} (=M_{A_F})$ considering  $\tilde{Z}_{ij}=0\,(i\neq j)$. The latter  have been varied from $260$ to $1000~{\rm GeV}$ with a step size of $10~{\rm GeV}$. 
We then take $v_s=1000$ GeV: 
such a large VEV produces a small production cross section $\sigma(pp\rightarrow H_F)$ and a correspondingly small partial width $\Gamma(H_F \rightarrow h h, ZZ)$, hence small (but non-negligible, for our
purposes) signal rates, however, this is necessary to comply with all theoretical and experimental limits. We display the cross section of the process $ pp \rightarrow  H_F$, $pp \to H_F\to h h~(h \to \gamma \gamma, h \to b \bar{b})$ and $pp \to H_F\to ZZ~(Z \to \ell \bar{\ell})$ on the left-hand-side of  Fig.~\ref{fig:crossMe1}, where the red line stands for $ \sigma(pp \rightarrow  H_F)$. 

One can thus understand the nature of the production and decay rates as follows.
The production cross sections of the heavy CP-even Flavon  $H_F$ (or pseudo scalar $A_F$, for that matter) mainly depends on the $g_{H_F t\bar{t}}= \frac{ c_\alpha v + s_\alpha
 v_s}{v_s}\, \frac{y_t}{\sqrt{2}}$ ($g_{A_F t\bar{t}}=\frac{v}{v_s}\, \frac{y_t}{\sqrt{2}} $) coupling, as the latter goes into the effective Higgs-to-two gluon vertex, $hgg$.  
The corresponding term in the Lagrangian  is given by~\cite{Plehn:2009nd}:
\begin{eqnarray}
\mathcal{L}_{\rm eff}&=&\frac{1}{v} \, g_{hgg} \, h \, G_{\mu\nu} G^{\mu\nu},\\
 g_{hgg} &=& -i \, \frac{\alpha_S}{8\pi}\, \tau (1+(1-\tau)\,f(\tau))~~~~~{\rm with}~~\tau = \frac{4 M_t^2}{M_h^2},\\
 f(\tau)&=& \begin{cases} 
(\sin^{-1}\sqrt{ \frac{1}{\tau} })^2, \quad\quad\quad\quad\quad\quad \tau\geq 1,\\ 
-\frac{1}{4}[\ln\frac{1+\sqrt{ 1-\tau}}{1-\sqrt{1-\tau}}-i\pi]^2\quad\quad\quad \tau<1.
\end{cases}
\end{eqnarray}
In this model, the $ggh$, 
$gg H_{F}$ and $gg A_{F}$ 
couplings take the following form:
$g_{hgg}=\left(\frac{c_\alpha v_s - s_\alpha v }{v_s}\right) \, g_{hgg}$,  $g_{H_F gg}=\left (\frac{c_\alpha v +s_\alpha v_s }{v_s}\right)\, g_{h gg}$ and $g_{A_F gg}=\frac{v}{v_s}\, g_{h gg}$, respectively. It is {to be} noted that, for $M_{H_F, A_F}>2\, M_t$, $f(\tau)=-\frac{1}{4}[\ln\frac{1+\sqrt{ 1-\tau}}{1-\sqrt{1-\tau}}-i\pi]^2$. Hence,  one can understand the shape of the plot by exploiting  these functions.
The  BRs of $H_F$ into various channels for 
$v_s=1000$ GeV are  shown on the right-hand-side   of Fig.~\ref{fig:crossMe1}. 
From the {\rm BR} plot, we can see that, for heavier 
$H_F$ masses, this state dominantly decays into $t{\bar t}$. For small masses, $H_F\to WW$ dominates. Yet,  $H_F\to hh$  is the third, while $H_F\to ZZ$ is the fourth largest decay channels. In the next subsections, we will focus on discussing the processes $H_F\to hh\,(h\to b\bar{b}, h\to\gamma\gamma)$ and $H_F\to ZZ \,(Z\to\ell \bar{\ell}) $ for the diagonal and $H_F\to tc \,(t\to b\ell\nu_{\ell}) $ for off-diagonal scenario, respectively. These processes are of particular interest because they are not as strongly suppressed by standard model backgrounds compared to the $H_F\to t\bar{t}$ and $H_F\to WW$ decays. 

\subsection{$pp \to H_F\to h h~(h\to \gamma \gamma, h\to b \bar{b})$}

The major SM backgrounds typically have the form  $hh + X$ 
(where $X$ is known SM particles),  which includes  SM Higgs pair $hh$ 
production, $h + X$ like $hZ, hb\bar{b} $ and $ht\bar{t} $, as well as  the 
non-Higgs processes which include $t\bar{t}$ and $  t\bar{t} \gamma$ 
(here, leptons may fake as photons) as well as $b\bar{b} \gamma \gamma$, $ c\bar{c} 
\gamma \gamma $ and $  jj \gamma \gamma$ (where $c$-jets and 
light-jets may fake  $b$-jets). The other relevant reducible backgrounds 
comprise $b\bar{b} j \gamma, c\bar{c} j \gamma$ and  $b\bar{b} jj$, 
where  $c$-jets may appear as $b$-jets and a light-jet may fake a photon.
The fake rate of a light-jet $j$ into a photon depends on the momentum of the 
jet, $p_T^j$~\cite{ATLAS:2013kpx}, as $9.3\times 10^{-3} {\rm exp}(-p_T^j/27.5 
~{\rm GeV})$. The $c$-jet is misidentified as a $b$-jet with a rate of $3.5\%$ 
whereas a  light-jet mimics a $b$-jet with a rate of $0.135\%$ \cite{CMS:2017wtu}.

\begin{table}[ht]
\begin{center}\scalebox{0.85}{
\begin{tabular}{|c|c|c|c|c|c|c|c|c|c|c|c|c|c|}
\hline
BPs  [GeV] &\multicolumn{1}{c|}{ The other input parameters }\\
\hline
\rule{0pt}{1ex}
BP1  ($M_{H_F}=800$)  &  $M_{A_F}=800$ GeV, $\lambda_1=0.36$, $\lambda_2=0.64$, $\lambda_3=0.25$ \\
\rule{0pt}{1ex}
BP2  ($M_{H_F}=900$)  &  $M_{A_F}=900$ GeV, $\lambda_1=0.39$, $\lambda_2=0.81$, $\lambda_3=0.32$  \\
\rule{0pt}{1ex}
BP3 ($M_{H_F}=1000$)   &  $M_{A_F}=1000$ GeV, $\lambda_1=0.42$, $\lambda_2=0.99$, $\lambda_3=0.40$ \\
\hline
\end{tabular}}
\end{center}
\caption{ The input parameters of the three BPs (BP1, BP2 and BP3) used in the remainder of the paper. We have 
 $M_{h}=125.5$ GeV, $\cos\alpha=0.995, v_s=1000$  GeV and 
$\Lambda_F=1$ TeV is this kept fixed for 
all  BPs.}
\label{tab:BPs}
\end{table}

\begin{table}[ht]
\begin{center}\scalebox{1.0}{
\begin{tabular}{|c|c|c|c|c|c|c|c|c|c|c|c|c|c|}
\hline
BPs  [GeV] &\multicolumn{3}{c|}{ {\rm BR}s and cross sections [pb] }\\
\hline
& ${\rm BR}(H_F \to h h)$ & $ \sigma( pp \rightarrow  H_F$) & $\sigma(pp \to H_F\to h h, h \to \gamma \gamma, h \to b \bar{b})$ \\
\cline{2-4}
\rule{0pt}{1ex}
BP1 ($M_{H_F}=800$)   &  $0.14$  & $0.41$ &  $7.18\times 10^{-5}$\\
\rule{0pt}{1ex}
BP2  ($M_{H_F}=900$)  &$0.15$  & $0.21$ &  $3.95\times 10^{-5}$ \\
\rule{0pt}{1ex}
BP3  ($M_{H_F}=1000$)  & $0.16$ & $0.11$ &   $2.27\times 10^{-5}$\\
\hline
\end{tabular}}
\end{center}
\caption{The ${\rm BR}(H_F \to h h)$ and cross sections for the processes $ pp \rightarrow  H_F$ and $\sigma(pp \to H_F$ $\to h h, h \to \gamma \gamma, h \to b \bar{b})$ for three BPs (BP1, BP2 and BP3) used in the remainder of the paper. }
\label{tab:cs}
\end{table}

\begin{table}[htpb!]
\begin{center}\scalebox{1.0}{
\begin{tabular}{|c|c|c|c|c|c|c|c|c|c|c|c|c|c|}
\hline
 SM backgrounds & Cross section [pb]  \\
\hline
\rule{0pt}{1ex}
$ pp \to b \bar{b} \gamma \gamma  $&    4.57    \\
$ pp \to Zh \, ( Z\to b \bar{b}, h\to \gamma \gamma)  $&       $1.40 \times 10^{-4}$ \\
\hline
$ pp \rightarrow b \bar{b} j \gamma  $ &    $7470.02$    \\
$ pp \rightarrow b \bar{b} j j  $ &    $5.03\times 10^6$    \\
($j$ mimic as photon)&\\
\hline
$ pp \rightarrow c \bar{c} \gamma \gamma  $ &    $6.21 $   \\
$ pp \rightarrow c \bar{c} j \gamma  $ &    $ 2085.01 $  \\
($c$ appear as $b$-tagged jets,&              \\
$j$ mimic as photon)&        \\
\hline
$ pp \rightarrow jj \gamma \gamma  $ &    $ 65.23 $  \\
($j$ appear as $b$-tagged jets)&              \\
\hline
$ pp \rightarrow t \bar{t}\, (t \to \bar{\ell} {\nu}_\ell b,  \bar{t}\to \ell \bar{\nu}_\ell \bar{b} )$ &   $6.17\times 10^{-5}$     \\
$ pp \rightarrow t \bar{t}\, (t \to jj b,  \bar{t}\to jj \bar{b} )$ &   $ 202.15   $ \\
($\ell,j$ mimic as photon)&     \\
\hline
\end{tabular}}
\end{center}
\caption{The cross sections for the most relevant SM background processes. (Note that these background rates will be multiplied by the fake rates during the analysis.)}
\label{tab:csBG}
\end{table}

We next present a detailed discussion of the collider search strategy employed to maximize the signal significance in the  search channel  $pp \to H_F\to h h ~(h \to \gamma \gamma, h \to b \bar{b})$. To start with, though, 
we show the production and decay cross section $pp \to H_F\to h h ~(h \to \gamma \gamma, h \to b \bar{b})$
for the three BPs presented in Tab.~\ref{tab:BPs} (with, in particular, $M_{H_F} =800,~900$ and $1000$ GeV, as seen  in Tab.~\ref{tab:cs}).
The corresponding dominant SM backgrounds are shown in Tab.~\ref{tab:csBG}.

Any charged objects  (leptons or  jets) or  photons produced in any hard 
scattering process at the LHC will be observed in the detector if and only
if they satisfy certain geometric criteria,  known as 
acceptance cuts. These  are the  same for both the signal and  background events
and reproduce the accessible region of the detector. We will then have to ask that both 
signal and background events pass these acceptance cuts, which are, in general, not sufficient 
to separate the two samples. However, eventually, we 
 will  construct various kinematic observables and study their 
distributions. Next, we will decide the final selection cuts after studying the 
distinguishing features of those distributions between signal and 
backgrounds, so as to increase the former and decrease the latter. We base this approach on a Monte Carlo (MC) analysis using the tools previously  
described. 
 
In our current scenario, an event is required to have exactly 
two $b$-tagged jets and two isolated photons $(\gamma)$ in the final state. However, we
 do not put any constraints on the number of light-jets. We then adopt the following acceptance cuts: 
\begin{itemize}
\item $p_T^\gamma > 20 $ GeV; 
\item $p_T^{e/\mu} > 20 $ GeV  (if  an electron/muon is present, for $b$-tagging purposes);
\item $p_T^j > 40 $ GeV, where $j$ stands for light-jets as well as
$b$-jets;
\item $\mid \eta_\ell \mid ~\leq 2.5 $ (again, $\ell = e/\mu $), $\mid \eta_\gamma \mid ~\leq 2.0 $ and $\mid \eta_j \mid ~\leq 2.0 $.
\end{itemize}
\begin{figure}[h!]
\begin{center}
\includegraphics[scale=0.17]{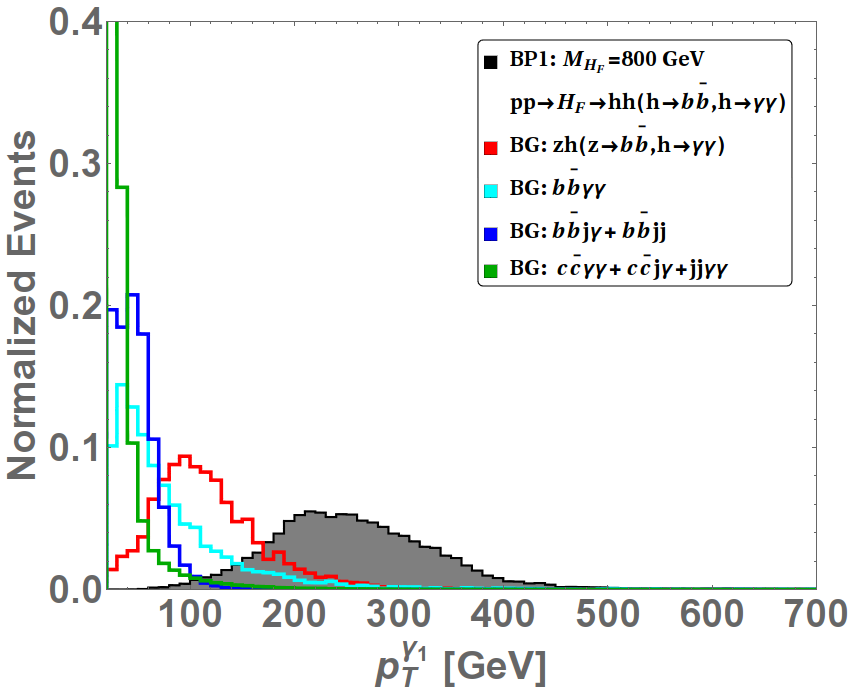}
\includegraphics[scale=0.17]{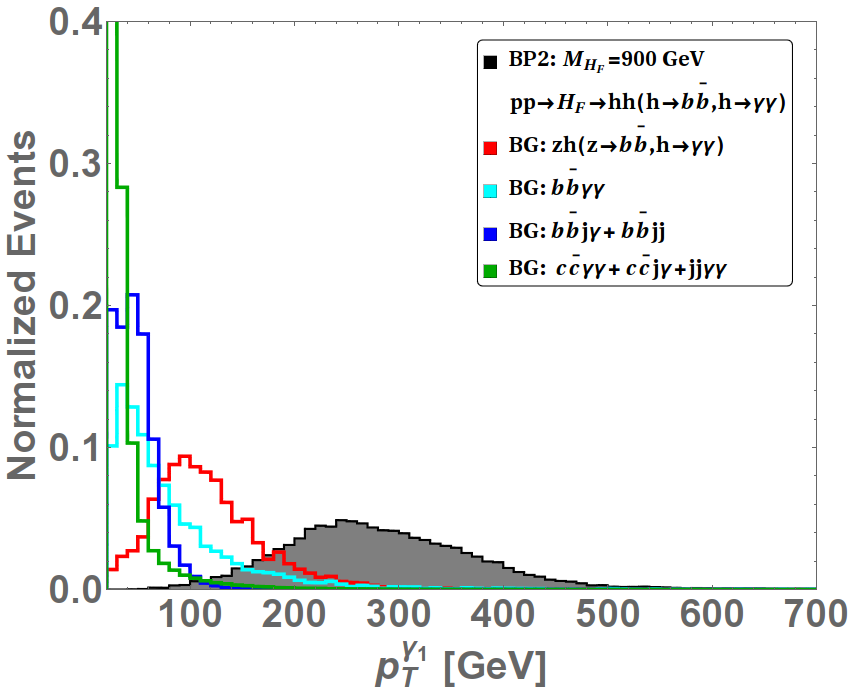} 
\includegraphics[scale=0.17]{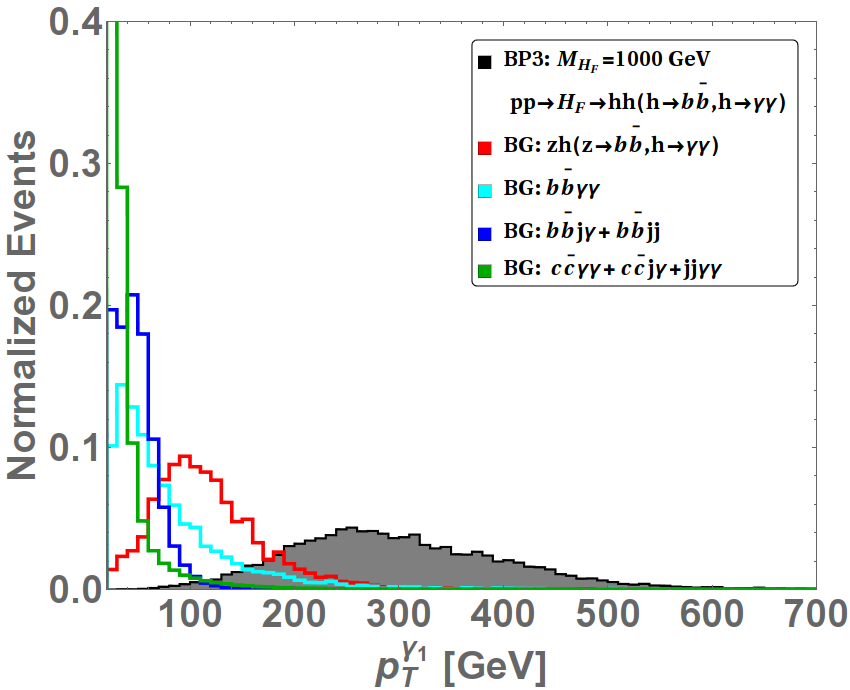}
\includegraphics[scale=0.17]{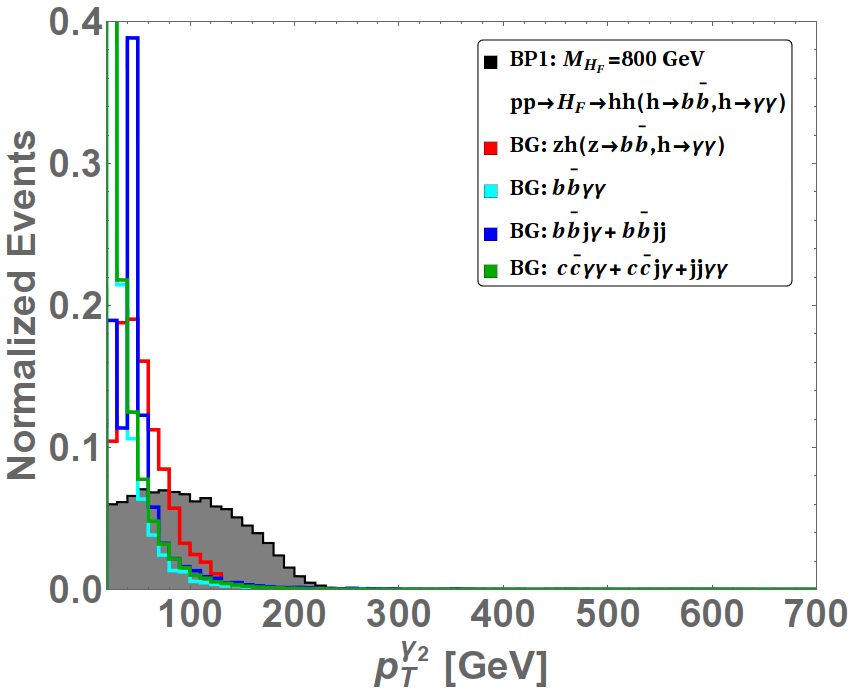} 
\includegraphics[scale=0.17]{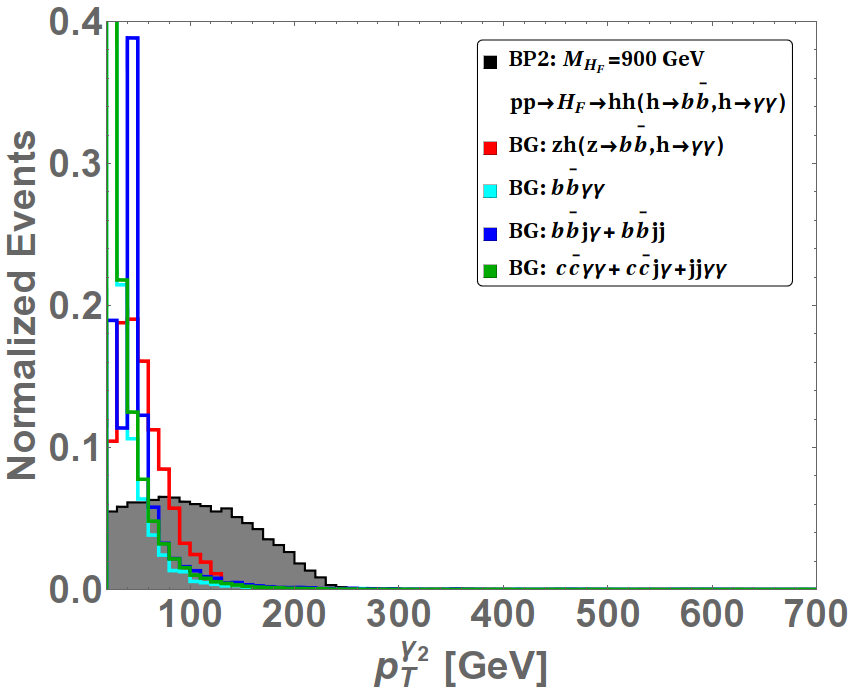}
\includegraphics[scale=0.17]{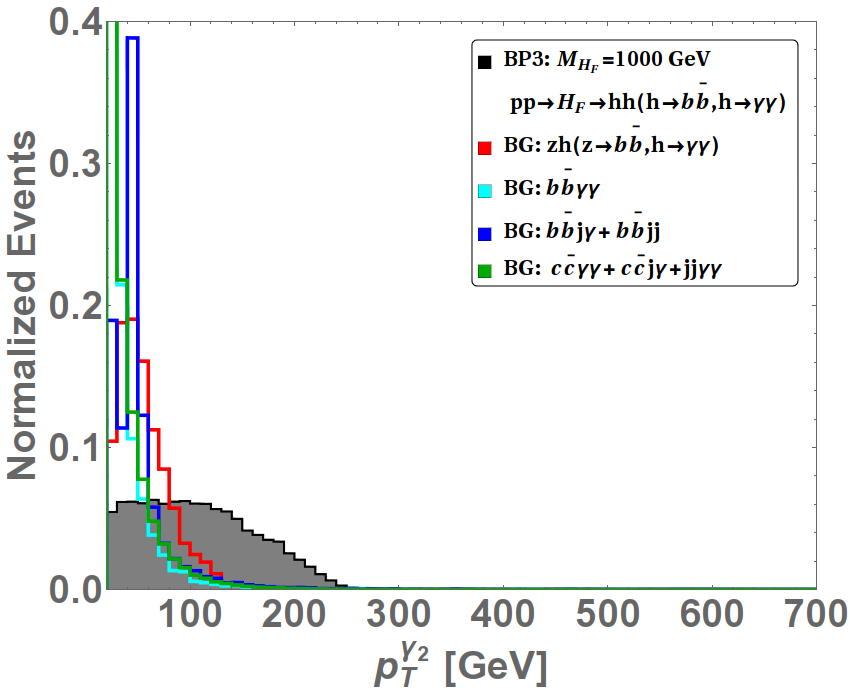}
\end{center}
\caption{Normalized distributions in photon transverse momentum for  signal and total background after the acceptance cuts.}
\label{fig:Dist1}
\end{figure}
After considering these basic requirements,
we apply a stronger selection (using additional  kinematic variables)  in order to enhance the signal-to-background ratio, as explained. A variety of such  observables   have been used to design the optimized Signal Region (SR), i.e., where the significance is maximized.
First and foremost, the transverse momentum of photons ($p_T^{\gamma_1} $, $p_T^{\gamma_2} $) and  $b$-jets  ($p_T^{b_1} $, $p_T^{b_2} $)\footnote{Here, ${1}$ and ${2} $ represents the $p_{T} $
 ordered leading and sub-leading photon and $b$-jet in the final state.} will be studied. 
 In addition, the separation between the two final state photons $\Delta R_{\gamma_1\gamma_2} $ and
 $b$-jets $\Delta R_{b_1 b_2} $ are also used.  
The separation between two detector objects, 
$\Delta R$,  is  defined as $\Delta R = \sqrt{\Delta \eta^{2}  + \Delta \phi^{2}}  $, where  $\Delta\eta$ and $\Delta\phi$
   are the differences in pseudorapidity and azimuthal angle, respectively.  
Then, the
 invariant mass of the final state photons ($M_{\gamma_1 \gamma_2} $)  and $b$-jets ($M_{b_1 b_2} $)  will also be used to discriminate between signal and 
 backgrounds, where we have  introduced $M_{ab}=\sqrt{(E^a+E^b)^2-\sum_{i=x,y,x} (p_i^a+p_i^b)^2 }$, with $ab=\gamma_1\gamma_2$ or $b_1b_2$. Finally,
we  use the invariant mass $M_{hh}$ for the final extraction. The $M_{hh}$ variable has been calculated as {$M_{hh}=\sqrt{(E^{\gamma_1}+E^{\gamma_2}+E^{b_1}+E^{b_2})^2 -  \sum_{i=x,y,z}  (p^{\gamma_1}_{i} +p^{\gamma_2}_{i} + p^{b_1}_{i} + p^{b_2}_{i})^2 }.$} In the above
formulae,  $E$ and $p_{i}~(i=x,y,z)$ stand for the energy and three-momentum component of the final state particles, respectively.
\begin{figure}[h!]
\begin{center}
\includegraphics[scale=0.17]{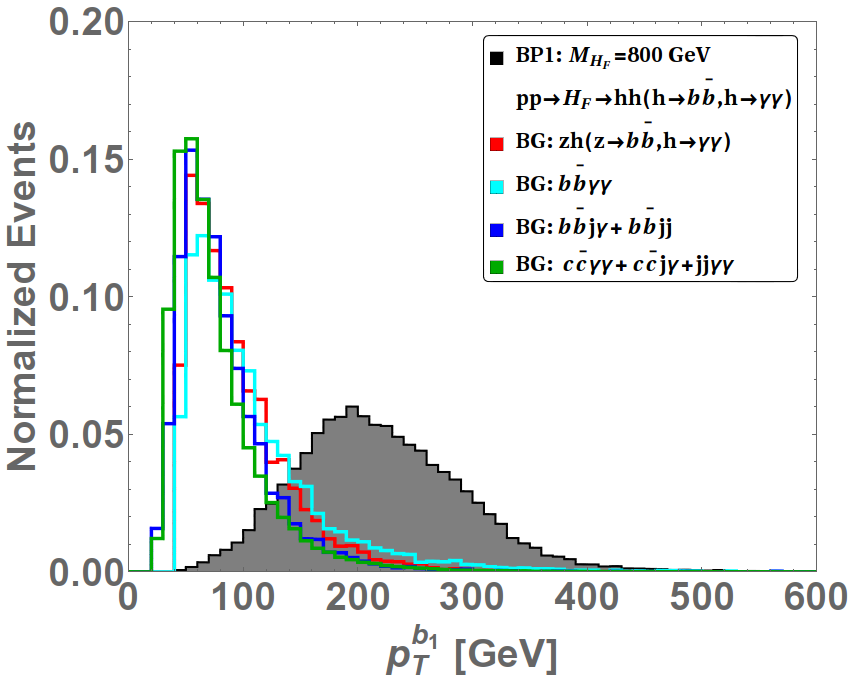} 
\includegraphics[scale=0.17]{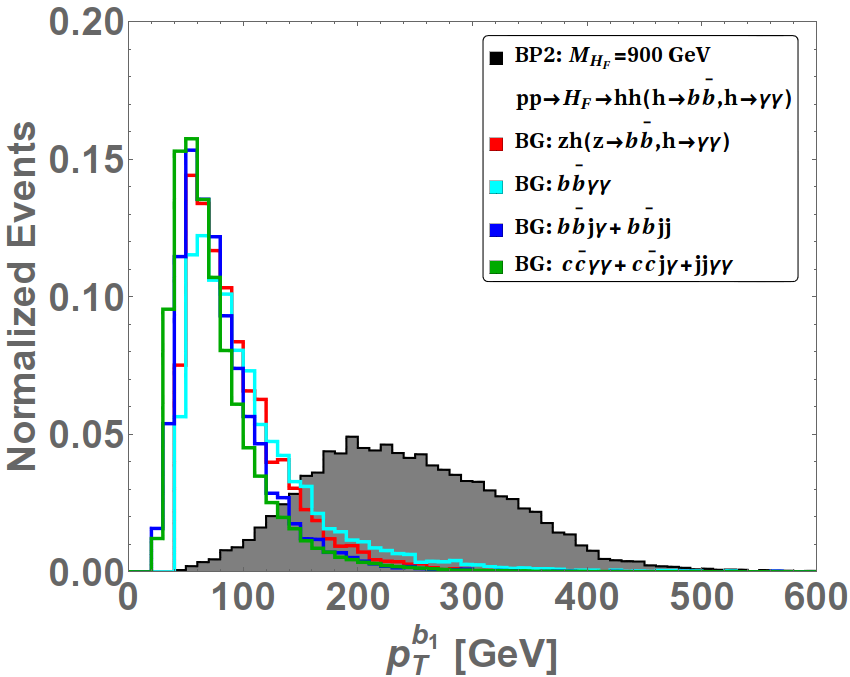}
\includegraphics[scale=0.17]{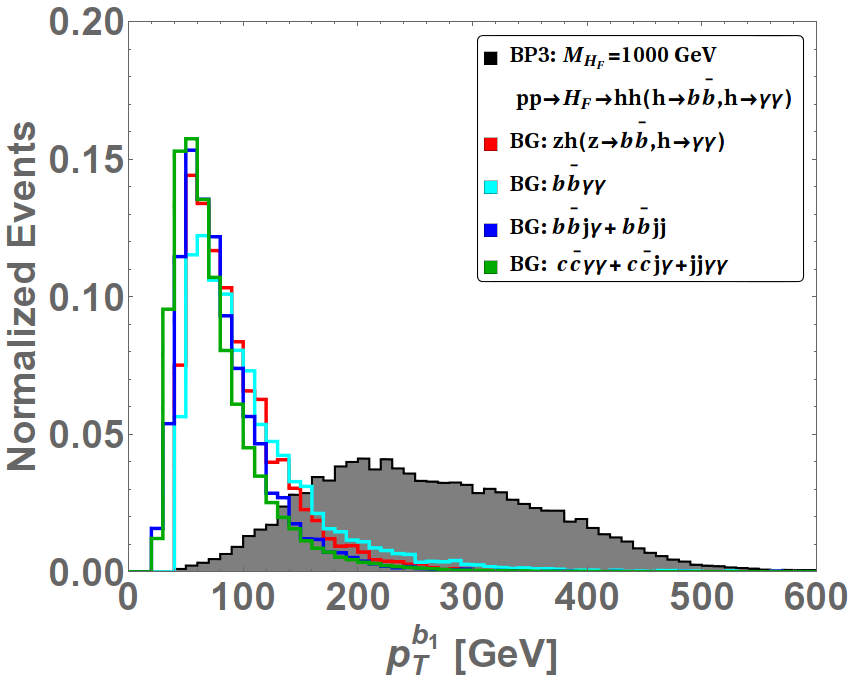} 
\includegraphics[scale=0.17]{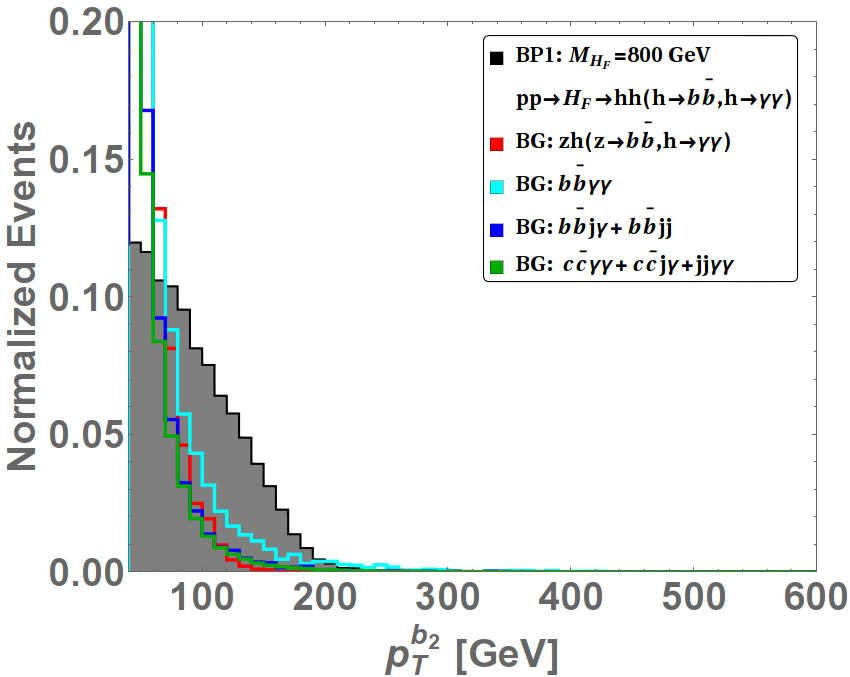}
\includegraphics[scale=0.17]{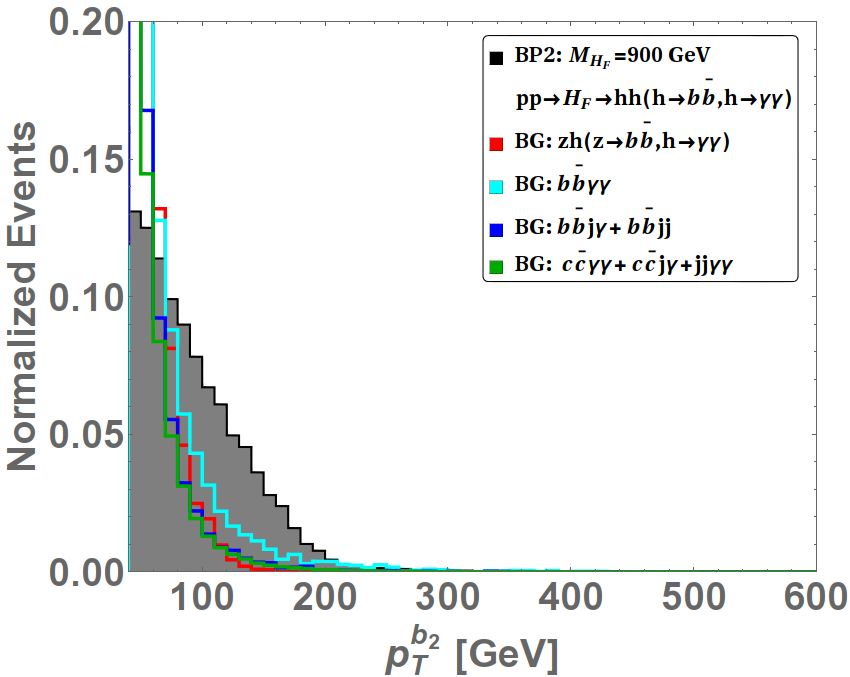}
\includegraphics[scale=0.17]{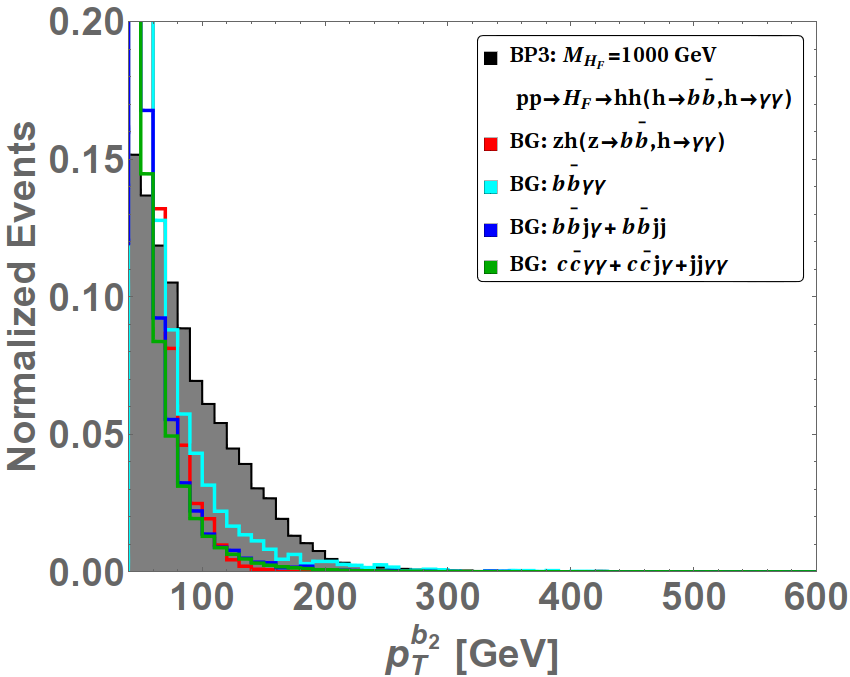}
\end{center}
\caption{Normalized distributions in $b$-jet transverse momentum for  signal and total background after the acceptance cuts.}
\label{fig:Dist2}
\end{figure}

\begin{figure}[h!]
\begin{center}
\includegraphics[scale=0.17]{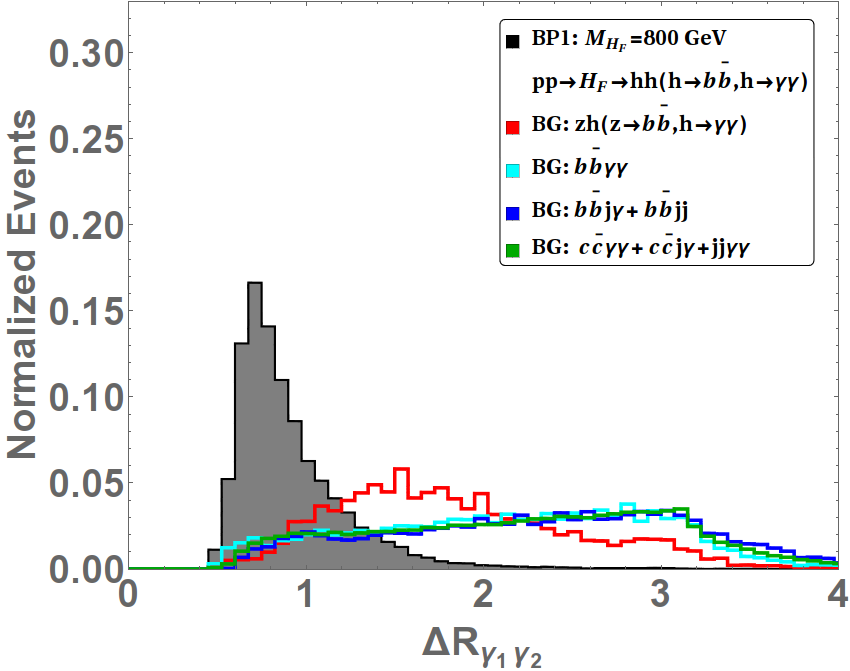}
\includegraphics[scale=0.17]{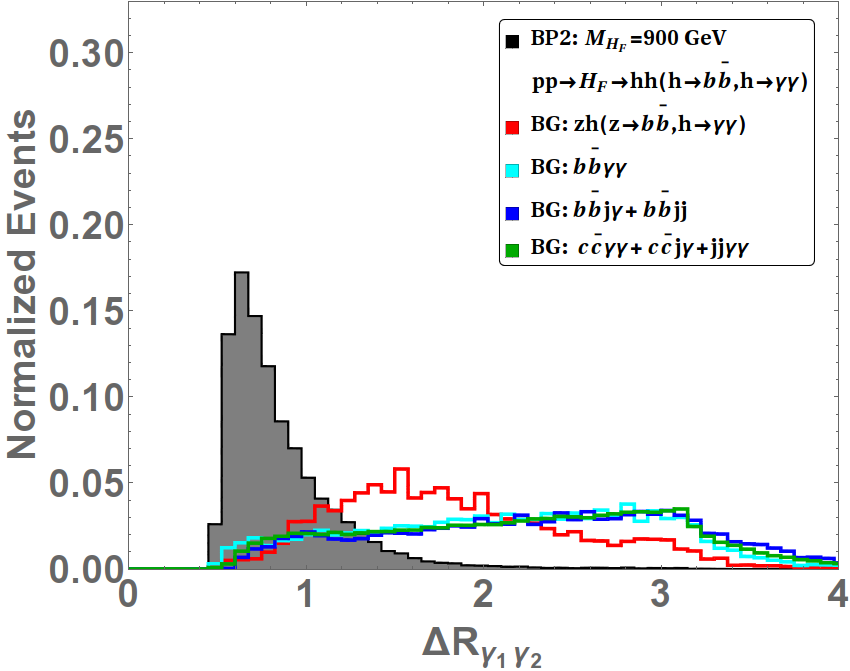} 
\includegraphics[scale=0.17]{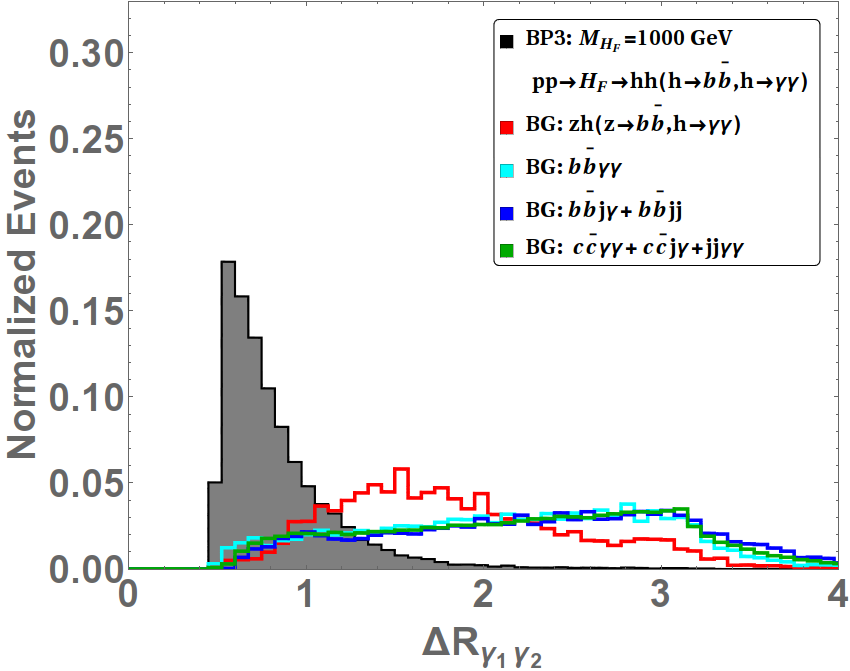}
\includegraphics[scale=0.17]{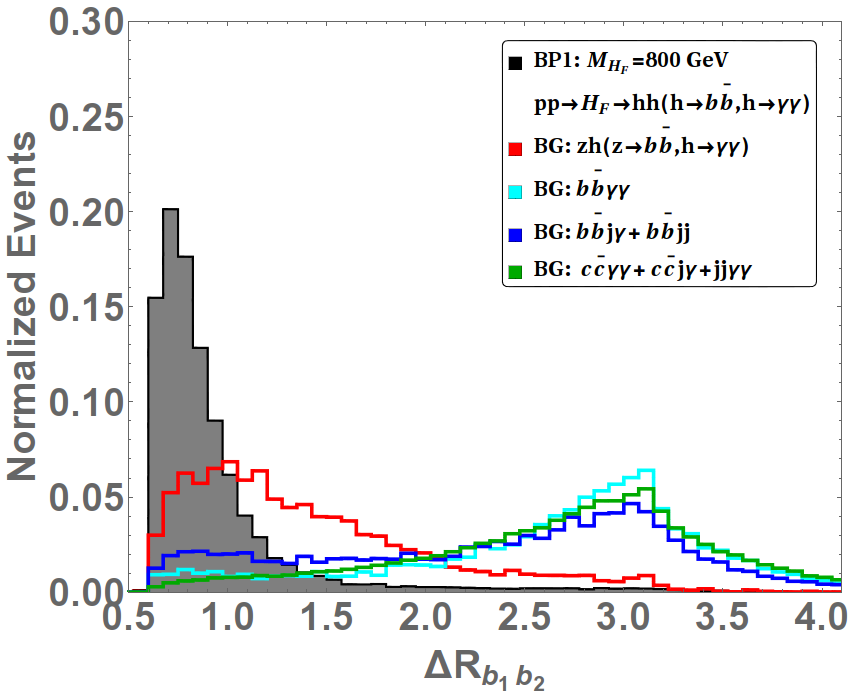} 
\includegraphics[scale=0.17]{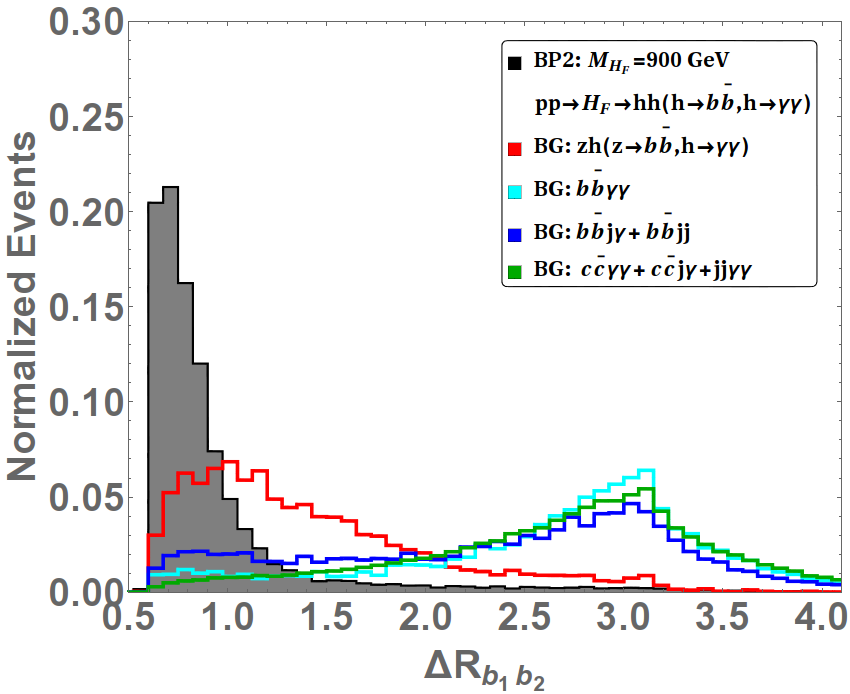}
\includegraphics[scale=0.17]{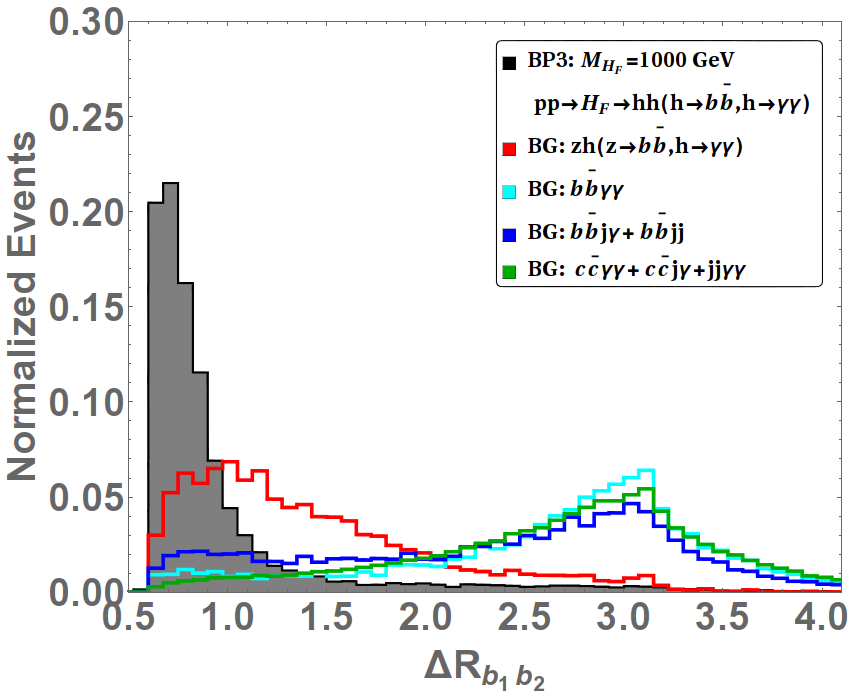}
\end{center}
\caption{Normalized distributions in di-photon and di-jet  separation for  signal and total background after the acceptance cuts.}
\label{fig:Dist3}
\end{figure}

\begin{figure}[h!]
\begin{center}
\includegraphics[scale=0.17]{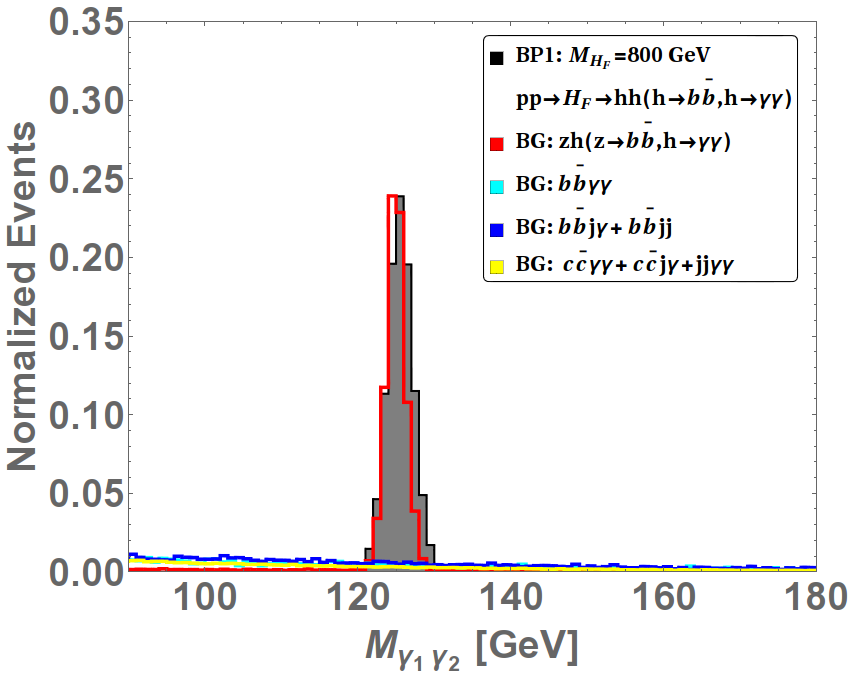} 
\includegraphics[scale=0.17]{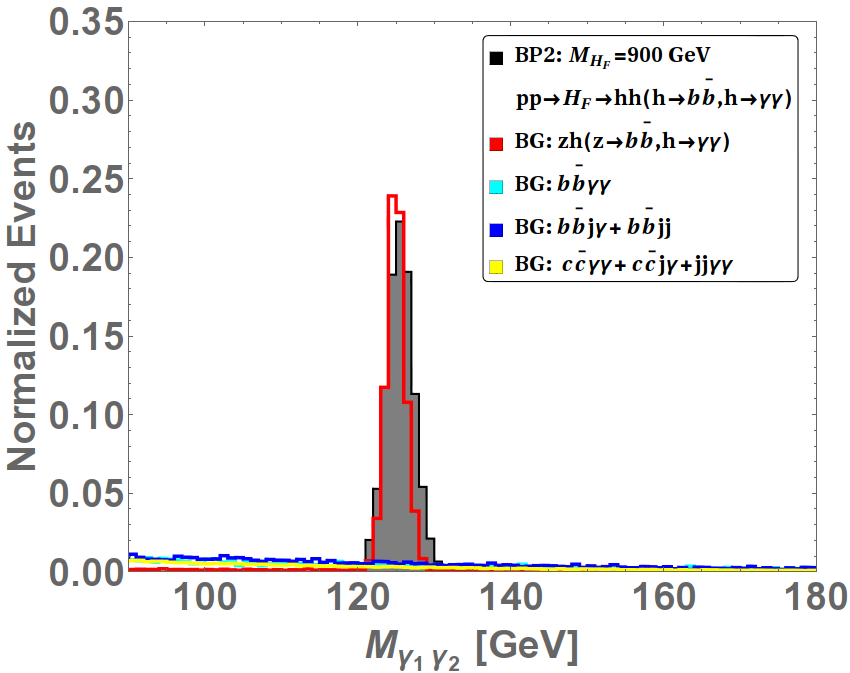}
\includegraphics[scale=0.17]{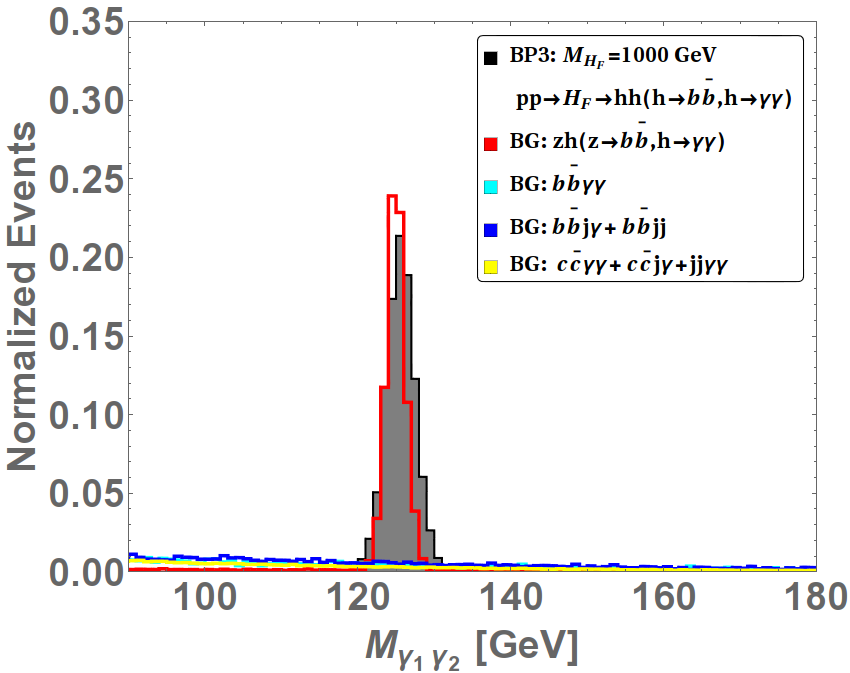}
\includegraphics[scale=0.17]{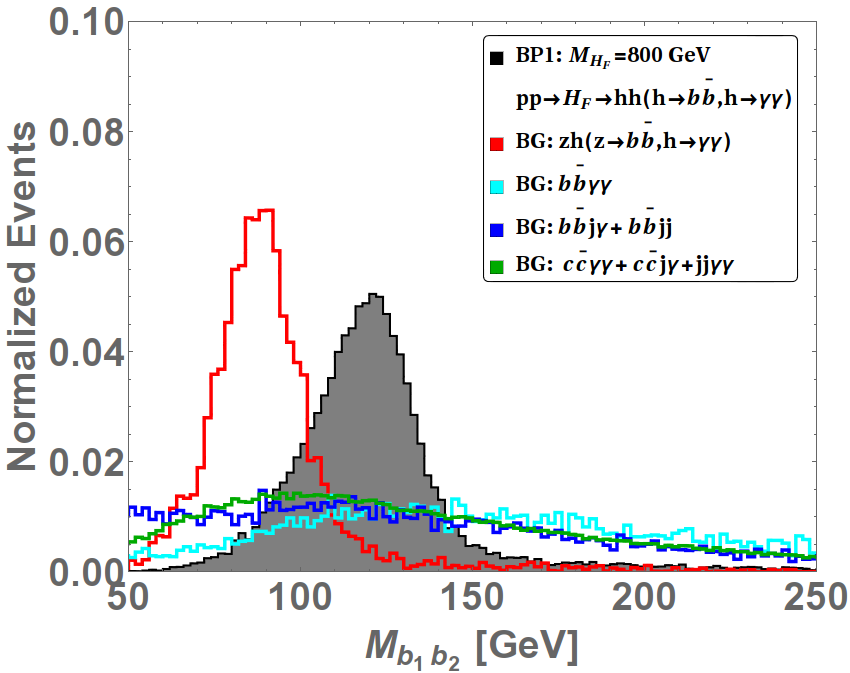}
\includegraphics[scale=0.17]{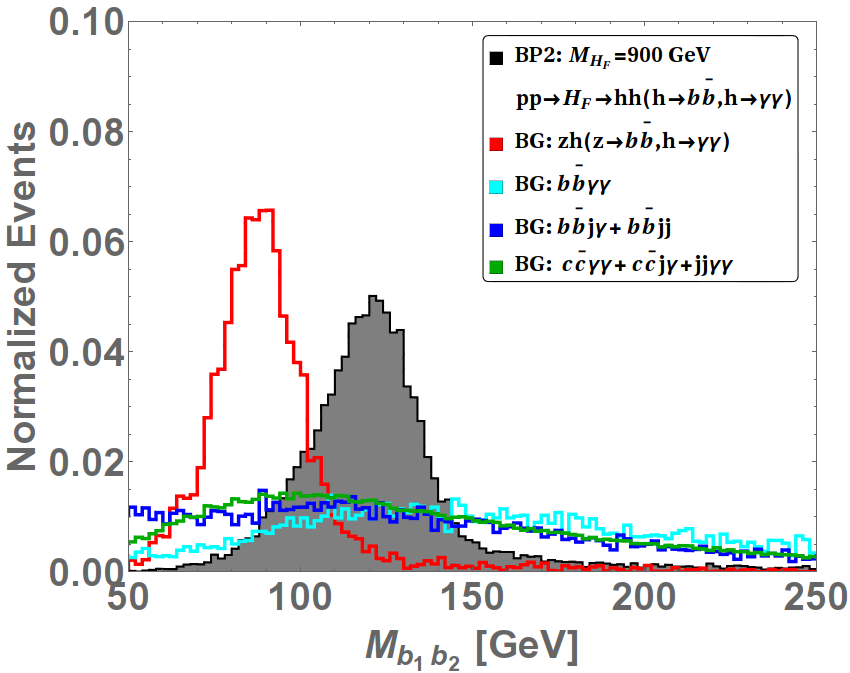} 
\includegraphics[scale=0.17]{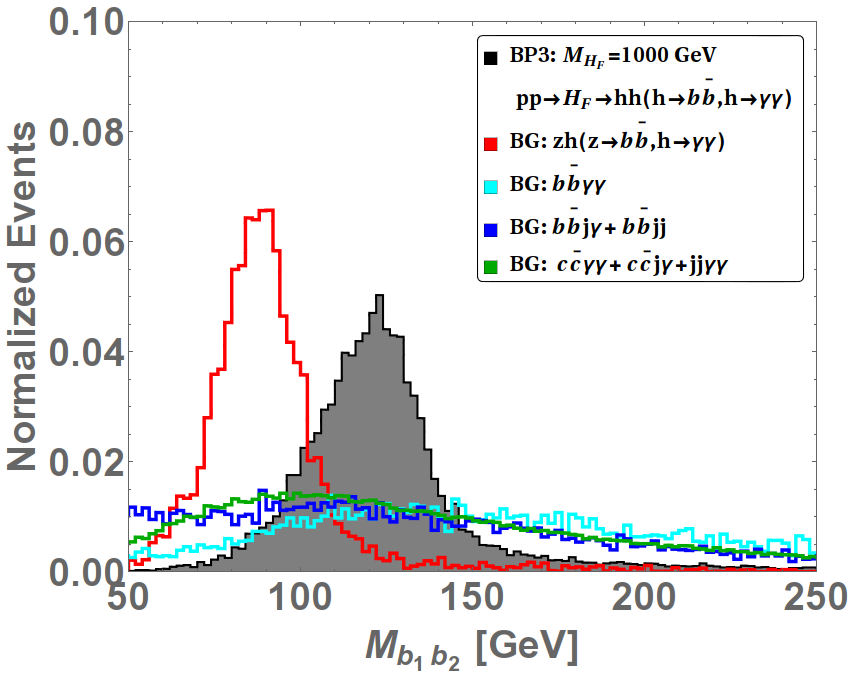}
\end{center}
\caption{Normalized distributions in di-photon and di-jet  invariant mass for  signal and total background after the acceptance cuts.}
\label{fig:Dist4}
\end{figure}
\begin{figure}[h!]
\begin{center}
\includegraphics[scale=0.16]{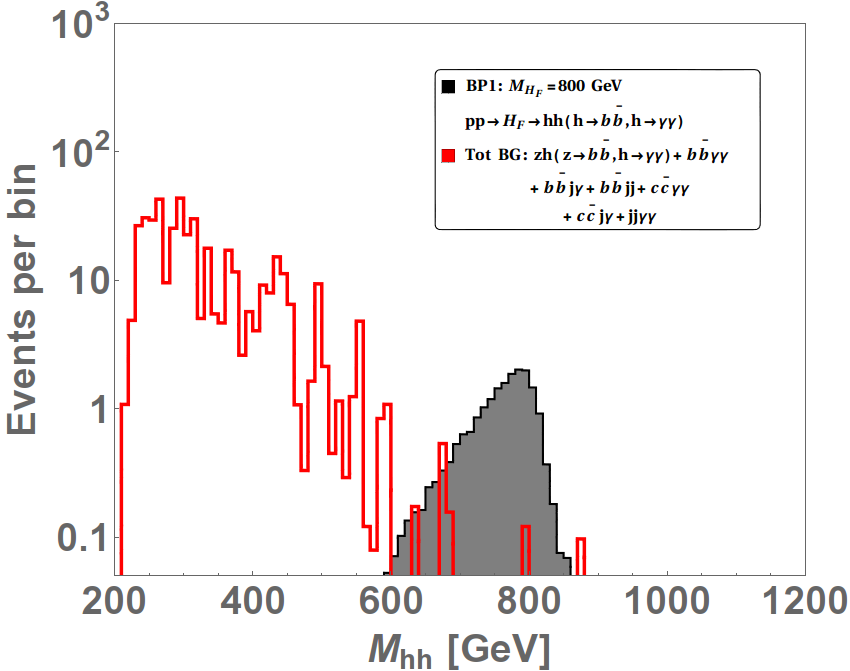}
\includegraphics[scale=0.16]{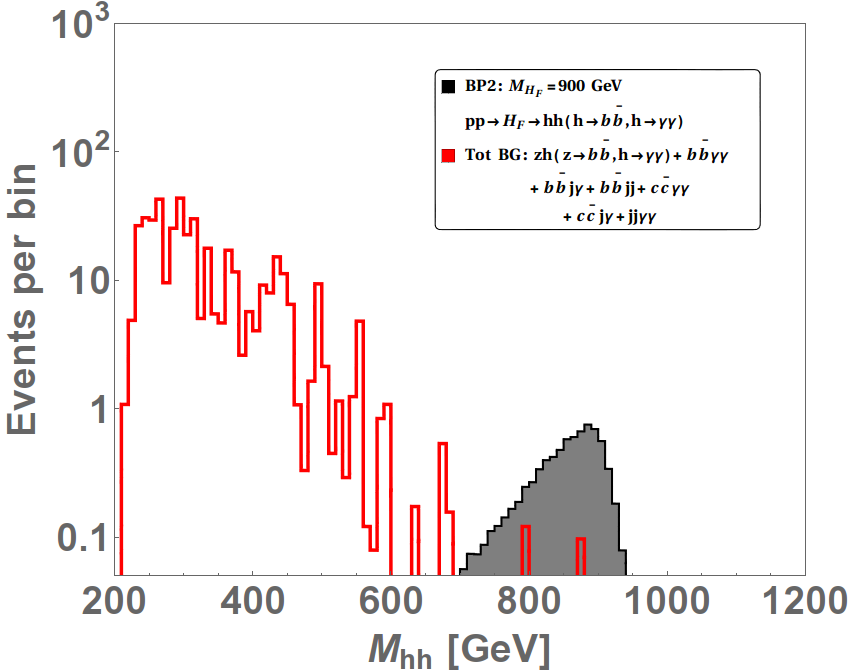}
\includegraphics[scale=0.16]{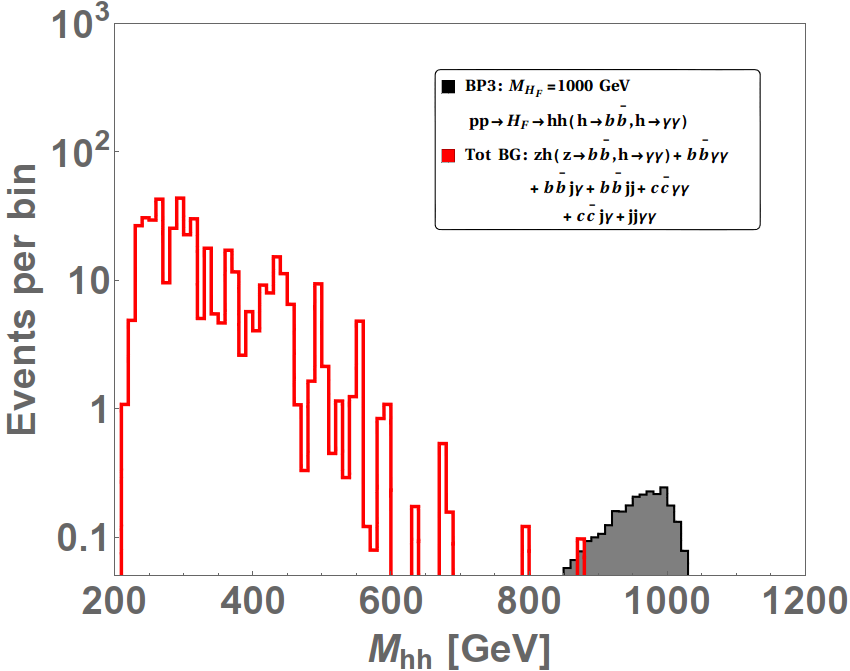} 
\end{center}
\caption{Distributions in the final state invariant mass for  signal and total background after the acceptance cuts as well as the selection ones on $p_T^{\gamma_{1,2}}$, $p_T^{b_{1,2}}$, $M_{\gamma_1 \gamma_2}$, $M_{b_1 b_2}$, $\Delta R_{\gamma_1 \gamma_2}$ and $\Delta R_{b_1 b_2}$, as  shown in Tab.~\ref{table:sr}.}
\label{fig:Dist1a}
\vspace*{1cm}
\end{figure}

The (arbitrarily) normalized distributions of all these kinematic variables for the three signal BPs and the total  background are  shown in 
Figs.~\ref{fig:Dist1}--\ref{fig:Dist1a}. Based on their inspection, as intimated, we then perform  a detailed cut-based analysis to maximize the signal significance against the  background.
\begin{table}[h!]
\begin{center}\scalebox{1.0}{
\begin{tabular}{|c|c|c|c|c|c|c|c|c|c|c|c|c|c|}
\hline
  &\multicolumn{2}{c|}{ Kinematic variables and cuts   }\\
\cline{2-3}
& ~~~~Observable~~~~& ~~~~~~~~~Value~~~~~~~~~\\
\cline{1-3}
\rule{0pt}{1ex}
  & $p_T^{\gamma_{1,2}}$ & $>$ 35.0 ~~~~~~~~(GeV)\\
\rule{0pt}{1ex}
  & $p_T^{b_{1,2}}$ &  $>$ 40.0 ~~~~~~~~(GeV) \\
\rule{0pt}{1ex}
SR  & $M_{\gamma_1 \gamma_2}$ &  $122.5-128.5$ ~~(GeV) \\
\rule{0pt}{1ex}
  & $M_{b_1 b_2}$ &  $70.0-135.0$ ~~(GeV) \\
\rule{0pt}{1ex}
  & $\Delta R_{\gamma_1 \gamma_2}$ & $0.4-4.6$ \\
\rule{0pt}{1ex}
  & $\Delta R_{b_1 b_2}$ & $0.4-3.6$ \\
\rule{0pt}{3ex}
  & $ M_{hh}$ (varied with $M_{H_F}$)& $0.7 M_{H_F} - 1.1 M_{H_F}$ \\
\hline
\end{tabular}}
\end{center}
\caption{The optimized SR as a function of the $H_F$ mass.}
\label{table:sr}
\vspace*{1cm}
\end{table}
The sequence of constraints adopted is shown in Tab.~\ref{table:sr}. Specifically, notice that, in applying the last requirement herein (on the $M_{hh}$
variable), one may assume that the $M_{H_F}$ value is a trial one, if it were not already known from previous analysis. 

The signal yields for BP1, BP2 and BP3, along with the
corresponding background ones,  obtained after the 
application of the acceptance and selection cuts defining the  SR, are shown in Tab.~\ref{table:signalsignificance} for $\sqrt s=14$ TeV and, e.g.,  
$\mathcal{L}=$  $3000 ~ {\rm fb^{-1}}$.
We initially calculate the signal significance using the relation $\sigma=\frac{S}{\sqrt{S+B}}$.
Here, $S$ and $B$ stand for the Signal and (total SM)  Background rates, respectively.
The number of $S$ and $B$  events is obtained  as $S,B=
\epsilon A \sigma_{S,B} \mathcal{L}$, where $\epsilon$ and $A$ stand for the selection and acceptance cut efficiency, respectively, $\sigma_{S,B}$ is the
$S$ or $B$ 
cross section  and  $\mathcal{L}$ is the luminosity.  Based on these definitions, it is clear from Tab.~\ref{table:signalsignificance}  that strong HL-LHC
sensitivity exists for all  $M_{H_F}$ choices, ranging from discovery (at small masses) to exclusion (at high masses). (It should
be appreciated  that these significances would be reduced by as much as $30\%$ in the absence of the final $M_{hh}$ selection.)
In fact, one can also consider {the systematic uncertainty 
in various SM background estimations while calculating the final signal
significance as}\footnote{To include the systematic uncertainty in $\sigma=\frac{S}{\sqrt{S+B}}$, one can replace $S+B$ in the denominator by the quadratic sum of $\sqrt{S+B}$ and use $\sigma_b=\kappa B$~\cite{SigForm}, i.e., $\sigma=\frac{S}{\sqrt{S+B+(\kappa B)^2}}$, with $\kappa$ being the percentage of systematic uncertainty of the total background.} 
$\sigma=\frac{S}{\sqrt{S+B+(\kappa B)^2}}$, where $\kappa$ is the 
percentage of systematic uncertainty \cite{SigForm}.
Upon adding $5\%$ for the latter, the significance in Tab.~\ref{table:signalsignificance} for BP1 decreases to $3.75$ while for 
BP2 and BP3 it  becomes $2.31$ and $1.24$, respectively. Hence, the HL-LHC sensitivity is very stable against unknowns affecting the data sample estimations, whatever the origin.

\begin{table}[h!]
\begin{center}\scalebox{0.750}{
\begin{tabular}{|c|c|c|c|c|c|c|c|c|c|c|c|c|c|}
\hline
\multicolumn{9}{|c|}{ Benchmark points: Signal and Significances  }\\
\cline{1-9}
 \multicolumn{3}{|c|}{  BP1~($M_{H_F}=800$ GeV) }&\multicolumn{3}{c|}{  BP2~($M_{H_F}=900$ GeV) }&\multicolumn{3}{c|}{ BP3~($M_{H_F}=1000$ GeV) }\\
\cline{1-9}
$\#$ Signal&$\#$ Background & Significance &$\#$ Signal&$\#$ Background & Significance&$\#$ Signal&$\#$ Background & Significance\\
\hline
18.45& 5.65&  3.81  &7.92&3.72&2.32  &3.10&$\sim 3$&1.25\\
\hline
\end{tabular}}
\end{center}
\caption{The signal significance $\sigma=\frac{S}{\sqrt{S+B}}$ for BP1, BP2 and BP3 corresponding to the optimized SR are shown. In addition, the total background yield and the total signal yield are also given at $\sqrt{s}=14$ TeV with integrated luminosity $\mathcal{L}=3000~{\rm fb^{-1}}$.}
\label{table:signalsignificance}
\end{table}

\begin{figure}[h!]
\begin{center}
\includegraphics[scale=0.28]{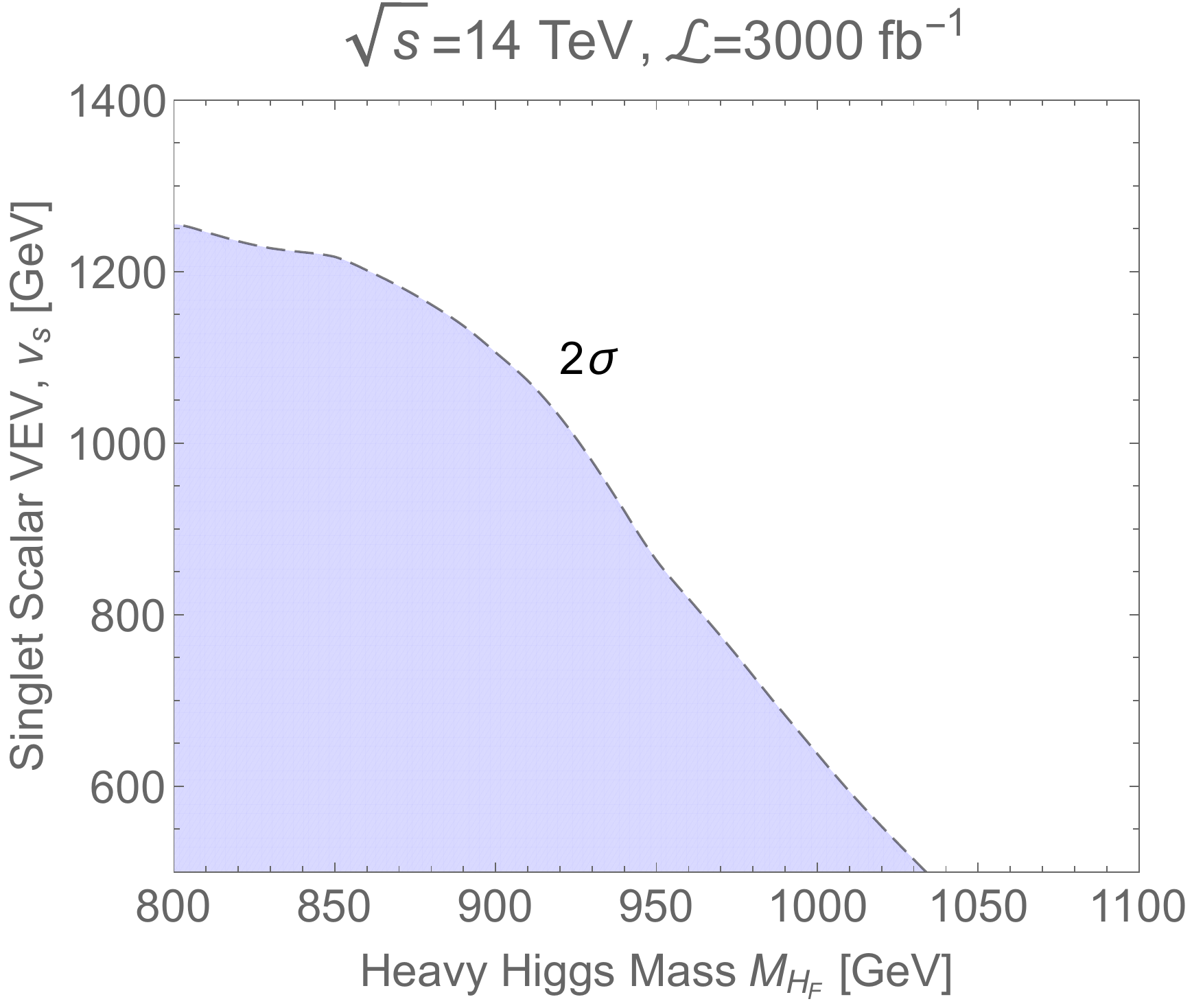}
\includegraphics[scale=0.28]{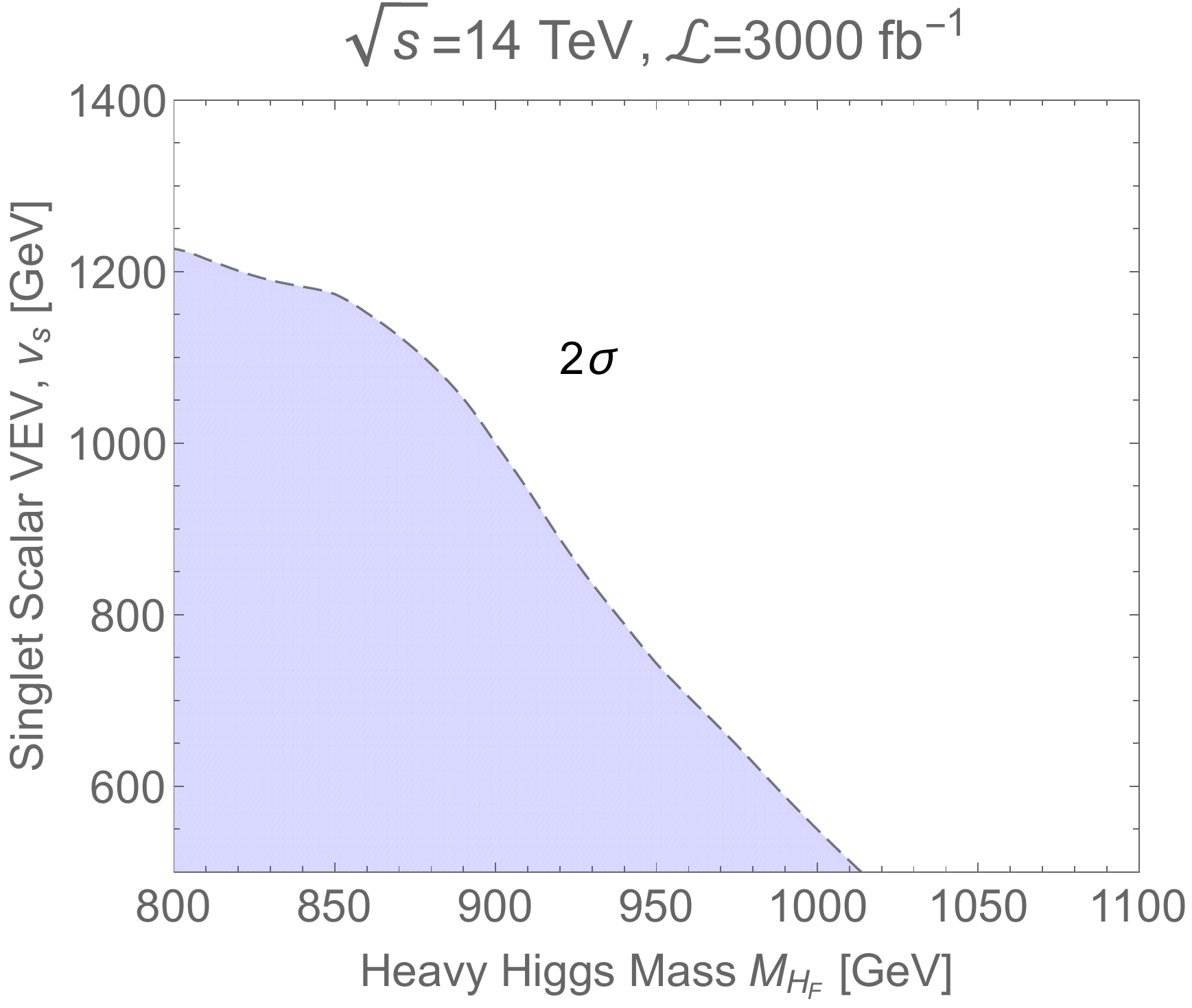}
\end{center}
\caption{The projected exclusion (light blue) and discovery (dark
blue) regions in the $M_{H_F}-$ $v_s$ plane. These plots are drawn for $\mathcal{L}=3000$ fb$^{-1}$. The right plot is drawn considering a systematic uncertainty $\kappa=5\%$. }
\label{fig:contour1}
\end{figure}
We now derive the various projected limits over the $M_{H_F}-v_s$ plane. It is to be noted that the variation of the  singlet scalar VEV $v_s$ will directly change the $H_F hh$ coupling and correspondingly the production cross section $\sigma(pp\rightarrow H_F\to hh)$.
In particular, the smaller the former  the larger the  latter. To accurately delineate sensitivity regions, we generate a large number of signal events for various combinations of heavy CP-even Flavon mass, $M_{H_F}$, and singlet scalar VEV, $v_s$. Specifically,  $M_{H_F}(\equiv M_{A_F})$  has been varied from $800~{\rm GeV}$ to $1000~{\rm GeV}$ with a step size of $5~{\rm GeV}$ while  $v_s$ has been varied between $500$ and $1000$ GeV with a step size of 25 GeV. 
The projected exclusion ($2\sigma$) region derived from the $\gamma\gamma b \bar{b}$ final state in the $M_{H_F}-v_s$ plane are given in Fig.~\ref{fig:contour1}.
The left plot is drawn for $\mathcal{L}=3000$  fb$^{-1}$ (HL-LHC).
Again, the left plot in Fig.~\ref{fig:contour1} is shown with no systematic uncertainty, i.e., $\kappa=0$, while the right plot is drawn based on a systematic uncertainty $\kappa=5\%$.
From the right plot, we should mention that the limits drop somewhat (by $5-10\%$) upon introducing a systematic uncertainty of $\kappa=5\%$, hence not too drastic a reduction of sensitivity in general (as already remarked for our BPs).

\subsection{$pp \to H_F\to ZZ~(Z\to  \ell \bar{\ell })$ }
In this section, we now discuss the signatures involving the final state with four leptons 
($2\bar{\ell}+2\ell$) in the context of HL-LHC. The primary contribution to these signatures 
typically arises from the process $pp \to H_F\to ZZ$, where each $Z$ boson further decays into 
a lepton-antilepton pair ($Z\to \ell \bar{\ell}$). To investigate the leptons' final state 
signatures, we have selected the same three benchmark points, which are $M_{H_F} = 800, 900$, and $1000$ GeV, respectively. 
The table~\ref{tab:cs2} displays the signal cross-sections for different processes. Among them, the primary background in the Standard Model is the production of two Z bosons accompanied by jets ($ZZ + jets$). In addition, there are other significant reducible backgrounds, such as the production of top quark pairs with jets ($t\bar{t} + jets$), the production of a Z boson and a Higgs boson with jets ($Zh + jets$), and so on. We have included all the relevant Standard Model backgrounds in the table~\ref{tab:csBG2}.
\begin{table}[h!]
\begin{center}\scalebox{1.0}{
\begin{tabular}{|c|c|c|c|c|c|c|c|c|c|c|c|c|c|}
\hline
BPs  [GeV] &\multicolumn{3}{c|}{ {\rm BR}s and cross sections [pb] }\\
\hline
& ${\rm BR}(H_F \to ZZ)$ & $ \sigma( pp \rightarrow  H_F$) & $\sigma(pp \to H_F$ $\to ZZ, Z \to \ell 
\bar{\ell})$ \\
\cline{2-4}
\rule{0pt}{1ex}
BP1 ($M_{H_F}=800$)   &  $0.10$  & $0.41$ &  $1.90\times 10^{-5}$\\
\rule{0pt}{1ex}
BP2  ($M_{H_F}=900$)  &$0.11$  & $0.21$ &  $1.07\times 10^{-5}$ \\
\rule{0pt}{1ex}
BP3  ($M_{H_F}=1000$)  & $0.12$ & $0.11$ &   $6.23\times 10^{-6}$\\
\hline
\end{tabular}}
\end{center}
\caption{The ${\rm BR}(H_F \to ZZ)$ and cross sections for the processes $ pp \rightarrow  H_F$ and  $\sigma(pp \to H_F$ $\to ZZ, Z \to \ell \bar{\ell})$ for three BPs (BP1, BP2 and BP3) used in the remainder of the paper. }
\label{tab:cs2}
\end{table}
\begin{table}[h!]
\begin{center}\scalebox{1.0}{
\begin{tabular}{|c|c|c|c|c|c|c|c|c|c|c|c|c|c|}
\hline
 SM backgrounds & Cross section [pb]  \\
\hline
\rule{0pt}{3ex}
$ pp \to ZZ jets$ (upto 3 jets)  &    11.64    \\
\rule{0pt}{3ex}
$ pp \to t \bar{t} Z jets$ (upto 2 jets)  &    0.76    \\
\rule{0pt}{3ex}
$ pp \to VVV (V=W/Z) jets$ (upto 2 jets)  &    1.04    \\
\rule{0pt}{3ex}
$ pp \to VH jets$ (upto 3 jets)  &    0.69    \\
\rule{0pt}{3ex}
$ pp \to WZ jets$ (upto 3 jets)  &    40.10    \\
\rule{0pt}{3ex}
$ pp \to WW jets$ (upto 3 jets)  &    89.20    \\
\rule{0pt}{3ex}
$ pp \to t \bar{t} jets$ (upto 2 jets)  &    915.10    \\
\hline
\end{tabular}}
\end{center}
\caption{The matched cross sections for the most relevant SM background processes. (Note that these background rates will be multiplied by the fake rates during the analysis.)}
\label{tab:csBG2}
\end{table}
 
In this particular scenario, the event must contain precisely four isolated leptons, consisting of two positively charged leptons and two negatively charged leptons. This requirement ensures the presence of same-flavor opposite-sign (SFOS) leptons (electron and/or muon) in the final state. However, no specific constraints are imposed on the number of light jets present in the event. We then adopt the following acceptance cuts: 
\begin{itemize}
\item $p_T^\gamma > 20 $ GeV; 
\item $p_T^{e/\mu} > 20 $ GeV;
\item $p_T^j > 40 $ GeV, where $j$ stands for light-jets as well as
$b$-jets;
\item $\mid \eta_\ell \mid ~\leq 2.5 $ (again, $\ell = e/\mu $), $\mid \eta_\gamma \mid ~\leq 2.0 $ and $\mid \eta_j \mid ~\leq 2.0 $.
\end{itemize}
After considering these basic requirements, we apply additional cuts using kinematic variables to enhance the signal-to-background ratio.
Various such kinematic variables have been used to design the optimized Signal Region (SR), i.e., where the significance is maximized.
First and foremost, the transverse momentum of the leptons ($p_T^{\ell_i}, i=1..4$) and the minimum invariant mass $M_{\ell \ell }^{min}$ out of four combinations ($M_{\ell_i \ell_j},~i,j=1..4$) and total 
transverse momentum of four leptons ($\sum p_T^{\ell_i}$) will be studied.

\begin{figure}[!htbp]
\begin{center}
\includegraphics[scale=0.17]{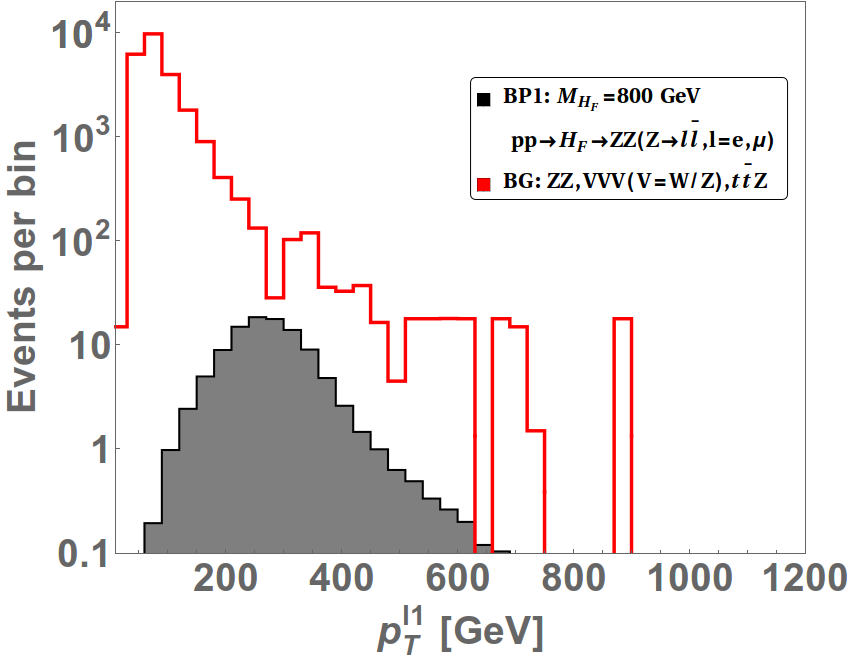} 
\includegraphics[scale=0.17]{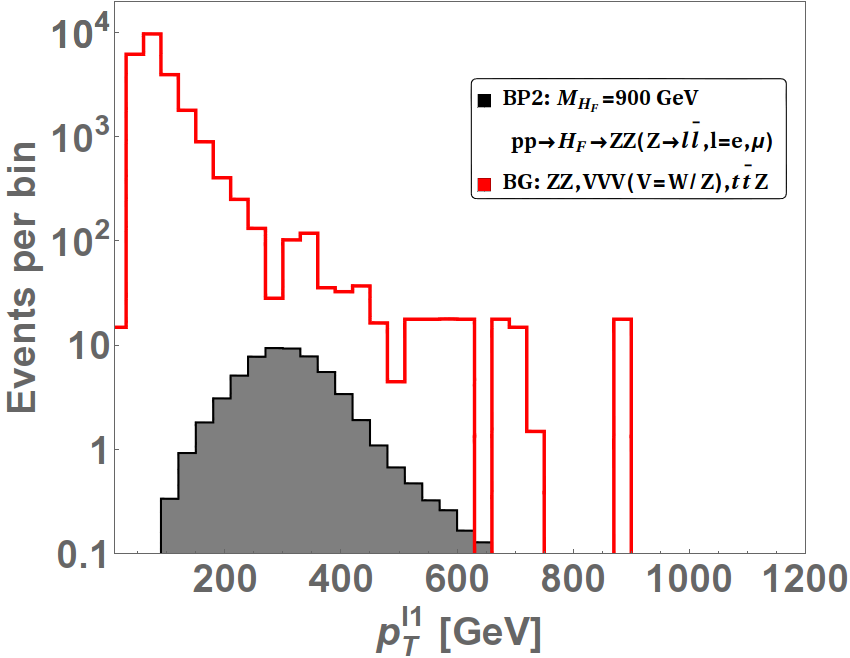}
\includegraphics[scale=0.17]{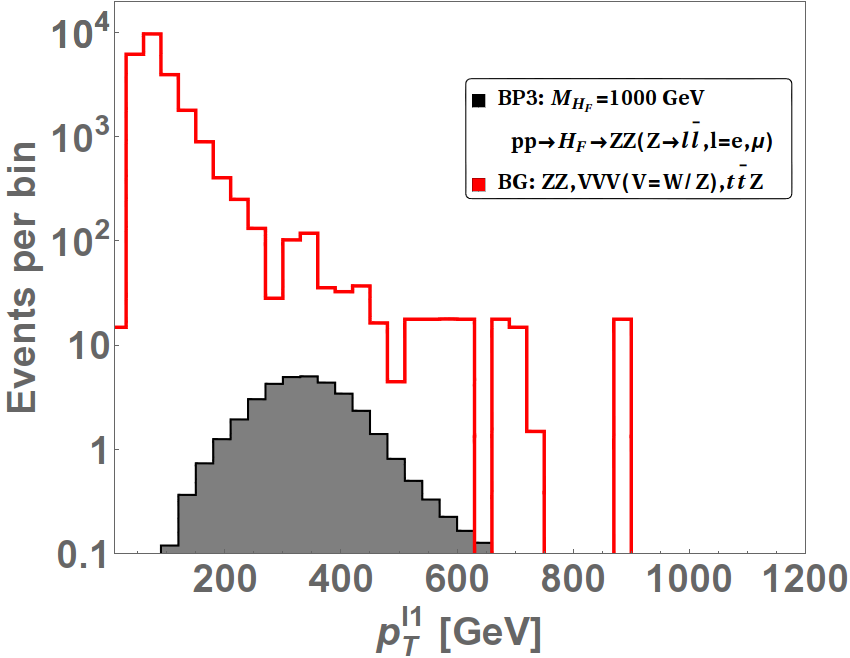}
\rule{0pt}{3ex}\\
\includegraphics[scale=0.17]{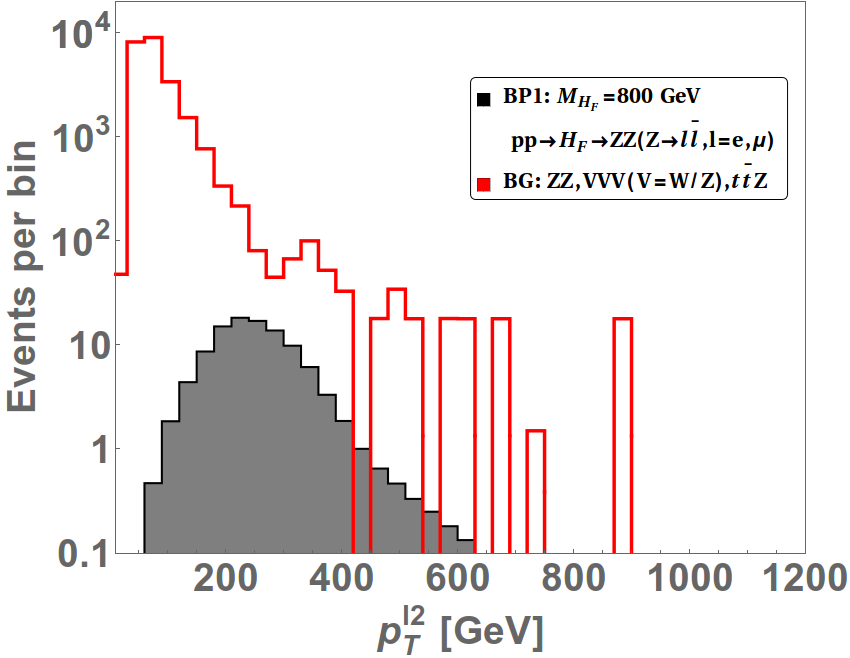} 
\includegraphics[scale=0.17]{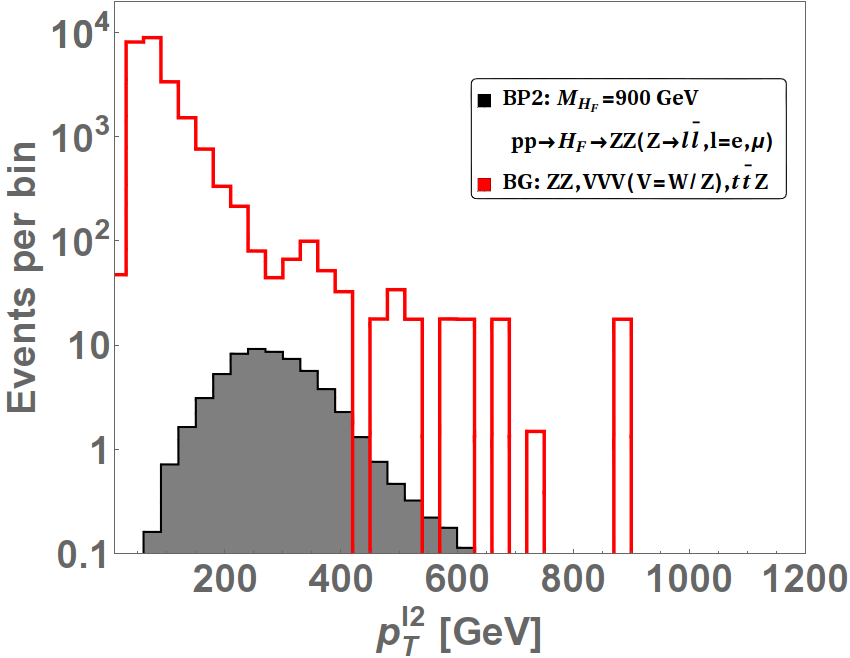}
\includegraphics[scale=0.17]{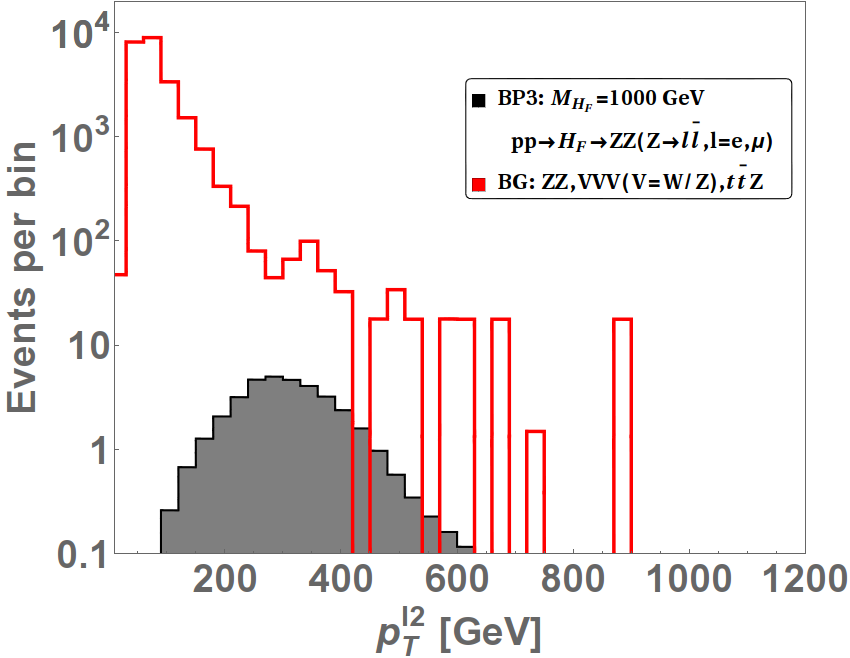}
\includegraphics[scale=0.17]{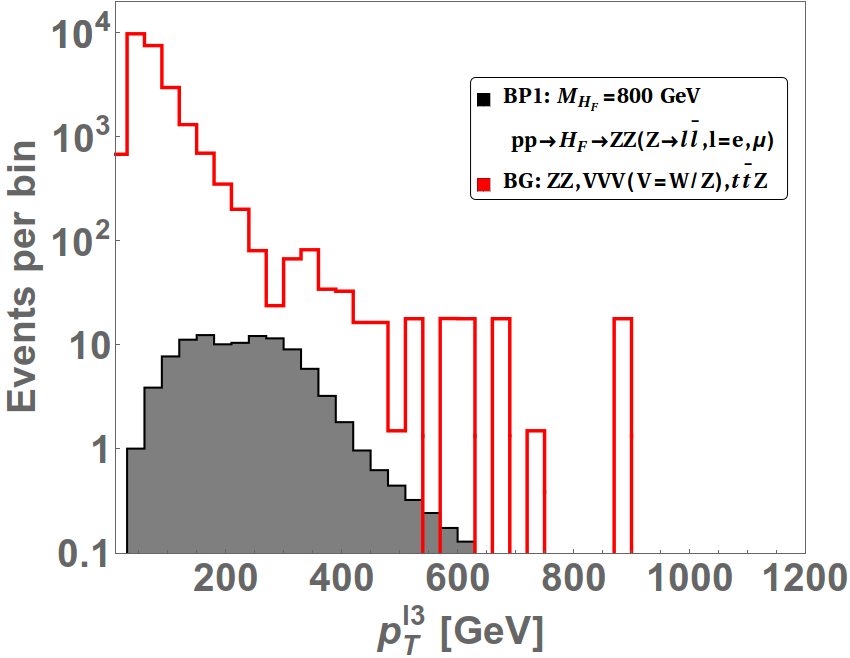} 
\includegraphics[scale=0.17]{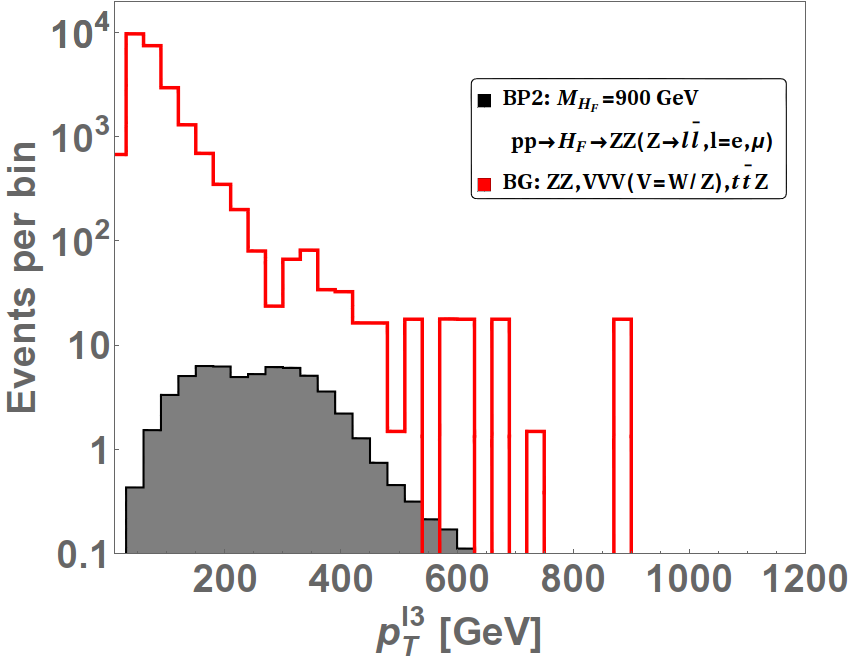}
\includegraphics[scale=0.17]{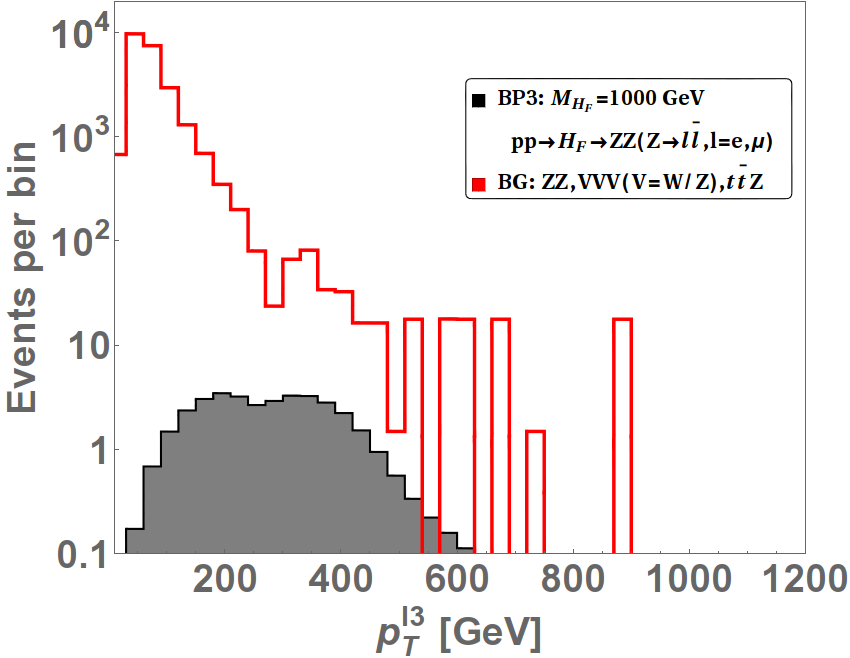}
\rule{0pt}{3ex}\\
\includegraphics[scale=0.17]{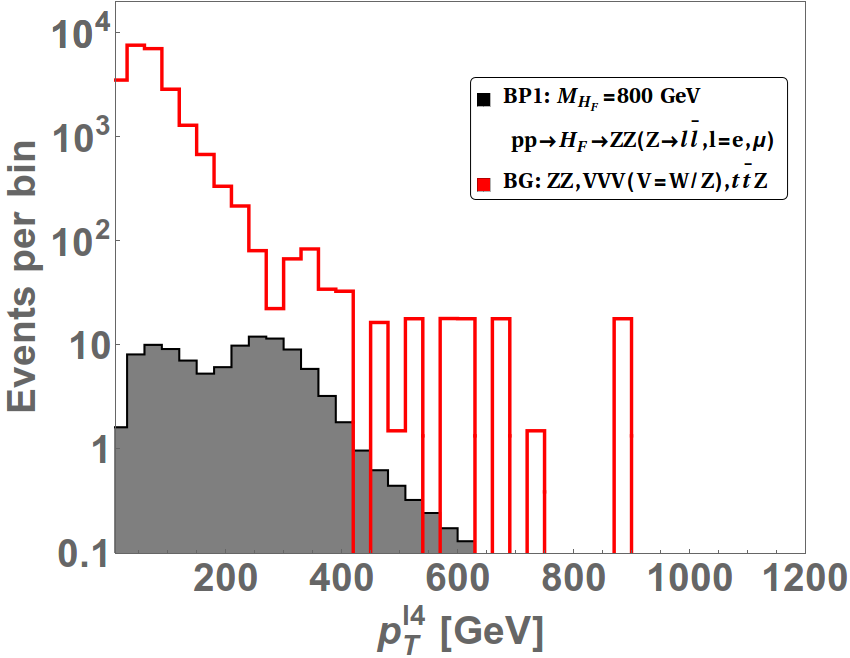} 
\includegraphics[scale=0.17]{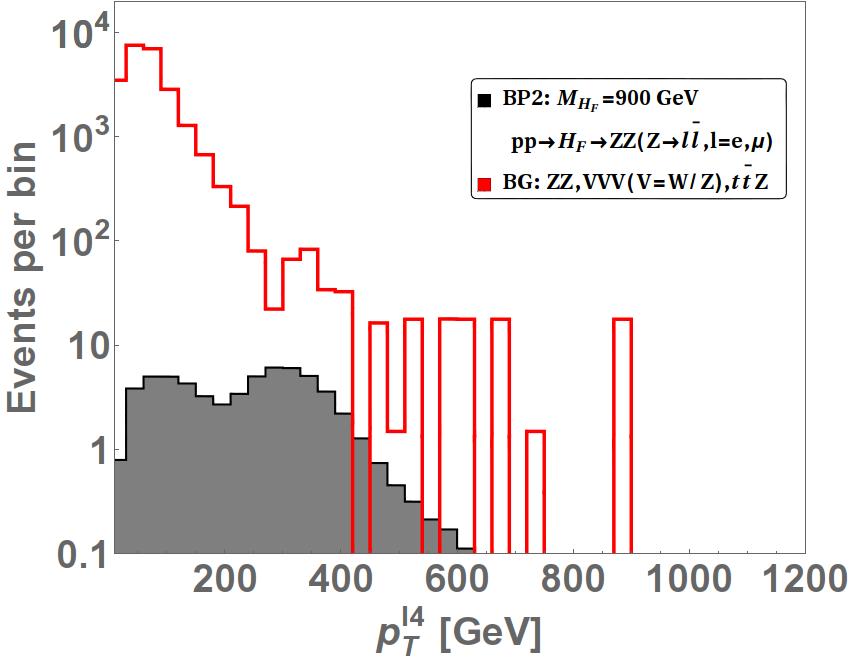}
\includegraphics[scale=0.17]{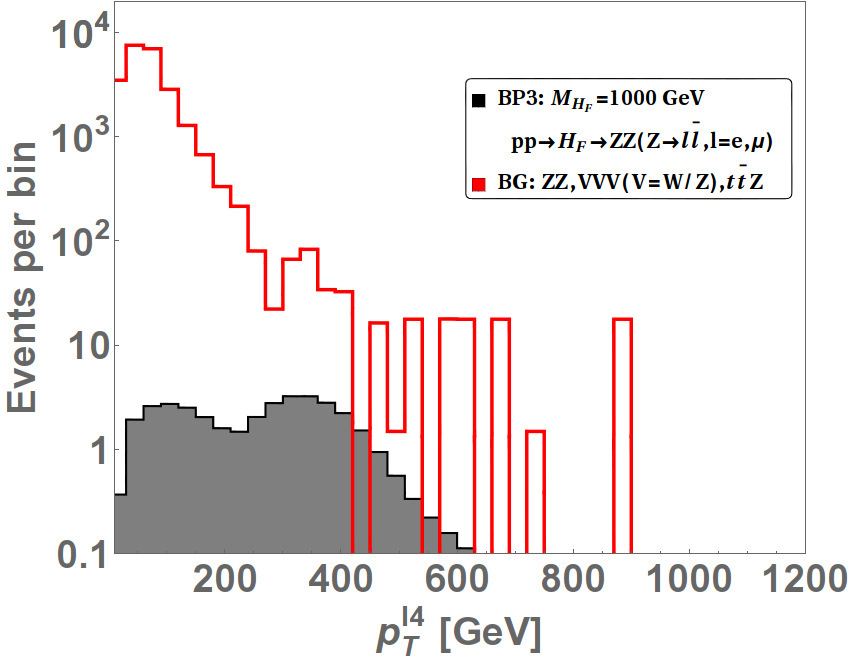}
\end{center}
\caption{Transverse momentum distributions for  signal and total background after the acceptance cuts.}
\label{fig:Dist4l}
\end{figure}
Finally, we use the invariant mass $M_{ZZ}$ for the final extraction. The $M_{ZZ}$ variable has been 
calculated as $M_{ZZ}=\sqrt{(E^{\ell_1}+E^{\ell_2}+E^{\ell_3}+E^{\ell_4})^2 -  \sum_{i=x,y,z}  
(p^{\ell_1}_{i} +p^{\ell_2}_{i} + p^{\ell_3}_{i} + p^{\ell_4}_{i})^2 }$.
Here $E$ and $p_{i}~(i=x,y,z)$ stand for the energy and three-momentum component of the final state leptons, respectively.

\begin{figure}[!htbp]
\begin{center}
\includegraphics[scale=0.17]{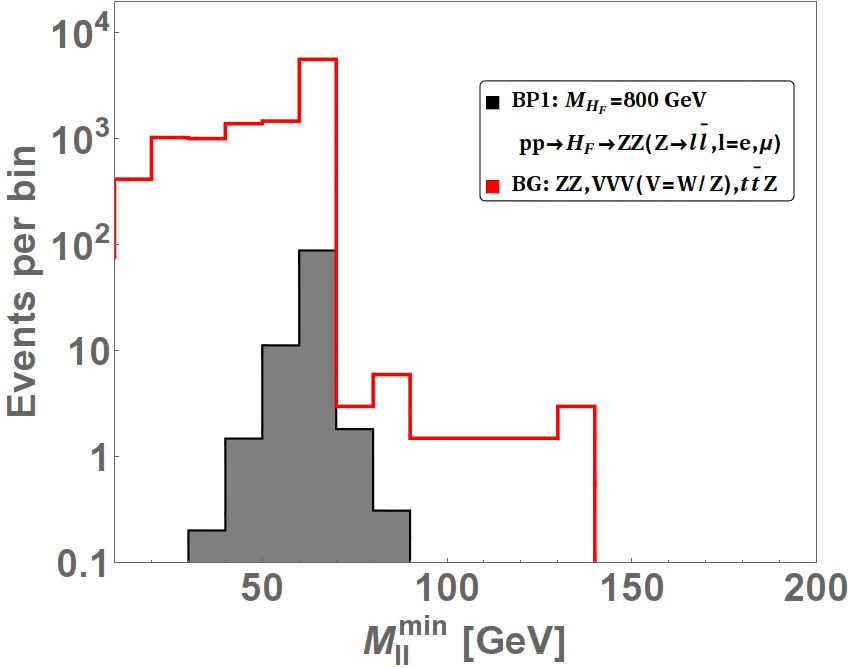} 
\includegraphics[scale=0.17]{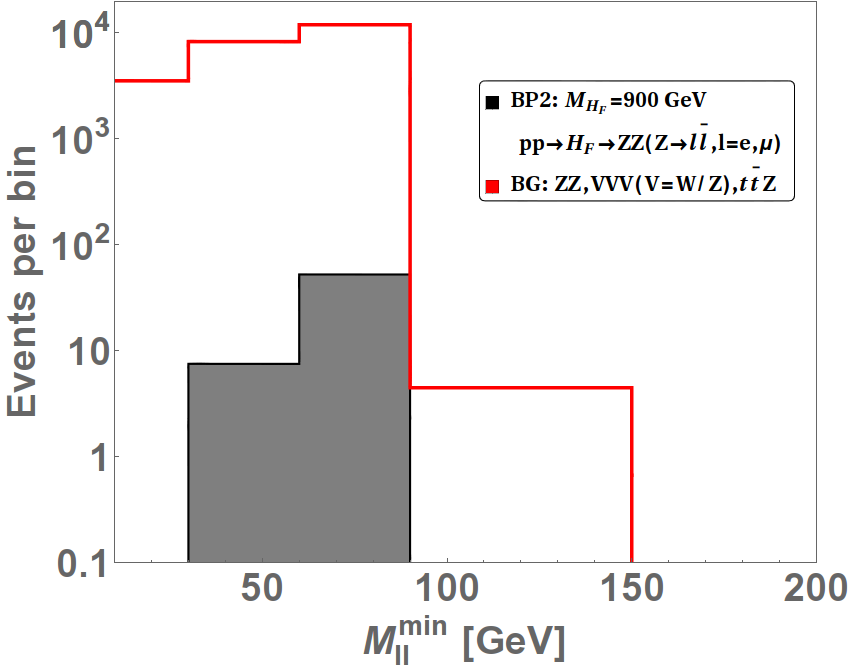}
\includegraphics[scale=0.17]{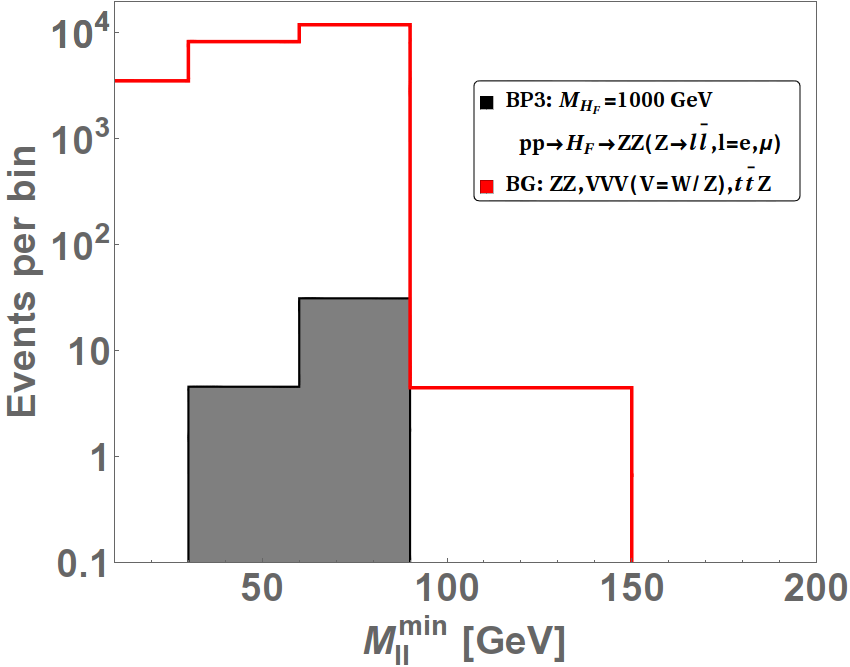}
\rule{0pt}{3ex}\\
\includegraphics[scale=0.17]{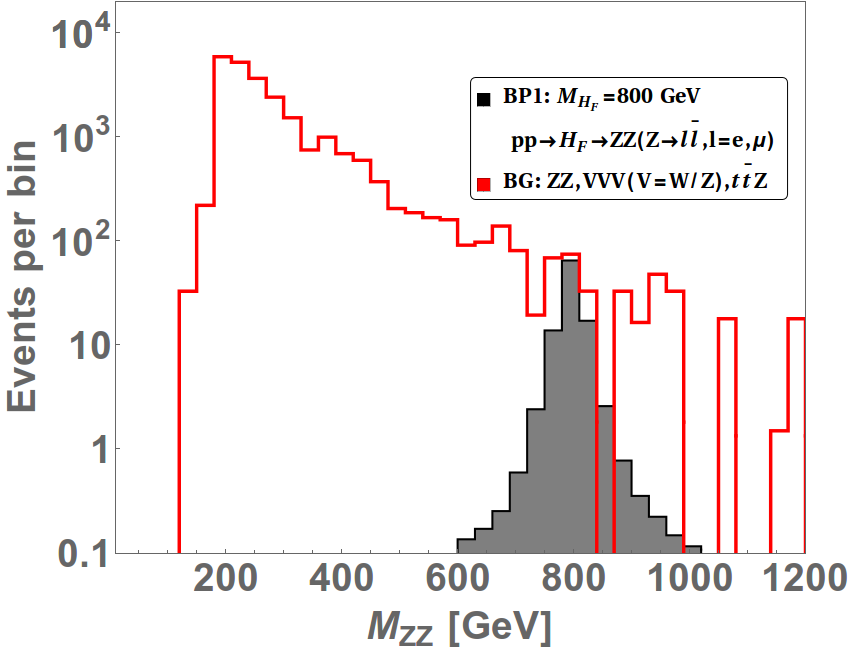} 
\includegraphics[scale=0.17]{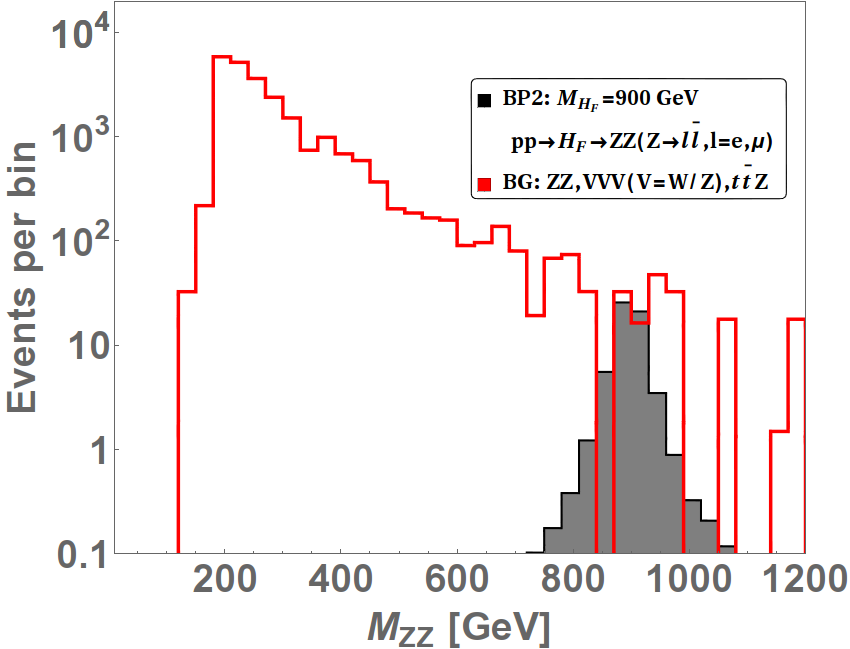}
\includegraphics[scale=0.17]{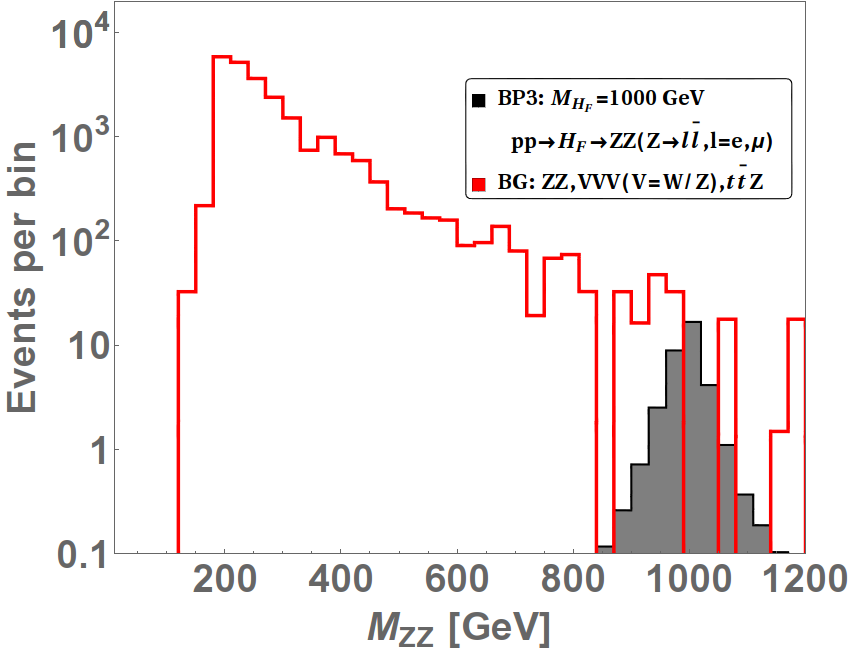}
\rule{0pt}{3ex}\\
\includegraphics[scale=0.17]{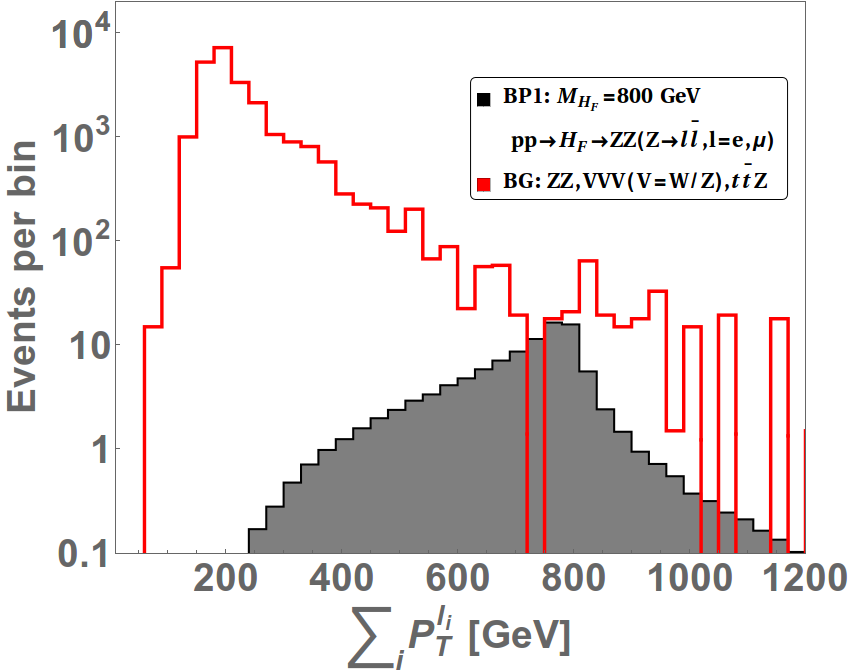} 
\includegraphics[scale=0.17]{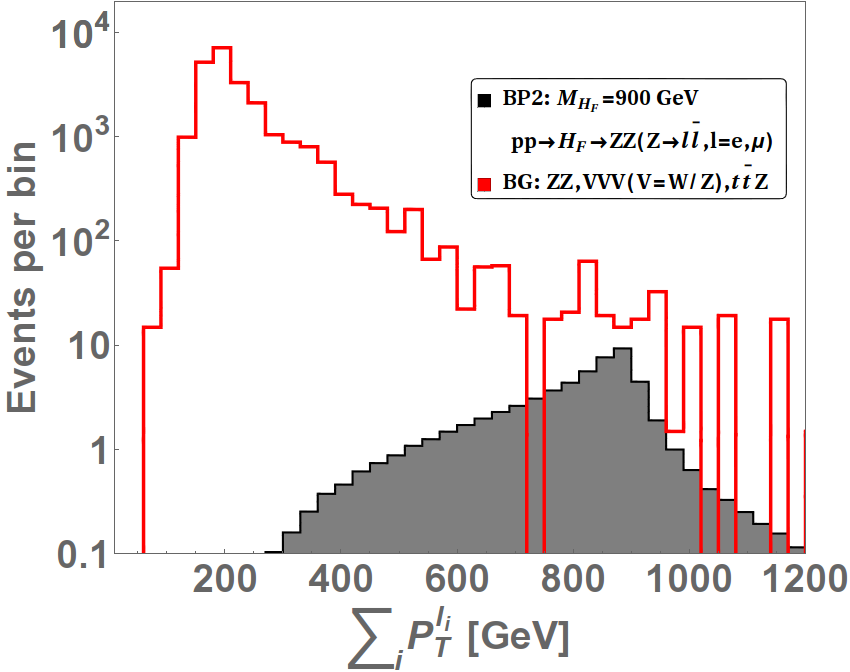}
\includegraphics[scale=0.17]{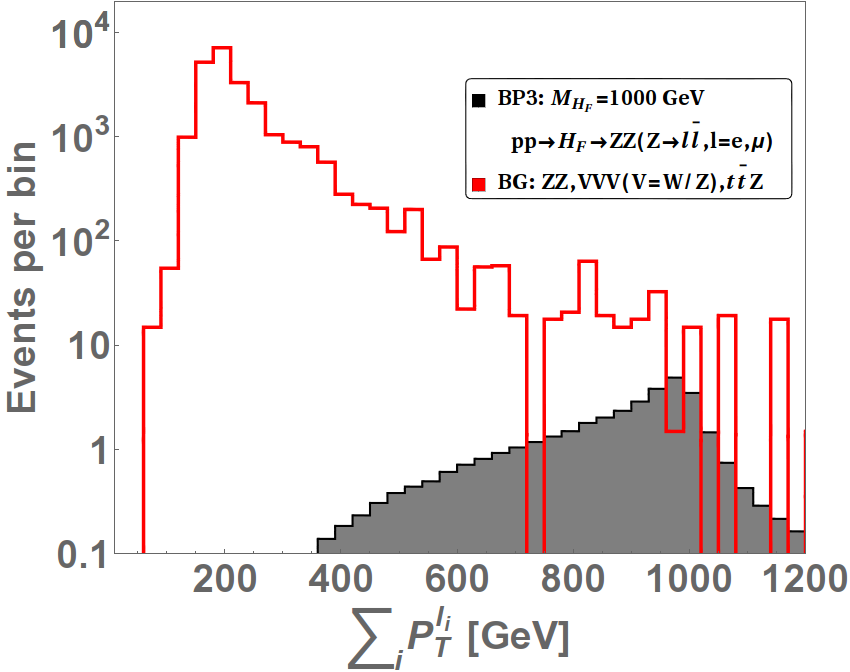}
\rule{0pt}{3ex}\\
\end{center}
\caption{Invariant mass of two leptons, four leptons, sum of all momentum distributions for  signal and total background after the acceptance cuts.}
\label{fig:Dist4n}
\end{figure}

The normalized distributions of all these kinematic variables for the three signal BPs and the total  background for this analysis are  shown in Figs.~\ref{fig:Dist4l}--\ref{fig:Dist4n}. We then perform  a detailed cut-based analysis to maximize the signal significance against the SM backgrounds. 
The figures labeled~\ref{fig:Dist4l} to~\ref{fig:Dist4n} illustrate the normalized distributions of various kinematic variables for the three signal benchmark points (BPs) as well as the total background in this analysis. Subsequently, we employ a thorough cut-based analysis technique to optimize the signal significance with respect to the Standard Model backgrounds. The specific sequence of cuts applied during this analysis is presented in Tab.~\ref{table:sr2}.
\begin{table}[!htbp]
\begin{center}\scalebox{1.0}{
\begin{tabular}{|c|c|c|c|c|c|c|c|c|c|c|c|c|c|}
\hline
  &\multicolumn{2}{c|}{ Kinematic variables and cuts   }\\
\cline{2-3}
& ~~~~Observable~~~~& ~~~~~~~~~Value~~~~~~~~~\\
\cline{1-3}
\rule{0pt}{1ex}
  & $p_T^{\ell_{1,2,3,4}}$ &  $>$ 35 ~~~~(GeV) \\
\rule{0pt}{1ex}
SR  & $\sum_i p_T^{\ell_{i}}$ &  $>$ 180 ~~(GeV) \\
\rule{0pt}{3ex}
  & $ M_{ZZ}$ (varied with $M_{H_F}$)& $0.95 M_{H_F} - 1.05 M_{H_F}$ \\
\hline
\end{tabular}}
\end{center}
\caption{The optimized SR as a function of the $H_F$ mass.}
\label{table:sr2}
\vspace*{1cm}
\end{table}

The Tab.~\ref{table:signalsignificance2} shows the signal yields for three benchmark points and the corresponding yields for the SM background. We obtained these numbers after applying acceptance and selection cuts that define the signal region (SR). The calculations were performed for a center-of-mass energy of $\sqrt s=14$ TeV and an integrated luminosity of $\mathcal{L}=3000 ~ {\rm fb^{-1}}$. We calculate the signal significance using the formula $\sigma = \frac{S}{\sqrt{S+B}}$, where $S$ represents the signal yield and $B$ represents the background yield.

\begin{table}[!htbp]
\begin{center}\scalebox{0.750}{
\begin{tabular}{|c|c|c|c|c|c|c|c|c|c|c|c|c|c|}
\hline
\multicolumn{9}{|c|}{ Benchmark points: Signal and Significances  }\\
\cline{1-9}
 \multicolumn{3}{|c|}{  BP1~($M_{H_F}=800$ GeV) }&\multicolumn{3}{c|}{  BP2~($M_{H_F}=900$ GeV) }&\multicolumn{3}{c|}{ BP3~($M_{H_F}=1000$ GeV) }\\
\cline{1-9}
$\#$ Signal&$\#$ Background & Significance &$\#$ Signal&$\#$ Background & Significance&$\#$ Signal&$\#$ Background & Significance\\
\hline
90.42& 83.67&  6.83  &51.48&27.98&5.78  & 51.77&66.95&4.75\\
\hline
\end{tabular}}
\end{center}
\caption{The signal significance $\sigma=\frac{S}{\sqrt{S+B}}$ for BP1, BP2 and BP3 corresponding to the optimized SR are shown. In addition, the total background yield and the total signal yield are also given at $\sqrt{s}=14$ TeV with integrated luminosity $\mathcal{L}=3000~{\rm fb^{-1}}$.}
\label{table:signalsignificance2}
\end{table}
\subsection{$pp\to H_F \to tc$ $(t\to b\ell\nu_{\ell})$}

The presence of non-zero $\tilde{Z}_{tc}$ allows for processes such as $H\to tc$, where the heavy Higgs decays into a top quark and an anti-charm quark, or a charm quark and an anti-top quark, respectively. These flavor-violating decays are possible due to the mixing between the top and charm quarks induced by the non-zero $\tilde{Z}_{tc}$.
The observation of such flavor-violating decays would have significant implications for our understanding of the $FN$ heavy Higgs sector. It would provide evidence for new physics beyond the SM, as the SM predicts negligible flavor violation in the Higgs sector. The presence of flavor-violating decays would suggest the existence of new particles or interactions that can induce such processes.

Studying the properties of the flavor-violating decays, such as their rates and kinematic distributions, can provide valuable information about the underlying physics responsible for the $FN$ heavy Higgs sector. It can help constrain the model's parameter space and provide insights into the flavor structure and dynamics of the theory. We present the analysis for the production of the $H_F$ via proton-proton collisions $pp\to H_F$, followed by the  FCNC decay  $H_F \to tc$ $(t\to b\ell\nu_{\ell})$ in the presence of non-zero $\tilde{Z}_{tc}$.
The model parameter values used in the simulation are shown in Table \ref{TablaParametros}.
\begin{table}[!htb]
\begin{centering}
\begin{tabular}{|c|c|}
\hline
\hline
Parameter & Value\tabularnewline
\hline
\hline
$c_{\alpha}$ & $0.995$\tabularnewline
$v_s$   & $600-1000$ (GeV)\tabularnewline
$\tilde{Z}_{tc}$ & $0.1$\tabularnewline
$M_{H_{F}}$ & $800-1000$ (GeV)\tabularnewline
\hline
\hline
\end{tabular}
\par\end{centering}
\caption{Model parameter values used in the Monte Carlo simulation.}
\label{TablaParametros}
\end{table}
\begin{figure}[!htb]	
\centering
\includegraphics[width=10cm]{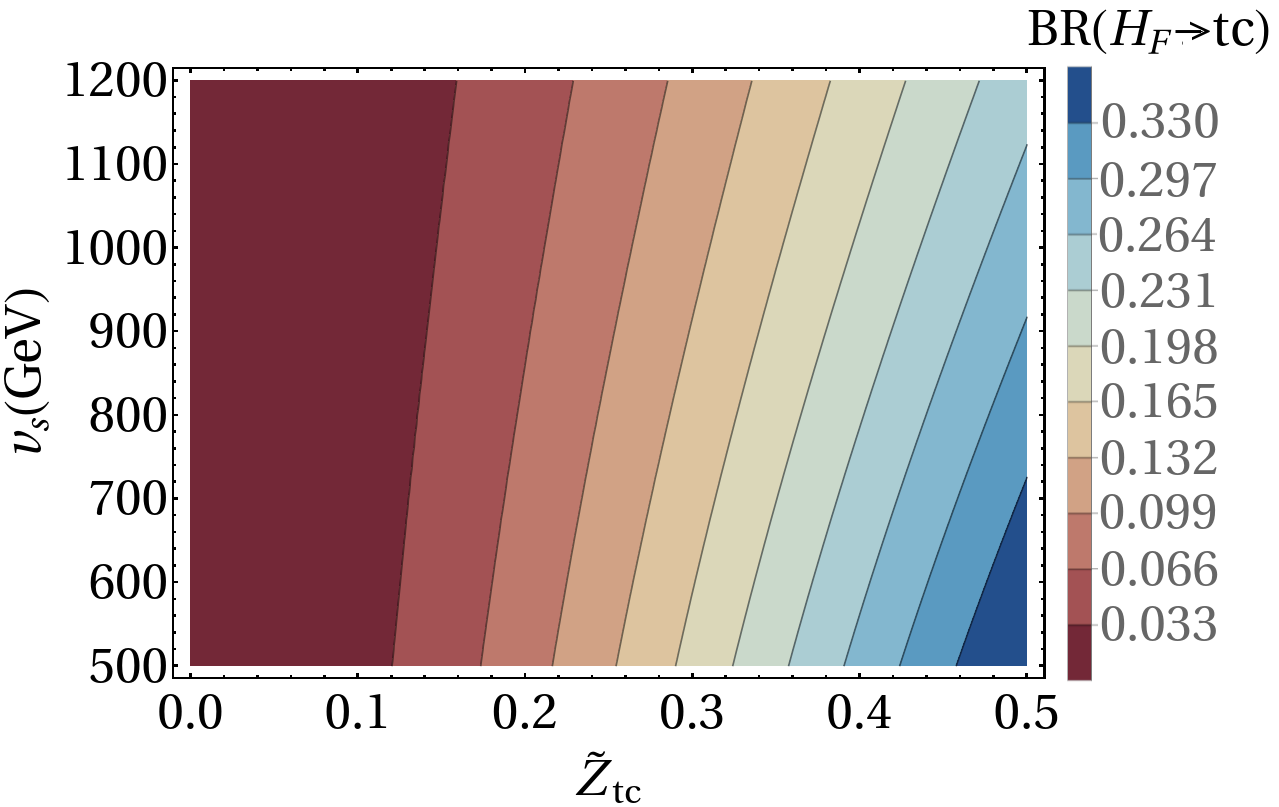}
\caption{${\rm BR}(H_F\to tc)$ as a function of the singlet VEV $v_s$ and the $\tilde{Z}_{tc}$ matrix element.}
\label{BRHF-tc}
\end{figure}

\begin{table}[ht]
\begin{center}\scalebox{1.0}{
\begin{tabular}{|c|c|c|c|c|c|c|c|c|c|c|c|c|c|}
\hline
BPs  [GeV] &\multicolumn{3}{c|}{ {\rm BR}s and cross sections [pb] }\\
\hline
& ${\rm BR}(H_F \to t c)$ & $ \sigma( pp \rightarrow  H_F$) & $\sigma(pp \to H_F\to t c, t \to  \ell\nu_{\ell} b)$ \\
\cline{2-4}
\rule{0pt}{1ex}
BP1 ($M_{H_F}=800$)   &  $0.0140$  & $0.41$ &  $1.34\times 10^{-3}$\\
\rule{0pt}{1ex}
BP2  ($M_{H_F}=900$)  &$0.0134$  & $0.21$ &  $6.23\times 10^{-4}$ \\
\rule{0pt}{1ex}
BP3  ($M_{H_F}=1000$)  & $0.0133$ & $0.11$ &   $3.32\times 10^{-4}$\\
\hline
\end{tabular}}
\end{center}
\caption{The ${\rm BR}(H_F \to tc)$ and cross sections for the processes $ pp \rightarrow  H_F$ and $\sigma(pp \to H_F$ $\to t c, t \to \ell\nu_{\ell}b)$ for three BPs (BP1, BP2 and BP3) used in the remainder of the paper. }
\label{XSHFtc}
\end{table}
The corresponding cross sections for the benchmark points used in this paper are presented in Table \ref{XSHFtc}. Meanwhile, the ${\rm BR}(H_F\to tc)$ as a function of the singlet VEV $v_s$ and the $\tilde{Z}_{tc}$ matrix element is shown in Fig \ref{BRHF-tc}.  We observe ${\rm BRs}(H_F\to tc)$ quite large $\mathcal{O}(0.1)$, which comes because the couplings $H_F WW$ and $H_F ZZ$ are suppresed, which allows the opening of the $tc$ channel.  

In this analysis, the main SM background comes from the final state of $bj\ell\nu_\ell$, whose source arises mainly from $Wjj+Wb\bar{b}$, $tb+tj$. Another important background is $t\bar{t}$ production, where either one of the two leptons is missed in the semi-leptonic top quark decays, or two of the four jets are missed when one of the top quarks decays semi-leptonically. The cross sections of the dominant SM background are shown in Table \ref{XSSMBGD}.


\begin{table}[ht]
\begin{centering}
\begin{tabular}{|c|c|}
\hline 
SM backgrounds & Cross section {[}pb{]}\tabularnewline
\hline 
\hline 
$pp\to Wjj+Wb\bar{b}\,(W\to\ell\nu_{\ell})$ & $3245$\tabularnewline
\hline 
$pp\to tb+tj\,(t\to\ell\nu_{\ell}b)$ & $1.61$\tabularnewline
\hline 
$pp\to t\bar{t}\,(t\to\ell\nu_{\ell}b,\,t\to q_{i}q_{j}b)$ & $65.50$\tabularnewline
\hline 
\end{tabular}
\par\end{centering}
\caption{Cross section of the dominant SM background processes.}
\label{XSSMBGD}
\end{table}

\begin{figure}[h!]	
 \centering
 \subfigure[ ]{\includegraphics[width=6.4cm]{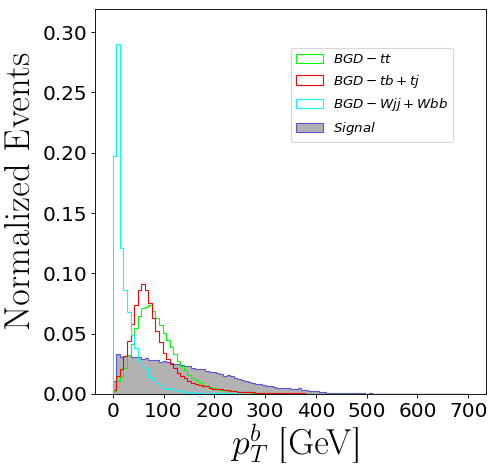}}
 \subfigure[ ] {\includegraphics[width=6.5cm]{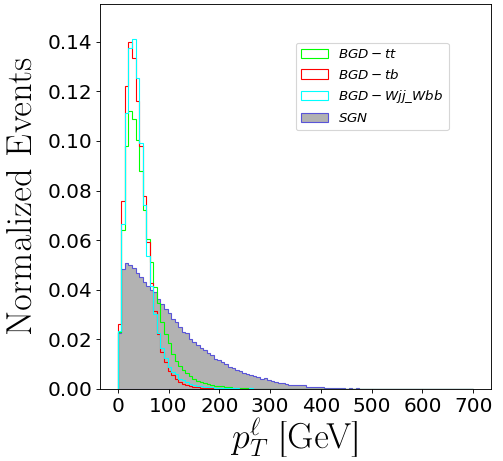}}\\
 \subfigure[ ] {\includegraphics[width=6.5cm]{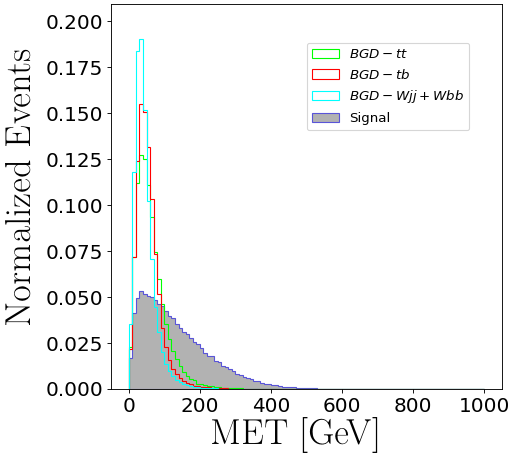}}
 \subfigure[ ] {\includegraphics[width=6.35cm]{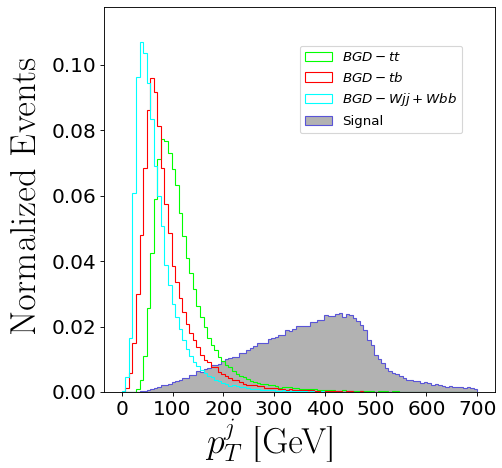}}
 \caption{Normalized transverse momentum distributions associated to the top decay: (a) leading b-jet, (b) leading charged lepton, (c) tranverse missing energy due to undetected neutrinos; (d) transverse momentum distribution of the c-jet.}\label{distributions}
\end{figure}
\begin{figure}[h!]	
\centering
\includegraphics[width=7cm]{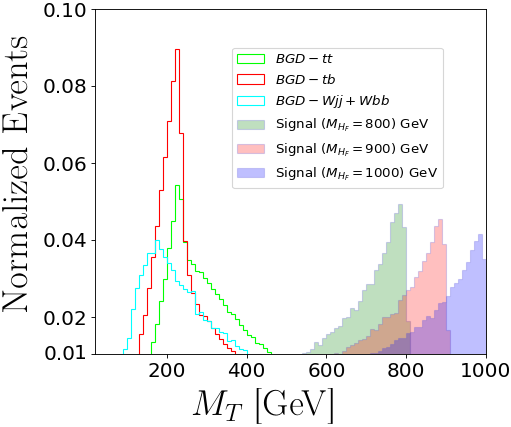}
\caption{Normalized transverse mass $M^T=\sqrt{2 p^T_\ell \slashed{E}_T (1-\cos \phi_{\ell \slashed{E}_T}) }$ for both background and signal processes. We have considered $M_{H_F}=800,\,900,\,1000$ GeV.}
\label{MT}
\end{figure}

Fig.~\ref{distributions} shows the kinematic distributions generated both by the signal (for $M_{H_F}=800$ GeV, $v_s=1000$ GeV) and background processes, namely, the transverse momentum of the particles produced by the decay of the top quark: (a) leading $b$-jet, (b) the charged lepton, (c) the missing energy transverse (MET) due to the neutrino in the final state. The transverse momentum of the leading light jet is shown in (d). A remarkable fact is the difference in transverse masses between the background and signal processes. So, we present in Fig \ref{MT} the transverse mass for both reactions,  which is the most important confirmation of the signal.

The following acceptance and kinematic cuts imposed to study possible evidence of the $H_F\to tc$ ($M_{H_F}=800$ GeV) at the LHC are as follows.
\begin{itemize}
\item  We requiere two jets with $|\eta^j|<2.5$ and $p_T^j>30$ GeV, one of them is tagged as a $b$-jet.
\item  We require one isolated lepton ($e\,\text{or}\,\mu$) with $|\eta^{\ell}|<2.5$ and $p_T^{\ell}>30$ GeV.
\item Since an undetected neutrino is included in the final state, we impose the cut  MET$>40$ GeV.
\end{itemize}
 
Finally, we impose a cut on the transverse mass $M^T \equiv M^T_{\ell \slashed{E}_T}=\sqrt{2 p^T_l \slashed{E}_T (1-\cos \phi_{\ell \slashed{E}_T}) }$ as $|M^T-M_{H_F}|<50$ GeV to enhance the signals.

\begin{figure}[H]
\centering
\subfigure[ ]{\includegraphics[scale = 0.15]{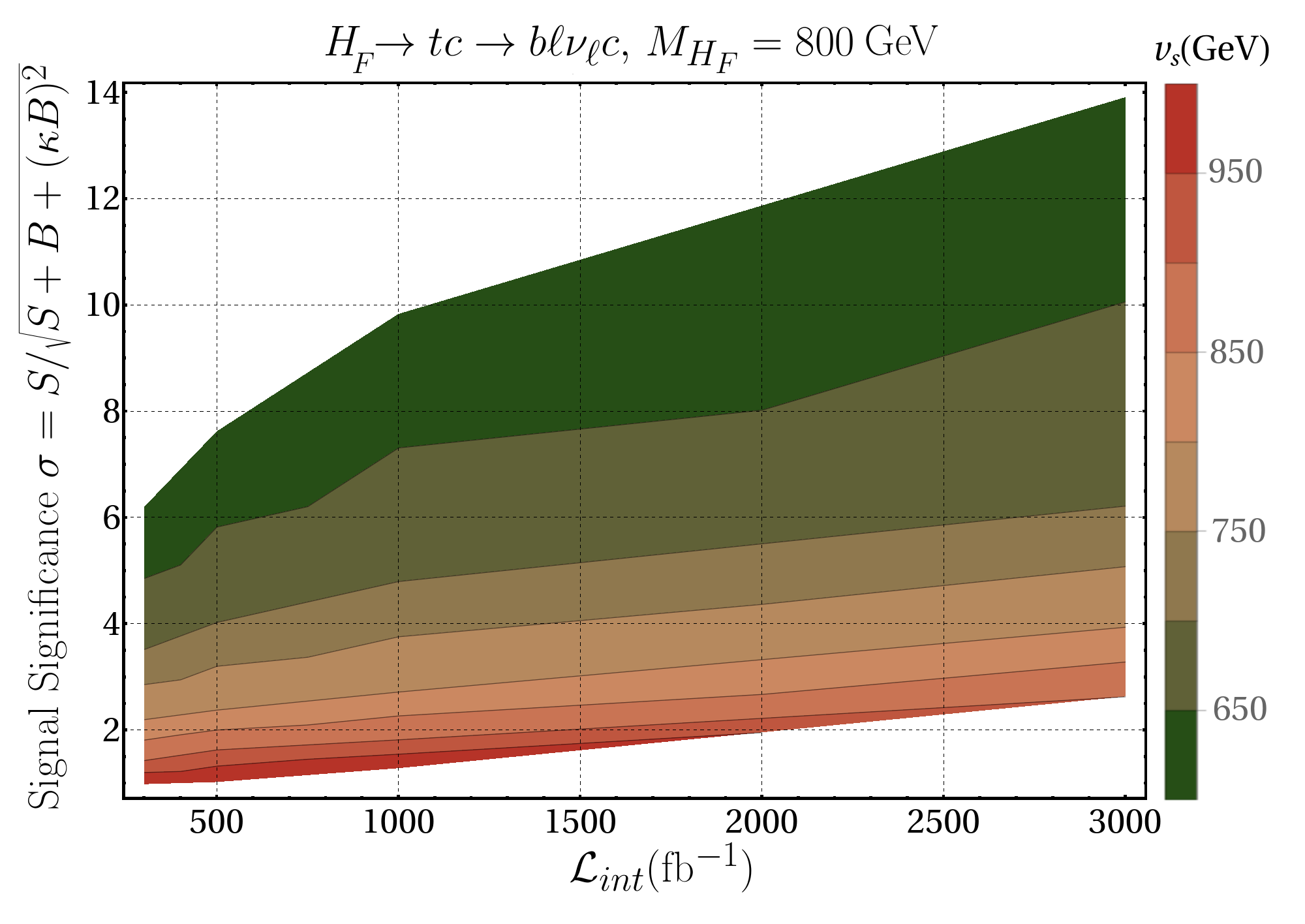}}
\subfigure[ ]{\includegraphics[scale = 0.15]{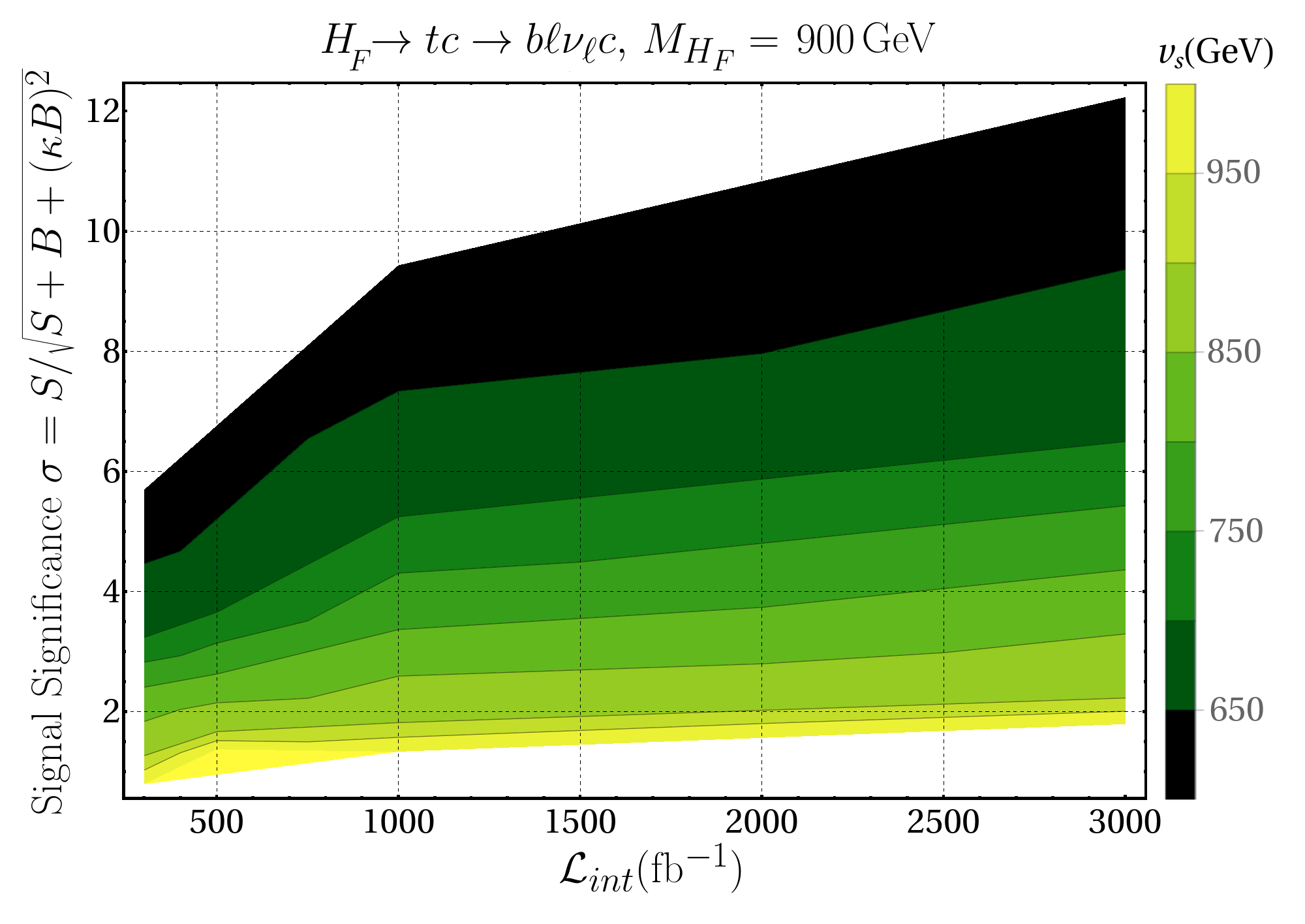}}
\subfigure[ ]{\includegraphics[scale = 0.15]{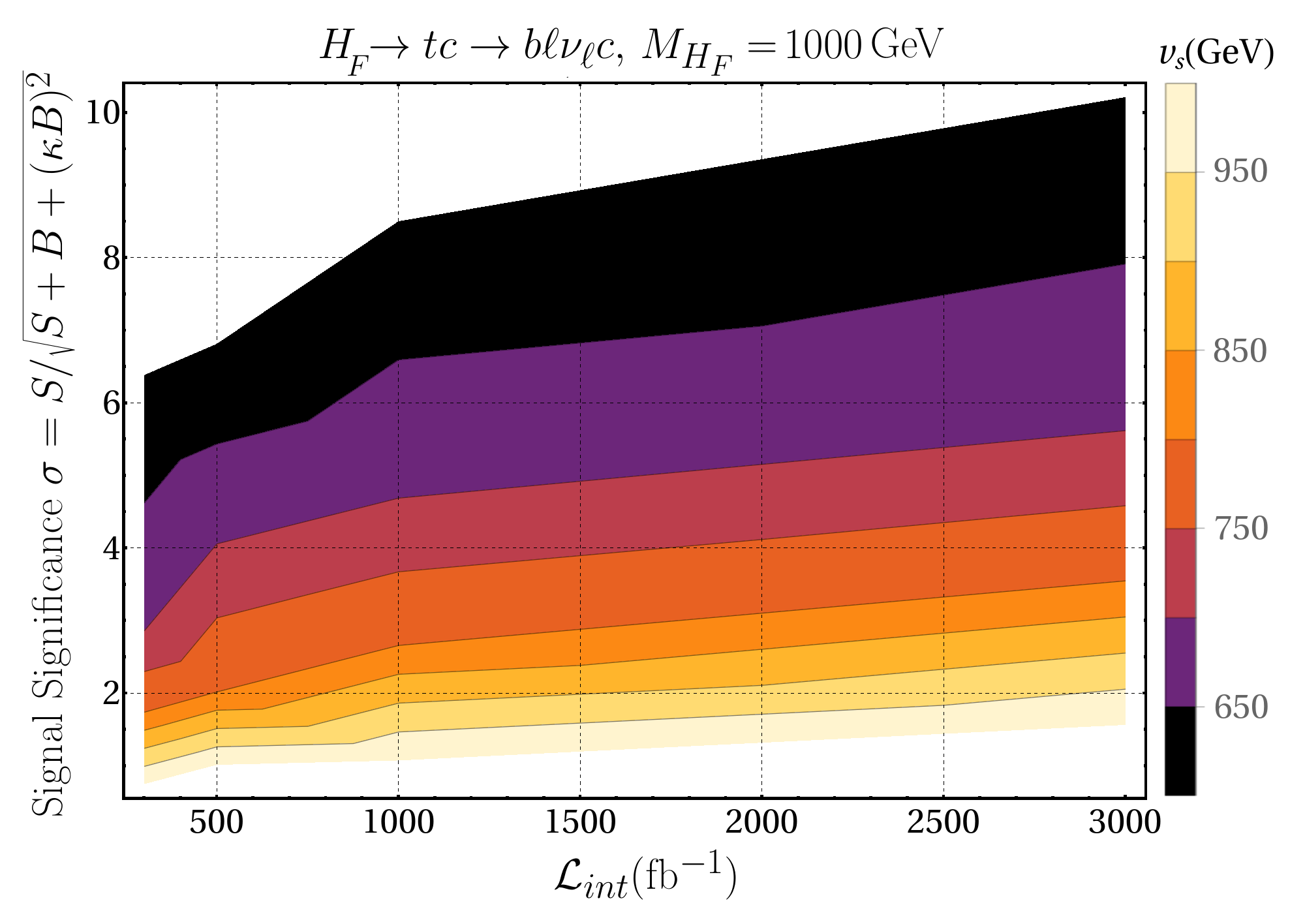}}
 \caption{Contour plots for the signal significance as a function of the integrated luminosity and 
the singlet VEV $v_s$. (a) $M_{H_F}=800$ GeV, (b) $M_{H_F}=900$ GeV, 
(c) $M_{H_F}=1000$ GeV. In these results we consider a systematic uncertainty $\kappa=5\%$}
\label{significance1}
\end{figure}

Fig.~\ref{significance1} displays the contour plots of the signal significance as a function of the integrated luminosity $\mathcal{L}_{\text{int}}$ and the singlet scalar VEV $v_s$, for $M_{H_F}=800,\,900,\,1000$ GeV. Once $\mathcal{L}_{\text{int}}=300 \rm fb^{-1}$ of accumulated data is achieved and assuming $v_s=640$ GeV (625 GeV, 620 GeV), we find that the LHC would have the possibility of exploring a detectable Flavon $H_F$ of mass 800 GeV (900 GeV, 1000 GeV). Even more promising results could be found in the HL-LHC era, which could corroborate the possible findings of the LHC regarding the $H_F\to tc$ process.
\section{Conclusions}
\label{se:concl}

The hierarchical structure and peculiar pattern of quark and lepton masses 
in the SM  have been a long standing  issue coined as the 
`flavor puzzle'. Various interesting beyond the SM proposals have been 
suggested to resolve this riddle. Among these, the one by  
 Froggatt and Nielsen is arguably one of the most
fascinating ones. Herein, the scalar sector predicts one singlet
complex scalar $S_F$ which is charged under a new $U(1)_F$ flavor symmetry
(which is softly broken). After EWSB and $U(1)_F$ breaking, the 
mixing between the SM Higgs doublet with the real part of the $S_F$ singlet produces two physical
scalars, $h$ and $H_F$, where $h$ is identified as the SM-like Higgs boson (discovered in 2012) 
while $H_F$ is an additional CP-even (so-called) Flavon with mass ${\cal O}$(1~TeV). (The imaginary 
part of $S_F$ is identified as the CP-odd heavy Flavon $A_F$.) The (pseudo)scalar sector
of this model is controlled by two parameters: the Flavon VEV $v_s$ 
and the mixing angle $\alpha $. The structure of various Yukawa couplings of this model
is such that one can have FCNCs involving the 
two new heavy (pseudo)scalars $(H_F \& A_F)$ even at tree-level. The corresponding contributions to FCNC 
processes thus attract severe constraints from various low energy
flavor physics data. Therefore, in our analysis of such a scenario, we have considered all possible experimental (as well as theoretical)  limits on the model parameters $v_s $
and $\alpha $. With the LHC currently running at CERN, it  is very tempting to utilize the ongoing (Run 3) and
future (HL-LHC) stages of the machine to explore the signature of such heavy flavons. 

{In this paper, our primary focus was on the CP-even heavy Flavon denoted as $H_F$. We explored its discovery potential at the LHC by investigating its production through gluon-gluon fusion followed by its subsequent decays. We considered various decay modes for it, including into two SM Higgs bosons and two SM (neutral) gauge bosons.
By studying these different decay channels and considering the corresponding signatures at HL-LHC (with $\sqrt{s} =14$ TeV),  assuming a  luminosity of 3000 fb$^{-1}$, we were able to confirm the discovery potential ($5\sigma$) of the CP-even heavy Flavon $H_F$ at the LHC through these SM signatures.} {In addition, we explored the flavor-changing $H_F\to tc$ decay, specific to our model, which is predicted to arise when $M_F\gtrsim m_t$. This decay can be as large as $\mathcal{O}(0.1)$  because the $H\to VV$ $(V=W,\,Z)$ decays are heavily suppressed once the $tc$ channel is opened. We thus showed that this non-SM channel offers an alternative opportunity to test our model, even at the standard LHC. Once a integrated luminosity of 300 fb$^{-1}$ is reached, we find in this channel a signal significance of up to $6\sigma$ for masses of the Flavon between $800-1000$ GeV.}

We have obtained such results following a thorough numerical 
analysis emulating both the aforementioned signal and the most relevant (ir)reducible backgrounds accounting for hard scattering, 
parton shower, hadronization and detector effects.   
We thus advocate that the experimental collaborations at the LHC, specifically, the multipurpose ones (ATLAS and CMS), tackle this search, as its results
can lead to a better understanding of the origin and solution of the flavor puzzle in  the SM. This should be facilitated by having implemented 
the advocated model in standard computational tools, which are available upon request.

%
%

\subsection*{Acknowledgments}
SM acknowledges funding  from the STFC Consolidated Grant ST/L000296/1 and is partially supported  through the NExT Institute. NK would like to acknowledge support from the DAE, Government of India and the Regional Centre for Accelerator-based Particle Physics (RECAPP), HRI (HRI-RECAPP-2023-05).
The work of Marco A. Arroyo-Ure\~na is supported by ``Estancias posdoctorales por M\'exico (CONACYT)". JLD-C acknowledges the support of SNI (M\'exico) and VIEP (BUAP). The work of AC is funded by the Department of Science and Technology, Government of India, under Grant No. IFA18PH224 (INSPIRE Faculty Award).

\bibliographystyle{apsrev4-1}
\bibliographystyle{utphys}
\bibliography{REFc}

\end{document}